\documentclass[pre,showpacs,showkeys,preprintnumbers,amsmath,amssymb,amsfonts,amsfontssuperscriptaddress,twocolumn]{revtex4}
\pdfoutput=0 
\usepackage{makeidx}
\usepackage{graphicx}
\usepackage{amssymb,amsfonts,amsmath}
\usepackage{verbatim}
\usepackage{epstopdf}
\usepackage{nameref,hyperref} 
\usepackage{color}
\usepackage{bm}

\def\Conv{{\rm Conv}}
\def\max{{\rm max}}
\def\diam{{\rm diam}}
\def\vol{{\rm vol}}

\def\E{{\mathbb E}}
\def\P{{\mathbb P}}

\def\x{\bm{x}}
\def\R{{\mathbb R}}

\begin{document}
\title{Unravelling intermittent features in single particle trajectories \\ by a local convex hull method}
\author{Yann Lanoisel\'ee}
  \email{yann.lanoiselee@polytechnique.edu}
\affiliation{Laboratoire de Physique de la Mati\`{e}re Condens\'{e}e (UMR 7643), \\ 
CNRS -- Ecole Polytechnique, University Paris-Saclay, 91128 Palaiseau, France}

\author{Denis~S.~Grebenkov}
 \email{denis.grebenkov@polytechnique.edu}
\affiliation{Laboratoire de Physique de la Mati\`{e}re Condens\'{e}e (UMR 7643), \\ 
CNRS -- Ecole Polytechnique, University Paris-Saclay, 91128 Palaiseau, France}

\affiliation{Interdisciplinary Scientific Center Poncelet (ISCP),%
\footnote{International Joint Research Unit -- UMI 2615 CNRS/ IUM/ IITP RAS/ Steklov MI RAS/ Skoltech/ HSE, Moscow, Russian Federation} \\
Bolshoy Vlasyevskiy Pereulok 11, 119002 Moscow, Russia}

\date{Received: \today / Revised version: }

\begin{abstract}
We propose a new model-free method to detect change points between
distinct phases in a single random trajectory of an intermittent
stochastic process.  The local convex hull (LCH) is constructed for
each trajectory point, while its geometric properties (e.g., the
diameter or the volume) are used as discriminators between phases.
The efficiency of the LCH method is validated for six models of
intermittent motion, including Brownian motion with different
diffusivities or drifts, fractional Brownian motion with different
Hurst exponents, and surface-mediated diffusion.  We discuss potential
applications of the method for detection of active and passive phases
in the intracellular transport, temporal trapping or binding of
diffusing molecules, alternating bulk and surface diffusion, run and
tumble (or search) phases in the motion of bacteria and foraging
animals, and instantaneous firing rates in neurons.
\end{abstract}





\maketitle

\section{Introduction}
	
Identifying transport mechanisms in complex systems and building their
mathematical models is a challenging problem of paramount importance,
especially in microbiology.  While numerous experimental observations
clearly revealed anomalous diffusion features in the intracellular
transport, their microscopic biophysical origins remain debatable
\cite{Tolic04,Golding06,Wilhelm08,Szymanski09,Metzler09,Sackmann10,Jeon11,Bertseva12,Bressloff13}.
The significant technological progress in optical microscopy over the
past decade has opened new opportunities in the statistical analysis
of single particle trajectories but also raised challenges in the
biophysical interpretation of a small number of random trajectories.
Most theoretical works remain focused on the simplest observable, the
time-averaged mean square displacement (TA MSD), which has been
studied for many anomalous diffusion models (see
\cite{Gal2013a,Metzler14,Kepten15} and references therein).  However,
since this observable does not discriminate between various models,
other observables have been proposed, e.g., first passage times
\cite{Condamin2008,Kenwright2012}, the maximum excursion \cite{Tejedor2010},
fundamental moments \cite{Thiel2013}, and the fractal dimension of the
explored space \cite{Meroz2013} (see also \cite{Meroz2015} and
references therein).  At the level of a single trajectory, elaborate
statistical tools have been developed to recognize fractional Brownian
motion (fBm) \cite{Burnecki2012d}, to distinguish between fBm and
Continuous Time Random Walk (CTRW)
\cite{Magdziarz2009,Weiss2013,Ernst2014}, and to reveal ergodicity
breaking \cite{Magdziarz2011,Lanoiselee2016}.  In addition, various
statistical tests from econometric studies can be employed in the
biophysical context, e.g., testing the Markov hypothesis
\cite{Chen2012}.  Once an appropriate model of anomalous diffusion is
identified, one can estimate its parameters that are related to the
intracellular transport, and then rely on the theoretical knowledge
about the model to predict biological implications, e.g., biochemical
reaction rates, translocation or transcription mechanisms, drug
delivery, etc. \cite{Bressloff13,Benichou14}.

The advantages of conventional models such as fBm or CTRW are the
reduced number of model parameters and a possibility to describe the
underlying process in a simplified but unified mathematical language.
This is also one of the major limitations of such effective
descriptions.  Many biological processes are intermittent, i.e., they
switch between two or several phases.  The most common examples are
the run-and-tumble motion of bacteria
\cite{Berg1972,Berg,Tailleur2008} and foraging strategies of animals,
e.g., predators who adopt hunting strategies by alternating between
slow careful search and fast displacement (see the review
\cite{Benichou2011} and references therein).  In microbiology,
examples include alternating phases of three-dimensional bulk
diffusion and one-dimensional sliding along DNA chains of DNA-binding
proteins \cite{Richter74,Berg81,vonHippel89,Loverdo09}, temporal
trapping of tracers in polymer cages \cite{Wong04,Fodor16},
binding/unbinding of macromolecules to form temporal ligand-receptor
pairs or to neutralize pathogens by antibodies
\cite{Shoup82,Zwanzig91,Saxton96,Bongrand99,Michelman09,Torreno-Pina14},
switching between distinct conformational states
\cite{Calderon10,Galanti16}, alternating phases of active and passive
vesicle transport by motor proteins
\cite{Saxton94,Caspi00,Arcizet08,Brangwynne09,Katrukha2017},
alternating phases of bulk and surface diffusion
\cite{Levitz05,Benichou10,Benichou11,Rupprecht12,Rojo13}, etc.  The
intermittent character of the motion makes the statistical analysis
and biophysical interpretation of single particle trajectories even
more challenging.  If the intermittence is ignored, switching between
several phases can be (mis)interpreted as a single effective phase
with peculiar properties.  For instance, switching between active
(ballistic, $\alpha = 2$) and passive (diffusive, $\alpha = 1$) motion
of a vesicle can be effectively understood as super-diffusion, with an
intermediate scaling exponent $1 < \alpha < 2$.  Moreover,
intermittency can be also misleadingly interpreted as non-stationarity
or non-ergodicity of the process according to statistical tests.
Identifying change points between distinct phases and their durations
from a single trajectory is therefore an important statistical problem
that can bring a finer, more accurate description of the underlying
biological process.

Different statistical methods have been developed to detect change
points of an intermittent stochastic process.  Traditionally, Bayesian
methods based on prior knowledge on the motion are preferred
\cite{Adams2007,Barber2011,Turkcan2013,Masson2014,Bosch2014,Hinsen2016,Ruggieri2016}.
The prior knowledge is represented by the probability
$P(\x_1,\ldots,\x_N ~|~ a_1,\ldots,a_K)$ of observing a trajectory
$\x_1,\ldots,\x_N$ with $N$ points, given the parameters
$a_1,\ldots,a_K$ of the model.  In other words, one defines a specific
model-based functional of the trajectory to ``process'' the observed
data.  Once the model is chosen, its parameters $a_1,\ldots,a_K$ can
be found by maximizing the likelihood of the observed trajectory
according to the Bayesian rules.  Although the Bayesian methods are
statistically efficient, their practical implementation is rather time
consuming, especially for long trajectories, while the results can be
strongly biased by the choice of the underlying model made by
researchers.  In addition, when dealing with intermittent processes,
one may need to probe many combinations of possible microscopic models
for each phase.  In this light, model-free methods can be preferred
for the analysis of complex dynamics such as the intracellular
transport.

In this paper, we address the challenging question of detection of
change points between distinct phases in a single random trajectory
without prior knowledge of the underlying stochastic model.  The
phases are distinguished by their dynamical properties such as
different (i) diffusivities, (ii) drifts, (iii) auto-correlations of
increments, (iv) distributions of increments, (v) dimensionalities
(e.g., bulk/surface), (vi) isotropic/anisotropic character, or (vii)
space accessibility (e.g., restricted character due to reflecting
obstacles or confinement).  The basic idea of such model-free methods
consists in considering a {\it local} functional of the trajectory,
$Q(n)$, which depends on a relatively small number of points around
the point $\x_n$.  When applied to successive points along the
trajectory, this local functional transforms the trajectory into a new
time series, which can then be used either to characterize the
dynamics (e.g., to get local drift or diffusivity), or to discriminate
between different phases of the motion.  For instance, the points
$\x_n$ with $Q(n)$ below some threshold can be assigned to one phase
while the remaining points are assigned to the other phase.  In
contrast to Bayesian methods, the functional $Q(n)$, which is used to
process the data, does not depend on the model, though the efficiency
of this binary classification clearly depends on the choice of the
functional $Q(n)$.  The simplest choices,
\begin{eqnarray}
Q_\mu(n) & =& \frac{1}{2\tau} (\x_{n+\tau} - \x_{n-\tau}),  \\  
\label{eq:Qsigma}
Q_\sigma(n) & =& \frac{1}{2\tau+1} \sum\limits_{k=-\tau}^{\tau} \|\x_{n+k+1} - \x_{n+k}\|^2 , 
\end{eqnarray}
are respectively the drift estimator and the variance estimator (over
a window of size $2\tau+1$) of isotropic Brownian motion.  Another
standard functional is the angle between successive increments of the
trajectory, which is often used to detect ballistic parts of the
motion \cite{Katrukha2017}.  Having a differential form (i.e.,
involving differences between points), these functionals are rather
sensitive to noise, exhibit large fluctuations and thus yield large
statistical uncertainties.  This drawback can be reduced by increasing
the window size $\tau$.  However, too large $\tau$ would make the
functional $Q(n)$ less sensitive to phase alternations, especially for
short phases.  Several improvements have been proposed, e.g., the root
mean square estimator for the variance \cite{Grebenkov2011} or the
scaling exponent extracted from the local TA MSD \cite{Arcizet08}.
Nevertheless, these improved estimators still have the differential
form and thus remain sensitive to noise.

To overcome this fundamental limitation, we propose the geometric
properties of a {\it local convex hull} (LCH) as robust discriminators
of different phases.  In contrast to earlier used differential-like
estimators, the convex hull is intrinsically an integral-like
characteristic that is thus less sensitive to noise, as shown below.
This new method is applicable for the analysis of single trajectories
of any dimensionality.  The method is model-free because it relies
exclusively on geometric properties of the trajectory.  We demonstrate
its efficiency in recognizing different intermittent dynamical
scenarios inspired from biology.

\section{Local Convex Hull Method}
\label{sec:Convex_Hull}

The convex hull of a finite set of points,
$\{\x_1,\ldots,\x_n\}\subset \R^d$, is the set of all convex
combinations based on these points:
\begin{equation}
\Conv(\x_1,\ldots,\x_n) = \left\{ \sum\limits_{k=1}^n \alpha_k \x_k ~|~ \alpha_k\geq 0, ~ \sum\limits_{k=1}^n \alpha_k = 1\right\}.
\end{equation}
In simpler terms, it is the minimal convex shape that encloses
all the points $\x_1,\ldots,\x_n$.  In the planar case, the convex
hull of a finite set of points $\{\x_1,\ldots,\x_n\}$ is a convex
polygon whose vertices are some of the points in this set.  The
construction of a convex hull is thus reduced to identifying these
points in a clockwise or counter-clockwise order.  In higher
dimensions, the convex hull is a convex polytope (e.g., a convex
polyhedron in three dimensions) whose vertices are some of the points
in the input set.  Among several efficient algorithms
\cite{ORourke,deBerg}, we chose the quickhull algorithm \cite{Barber96}
because of its available implementation in Matlab as functions
``convhull'' (in two dimensions) and ``convhulln'' (in higher
dimensions).  The convex hull has been used for home range estimations
of animal territories \cite{Worton1995,Getz2004a,Randon-Furling2009a}
and for computation of fractal dimensions \cite{Normant1991a}.  Some
basic properties of the convex hull applied to stochastic processes
are summarized in Appendix \ref{sec:Aproperties}.

\begin{figure}
\begin{center}  
\includegraphics[width=55mm]{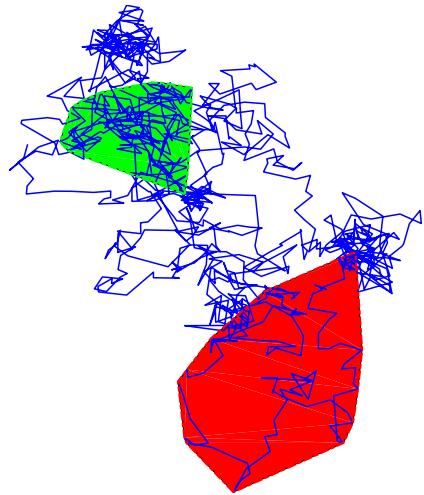}  
\end{center}
\caption{
Illustration of the local convex hull method applied to an
intermittent planar trajectory, alternating a ``fast'' phase of
Brownian motion (with $D = 1/2$) and a ``slow'' phase of an
Ornstein-Uhlenbeck process (with $D = 1/2$ and $k = 0.1$, see
Sec. \ref{sec:OU}), of equal durations $T_1 = T_2 = 40$.  Only two
LCHs (based on $2\tau+1 = 41$ points) are shown by shadowed regions
that correspond to the fast (dark red) and slow (light green) phases.}
\label{fig:conv_hull}
\end{figure}

We propose to consider the diameter and the volume of the local convex
hull, computed over $2\tau+1$ trajectory points $\x_{n-\tau},
\x_{n-\tau+1}, \ldots, \x_{n+\tau}$, as two functionals:
\begin{eqnarray}  
\label{eq:Qd}
Q_d(n) & = & \diam(\Conv(\x_{n-\tau},\ldots,\x_{n+\tau})) , \\
\label{eq:Qv}
Q_v(n) & = & \vol_d(\Conv(\x_{n-\tau},\ldots,\x_{n+\tau}))   
\end{eqnarray}
(the volume is replaced by the area in two dimensions).  Note that the
diameter of the LCH is simply the largest distance between any two
points in the sequence $\{\x_{n-\tau}, \ldots, \x_{n+\tau}\}$, i.e.,
it is particularly easy to compute.  Any significant change in the
dynamics (i.e., switching between phases) is expected to be reflected
in a notable change in the geometric form of the trajectory that is
captured here by the local convex hull (Fig. \ref{fig:conv_hull}).
For instance, an increase in the diffusion coefficient or in the drift
leads to larger increments and thus a larger LCH, with larger $Q_d(n)$
and $Q_v(n)$.  In addition, anisotropic features of the motion, such
as dimensionality reduction or presence of a reflecting wall, would be
reflected in an anisotropic shape and in a reduced volume of the LCH.
Depending on the observable chosen, the identification will be more
sensitive to a particular aspect of the motion.  The diameter is more
sensitive to correlations and diffusion coefficients, whereas the
volume is more sensitive to changes in the dimensionality and
anisotropy.  The window size $\tau$ is a parameter of the method that
controls the compromise between reactivity and robustness in change
points detection: smaller $\tau$ facilitates identification of short
phases (the method is more reactive) but makes it less robust, as the
estimators become more sensitive to fluctuations, noises, and
outliers.

According to Eqs. (\ref{eq:Qd}, \ref{eq:Qv}), the point $\x_n$
contributes to $2\tau+1$ estimators $Q_d(k)$ (or $Q_v(k)$) with $k\in
[n-\tau,n+\tau]$.  It is therefore natural to classify the point
$\x_n$ by using all the estimators $Q_d(k)$ (or $Q_v(k)$), to which it
contributed.  We define thus the discriminator $S_d(n)$ as a weighted
sum of $Q_d(k)$ over $\tau$ left and $\tau$ right neighbors of $\x_n$
(similar for $S_v(n)$).  The simplest choice consists in setting equal
weights to each contribution:
\begin{eqnarray}
\label{eq:Sd}
S_d(n) &=& \frac{1}{2\tau+1}\sum\limits_{k=n-\tau}^{n+\tau} Q_d(k)  ,\\
\label{eq:Sv}
S_v(n) &=& \frac{1}{2\tau+1}\sum\limits_{k=n-\tau}^{n+\tau} Q_v(k)  .
\end{eqnarray}
Other weighting schemes (e.g., exponential) are as well possible.

The LCH method for change points detection consists in two steps.  In
the first step, for a given trajectory with $N$ points, discriminators
$S_d(n)$ and $S_v(n)$ are computed for all points $\x_n$ with $n$ from
$2\tau+1$ to $N-2\tau$ (the first and last $2\tau$ points of the
trajectory are discarded and remain unclassified).  In other words,
these discriminators transform the trajectory in $\R^d$ into two new
one-dimensional time series.  Due to the integral-like nature of the
LCH, these times series are expected to be less erratic than
trajectories and more sensitive to changes in the dynamics.  In the
second step, these time series are used to detect change points
between two phases.  In the basic setting, the points $\x_n$ with
$S_d(n) > S_d$ (or $S_v(n) > S_v$) are classified as belonging to the
``fast'' phase whereas the points $\x_n$ with $S_d(n) \leq S_d$ (or
$S_v(n) \leq S_v$) as belonging to the ``slow'' phase.  The choice of
the threshold value $S_d$ (resp. $S_v$) is crucial for an efficient
detection of change points.  Without {\it a priori} knowledge of the
underlying stochastic process, we set $S_d$ (resp. $S_v$) to be the
empirical mean of $S_d(n)$ (resp., of $S_v(n)$) over the trajectory,
i.e.,
\begin{equation}
\label{eq:threshold}
\begin{split}
S_d & = \frac{1}{N-4\tau} \sum\limits_{n=2\tau+1}^{N-2\tau} S_d(n) , \\
S_v & = \frac{1}{N-4\tau} \sum\limits_{n=2\tau+1}^{N-2\tau} S_v(n) . \\
\end{split}
\end{equation}
In Sec. \ref{sec:parameters}, we discuss this point and possible
improvements.

We emphasize that the LCH method can be applied to a {\it single}
trajectory of essentially {\it any} intermittent stochastic process.
In turn, the quality of change points detection depends on how
distinct the geometric properties of two phases are, as well as on
durations of these phases.  We also stress that the method does not
rely on specific properties of the underlying stochastic model, nor
does it identify the properties of each phase.  This analysis can be
done by conventional methods after the change points detection.
Finally, the same two phases do not need to be repeated along the
trajectory.  Although many distinct phases could in principle be
present in a single trajectory, we focus on the particular case of two
alternating phases (that we qualitatively called ``fast'' and
``slow'').

Note that the convex hull diameter can be considered as an alternative
to the maximal excursion, i.e., the greatest distance from the origin
that a particle reaches until time $t$ \cite{Bidaux99}.  The latter
has been shown to have a narrower distribution than the conventional
TA MSD, and thus proposed as an estimator of the scaling exponent that
can be applied to single particle trajectories \cite{Tejedor2010}.

\section{Numerical validation}
\label{sec:Models_to_test}

In this section we investigate the efficiency of the LCH method by
simulating six models of intermittent processes with the following
phases:
\begin{enumerate}
\item
two planar Brownian motions with distinct diffusion coefficients $D_1$
and $D_2$ (change in diffusivity);

\item
two planar Brownian motions with the same diffusion coefficient $D$,
with and without drift (change in directionality);

\item
two planar fractional Brownian motions with distinct Hurst exponents
$H_1$ and $H_2$ (change in auto-correlations);

\item
planar Brownian motion and Ornstein-Uhlenbeck process (change in
auto-correlations);

\item
planar Brownian motion and exponential flights (change in the
distribution of increments);

\item
surface-mediated diffusion, with alternating phases of
three-dimensional bulk diffusion and two-dimensional surface diffusion
(change in dimensionality).

\end{enumerate}
Although many other intermittent processes could be considered, we
focus on the above cases as representative examples.

The simulations are performed as follows.  First, we generate a
thousand random trajectories, each with a thousand points (i.e., $N =
1000$).  The intermittence is implemented by partitioning these points
into two alternating phases by assigning random, exponentially
distributed durations, with prescribed mean durations $T_1$ and $T_2$
of two phases.  Throughout this section, we consider two phases of
equal mean duration, $T_1 = T_2 = T$, while the situation of unequal
phases is discussed in Sec. \ref{sec:phase_duration}.  A white
Gaussian noise of standard deviation $\sigma_n$ is added to each
trajectory point in order to check the robustness of the LCH method to
measurement noises.  The noise level $\sigma_n$ is set to be
proportional to the empirical standard deviation $\sigma$ of
increments, computed for each trajectory.  Note that the noise has a
stronger impact onto the ``slow'' phase than onto the ``fast'' phase.
Second, we fix the window size $\tau$ and compute the diameter,
$Q_d(n)$, and the volume, $Q_v(n)$, of the LCH at all time steps $n$
from $\tau+1$ to $N-\tau$ for each trajectory.  These estimators are
then transformed into the weighted discriminators $S_d(n)$ and
$S_v(n)$ according to Eqs. (\ref{eq:Sd}, \ref{eq:Sv}) for $n$ from
$2\tau+1$ to $N-2\tau$.  The default value of the window size $\tau$
is $10$, other sizes being considered in Sec. \ref{sec:parameters}.
Third, for each trajectory, we assign the points with $S_d(n) > S_d$
(resp. $S_v(n) > S_v$) to the ``fast'' phase and the points with
$S_d(n) \leq S_d$ (resp. $S_v(n) \leq S_v$) to the ``slow'' phase,
where $S_d$ (resp. $S_v$) is the empirical mean diameter (resp.,
volume) of the LCH given by Eq. (\ref{eq:threshold}).  We stress that
the threshold value is computed separately for each trajectory.  In
this way, we obtain two {\it unrelated} binary classifications, the
one based on the diameter and the other based on the volume.  {\it A
priori}, it is unclear which classification results in a better
detection of change points.  In practice, one would use one of these
classifications, depending on the anticipated properties of the
phases.  Finally, to quantify the efficiency of the LCH method, we
introduce the recognition score, $R$, as the fraction of points that
have been correctly classified.  This score is computed for each
trajectory and then averaged over all simulated trajectories.  The
score $0.5$ would be obtained by a completely random classification,
whereas the score $1$ corresponds to the perfect classification.  We
will analyze how the efficiency of the LCH method is affected by the
phase duration $T$, the window size $\tau$, and noise level
$\sigma_n$.  We emphasize that the phase classification is performed
{\it individually} for each trajectory, whereas the ensemble average
appears only at the last step to assess the quality of the method
through the recognition score.

Throughout the paper, we employ dimensionless units.  In particular,
time is identified with the point index $n$.

\subsection{Two Brownian motions}

We start with heterogeneous diffusion, in which a particle diffuses in
a composite medium with high and low diffusivities.  A simplified
model of this dynamics is a Brownian motion switching between two
diffusion coefficients $D_1$ and $D_2$.  This model can also represent
the motion of a polymer that switches between two conformational states
having distinct hydrodynamic radii (e.g., a compact globular structure
versus an extended fibrous one).

\begin{figure*}
\begin{center}
\includegraphics[width=75mm]{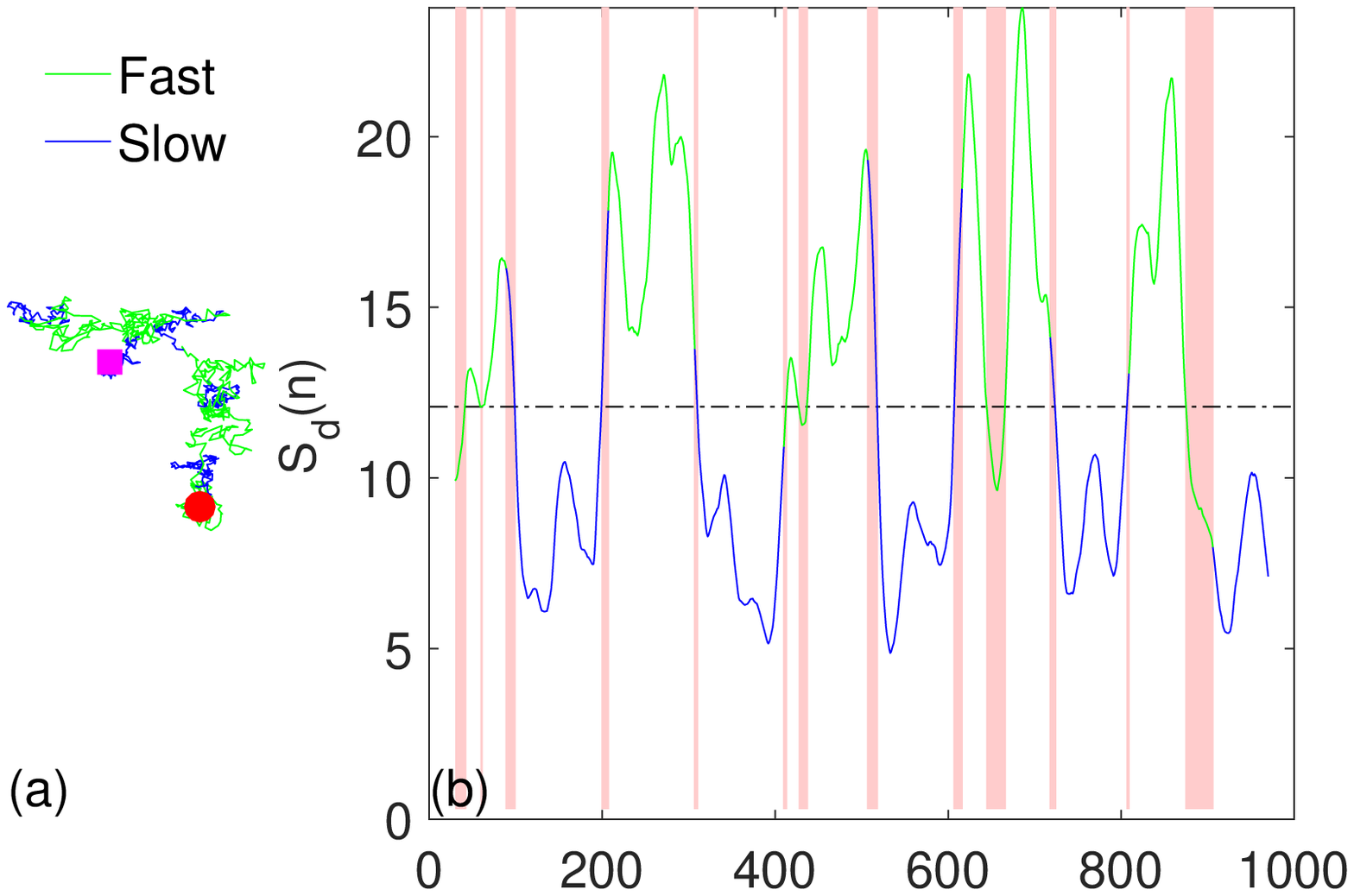}  
\includegraphics[width=75mm]{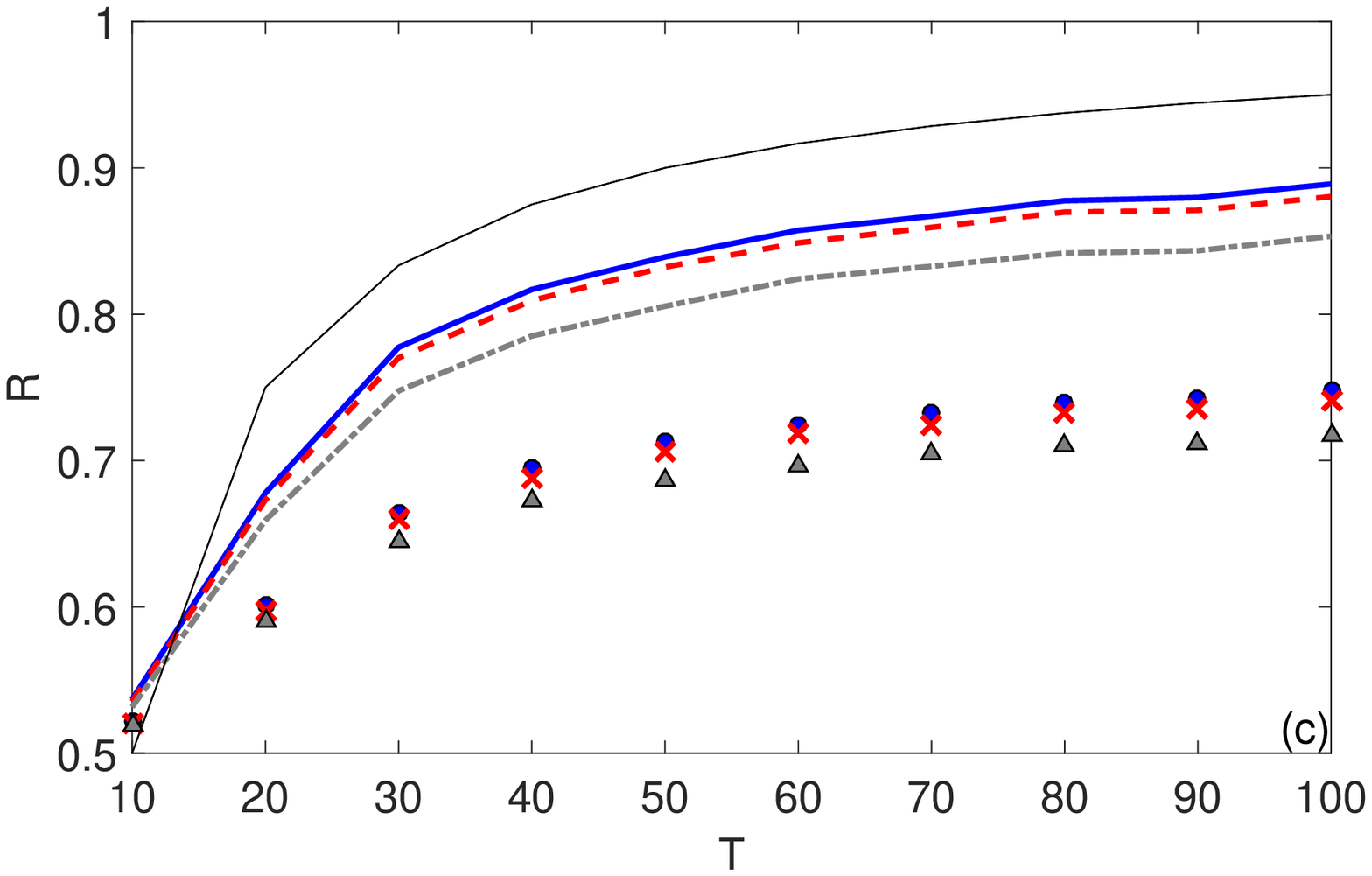}   
\includegraphics[width=75mm]{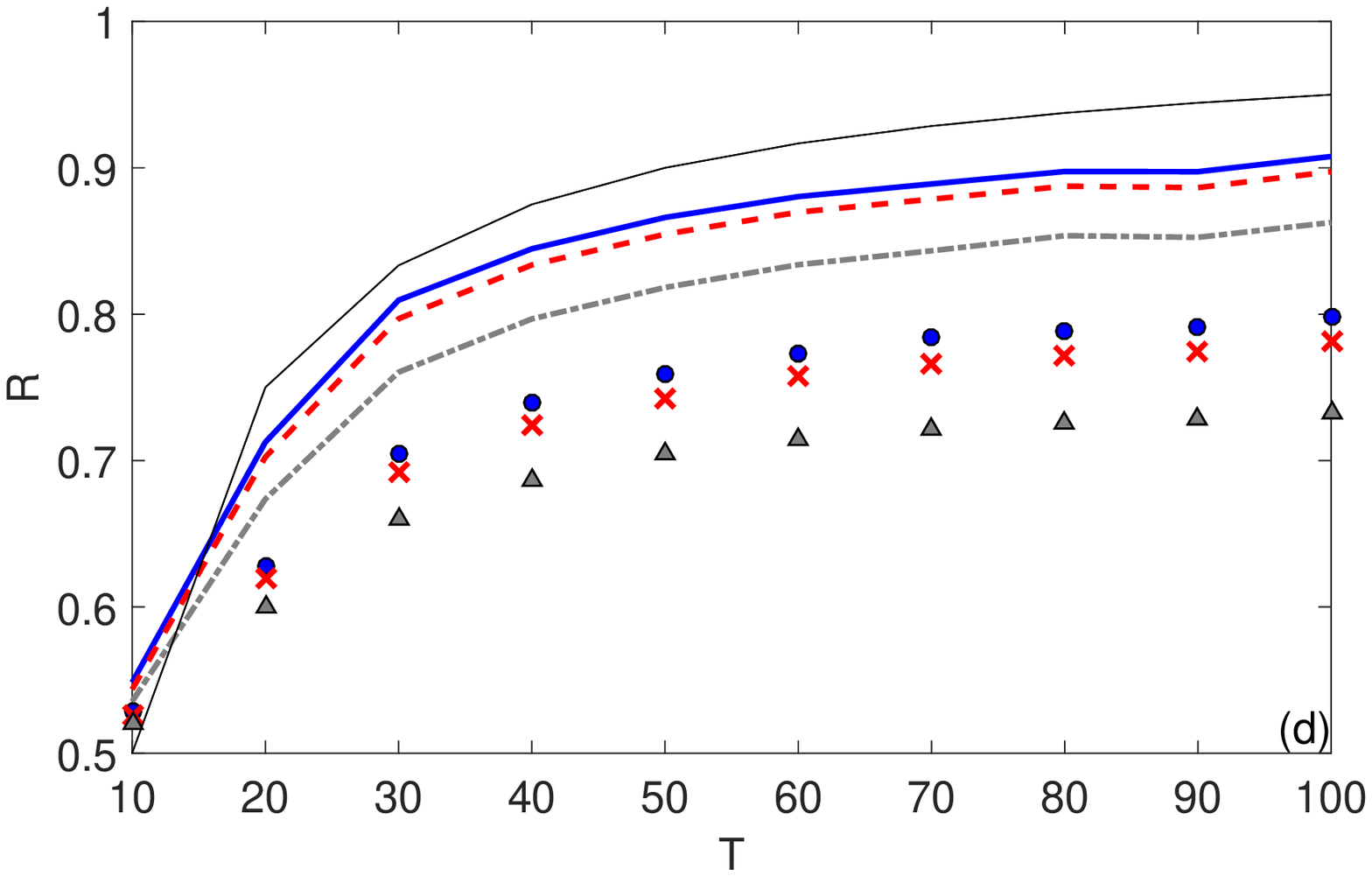}   
\includegraphics[width=75mm]{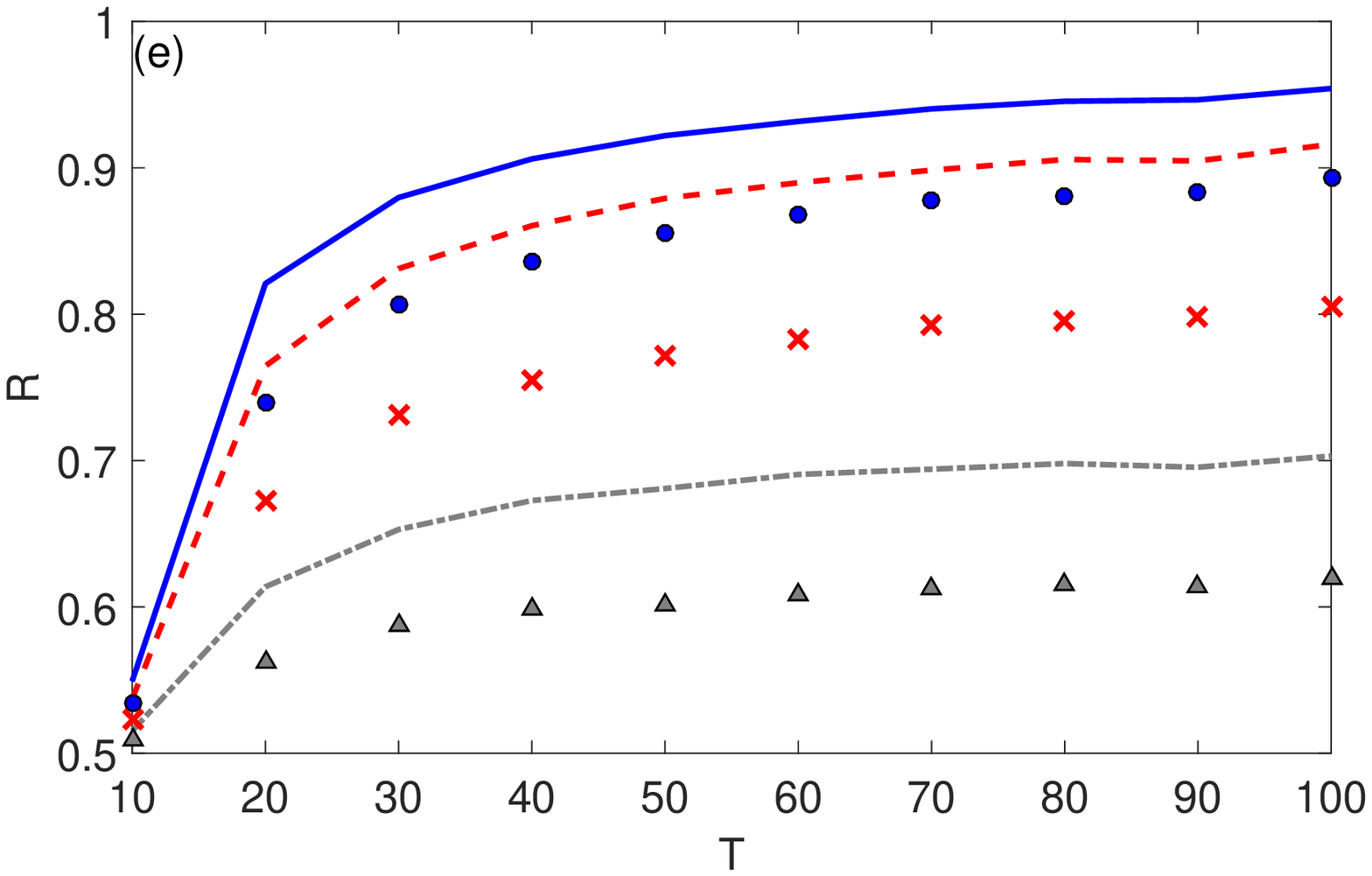}   
\end{center}
\caption{
(Color online) {\bf Model 1} {\bf (a)} A single trajectory of planar
Brownian motion alternating a ``slow'' phase ($D_1 = 1/2$, dark blue)
and a ``fast'' phase ($D_2 = 2$, light green), of the mean phase
duration $T=100$.  Circle and square indicate the starting and ending
points.  {\bf (b)} The weighted LCH diameter $S_d(n)$ with the window
size $\tau = 10$, applied to this trajectory.  Pink shadow highlights
the false classification zones.  Dashed horizontal line shows the
empirical mean $S_d$ over that trajectory.  {\bf (c,d,e)} Recognition
score $R$ of the diameter-based discriminator $S_d(n)$ {\bf (c)}, the
volume-based discriminator $S_v(n)$ {\bf (d)}, and the TA MSD-based
discriminator {\bf (e)} as a function of the mean phase duration $T$.
Lines show the results for the case $D_1 = 1/2$, $D_2 = 2$ with three
noise levels $\sigma_n$: $0$ (blue solid), $0.5
\sigma$ (red dashed), and $\sigma$ (gray dash-dotted) ($\sigma$ being
the empirical standard deviation of increment calculated for each
trajectory).  Symbols correspond to the case $D_1 = 1/2$, $D_2 = 1$
with the same levels of noise $\sigma_n$: $0$ (circles), $0.5\sigma$
(crosses), and $\sigma$ (triangles).  Thin black solid line shows the
hypothetical curve $1 - \tau/(2T)$ for qualitative comparison. }
\label{Result_Bm_Bm}
\end{figure*}

Figure \ref{Result_Bm_Bm}a shows a single planar trajectory with two
alternating phases of slow ($D_1 = 1/2$) and fast ($D_2 = 2$)
diffusion, each phase of the mean duration $T = 100$.  The weighted
LCH diameter $S_d(n)$ with the window size $\tau = 10$ is shown in
Fig. \ref{Result_Bm_Bm}b.  The change points are identified as
crossings of $S_d(n)$ by the dashed horizontal line which indicates
the empirical mean $S_d$ over that trajectory.  As expected, the
detected change points are slightly delayed with respect to the actual
change points (shown by changes of color/brightness in the curve).
This delay is caused by the fact that the LCH at the actual change
point, $n_c$, includes half points of one phase and half points of the
other phase.  Only when $n = n_c + 2\tau + 1$, the LCH gets rid off
the points of the previous phase.  As a consequence, the delay is
expected to be of the order of $2\tau$, as qualitatively confirmed by
this figure.  The delay results in a false classification of some
points, as illustrated by pink shadowed regions in
Fig. \ref{Result_Bm_Bm}b.  The false classification can also result
from spontaneous crossings by $S_d(n)$ of the dashed horizontal line
(e.g., see the shadowed region at the time step around 640).  This is
just a random nature of the motion: the particle in the ``fast''
(resp., ``slow'') phase starts to explore the space slower (resp.,
faster) than usual due to stochastic fluctuations.

Figure \ref{Result_Bm_Bm}c shows the recognition score $R$ as a
function of the mean phase duration $T$ for the diameter-based
discriminator $S_d(n)$.  When the phase duration is too short (say, $T
= 10$), the LCH of window size $\tau = 10$ includes 21 consecutive
points and thus almost always contains points from both phases.  In
this extreme case, the method is clearly unable to detect change
points, in agreement with the obtained recognition score close to
$0.5$.  When the phase duration $T$ is comparable to $2\tau+1 = 21$
(the number of points used to construct the LCH) and two phases are
quite distinct (the example $D_1 = 1/2$, $D_2 = 2$), the recognition
score is around $0.67$, i.e., two thirds of points are correctly
classified.  The recognition score further increases up to $88\%$ as
the mean phase duration grows up to $T = 100$.  If the delay in
detection was equal on average to $c \tau$, the curve would be $(T - c
\tau)/T = 1 - c \tau/T$, as illustrated by thin black line for $c = 1/2$.
One can see that this hypothetical curve over-estimates the
recognition score, probably because of additional spontaneous false
classifications.  

Figure \ref{Result_Bm_Bm}d presents the recognition score for the
volume-based classification.  While this classification slightly
outperforms the diameter-based one, the behavior is very similar to
that shown in Fig. \ref{Result_Bm_Bm}c.  This is not surprising
because both phases and thus the shapes of the LCHs are isotropic so
that the volume and the diameter of the LCH bear essentially the same
information.

Figures \ref{Result_Bm_Bm}c,d also show that the LCH estimators are
robust against measurement noises.  In fact, adding the white Gaussian
noise of amplitude $\sigma_n$ which is comparable to the amplitude of
one-step increments, has almost no effect on the recognition score.
This is an important advantage of the LCH method, which is based on
integral-like characteristics of the trajectory, as compared to
conventional techniques based on differential-like estimators such as
local TA MSD, which are more sensitive to noise.  This point is
illustrated in Fig. \ref{Result_Bm_Bm}e, which shows the recognition
score of the classification scheme, in which the LCH-based estimator
$Q_d(n)$ is replaced by the local TA MSD estimator $Q_\sigma(n)$ from
Eq. (\ref{eq:Qsigma}).  We use the same window size $\tau = 10$ as for
the LCH estimators.  When there is no noise, the local TA MSD
estimator outperforms the LCH-based estimators $Q_d(n)$ and $Q_v(n)$.
This is not surprising as the diffusivity estimator $Q_\sigma(n)$ is
known to be optimal for Brownian motion
\cite{Voisinne10,Berglund10,Grebenkov2011}.  However, the presence of
noise drastically deteriorates the quality of change points detection
by TA MSD.  In turn, the effect of the same level of noise onto the
LCH-based estimator is much weaker.

When the diffusion coefficients of two phases become closer, the
recognition score is reduced.  This is illustrated in
Fig. \ref{Result_Bm_Bm}c,d,e by symbols that show the recognition
score for the case $D_1 =1/2$ and $D_2 = 1$.  In the ultimate case
$D_1 = D_2$, the two phases become identical, and any phase detection
is meaningless, yielding the recognition score close to $0.5$ (not
shown).

\subsection{Brownian motion with a drift}

In the second example, we consider a planar Brownian motion
alternating two phases (of the same diffusivity), without and with a
drift $\mu$ in a fixed direction.  This is a very basic model for
active intracellular transport, in which cargos can attach to motor
proteins, be transported ballistically along microtubules,
spontaneously detach, and resume diffusion
\cite{Brangwynne09,Bressloff13}.  If the intermittent character of
this process was ignored, switching between two phases would
effectively look like a superdiffusive motion with a scaling exponent
$\alpha$ between $1$ and $2$.

\begin{figure*}
\begin{center}
\includegraphics[width=75mm]{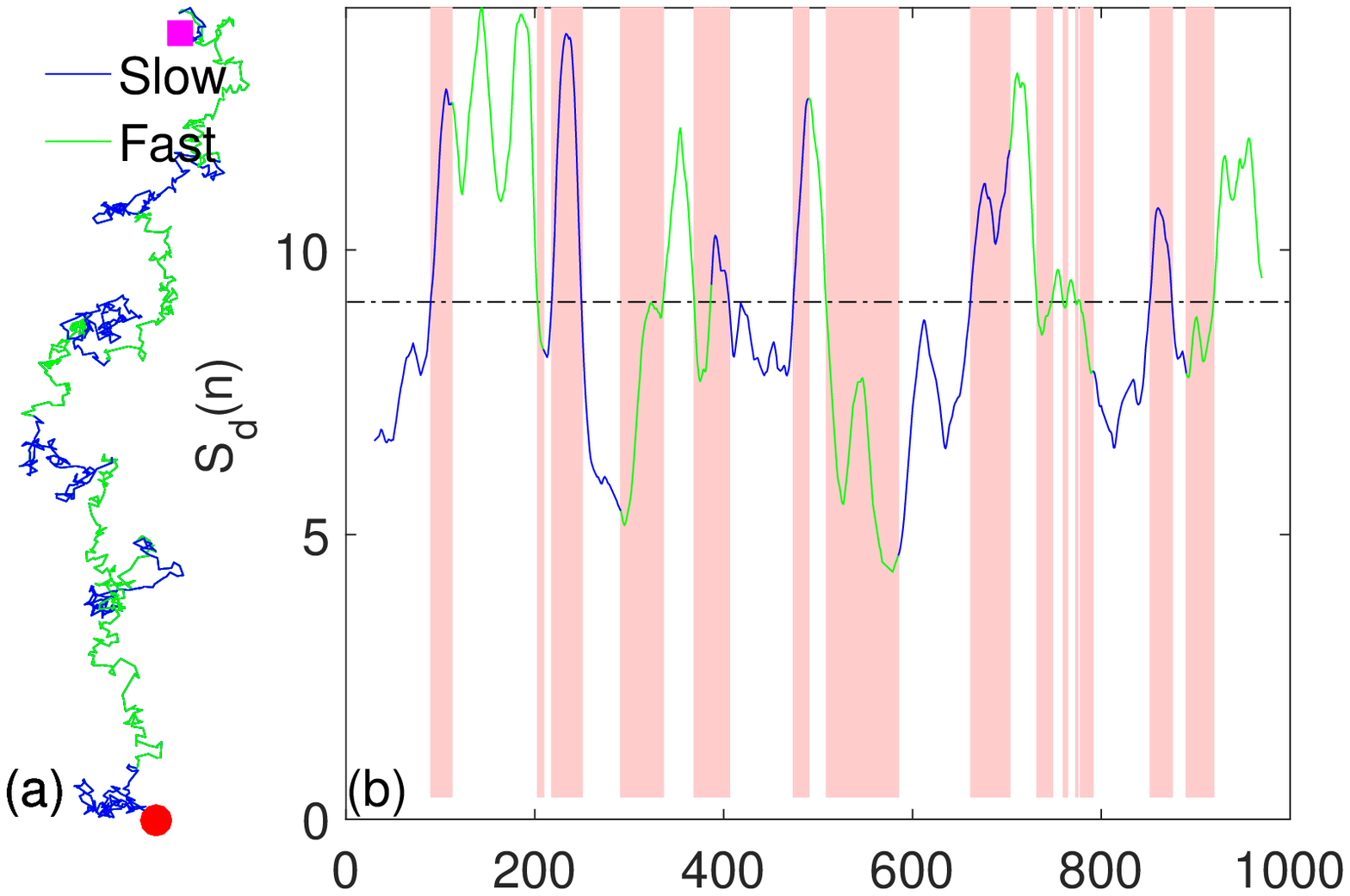} 
\includegraphics[width=75mm]{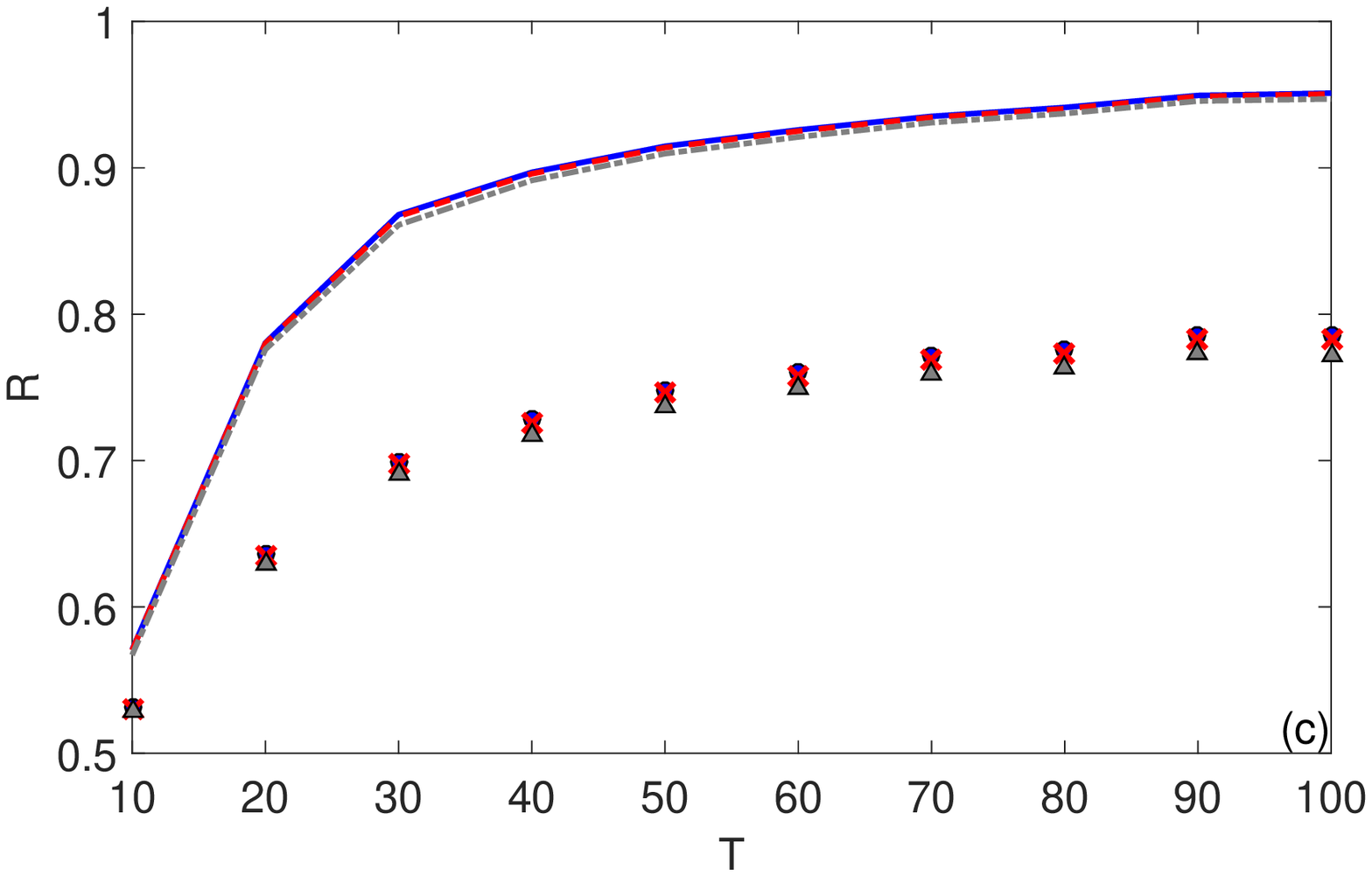}  
\includegraphics[width=75mm]{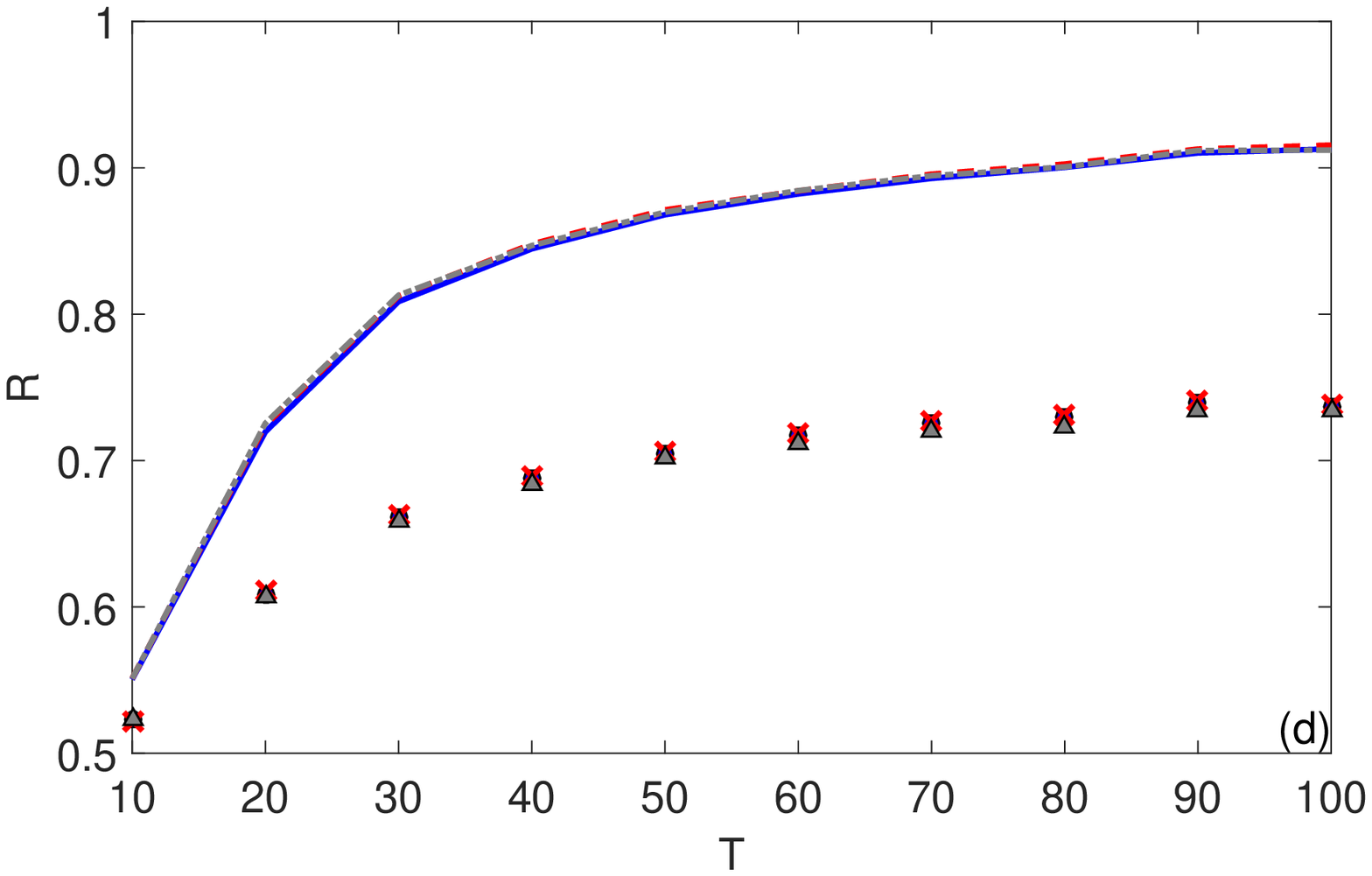}  
\includegraphics[width=75mm]{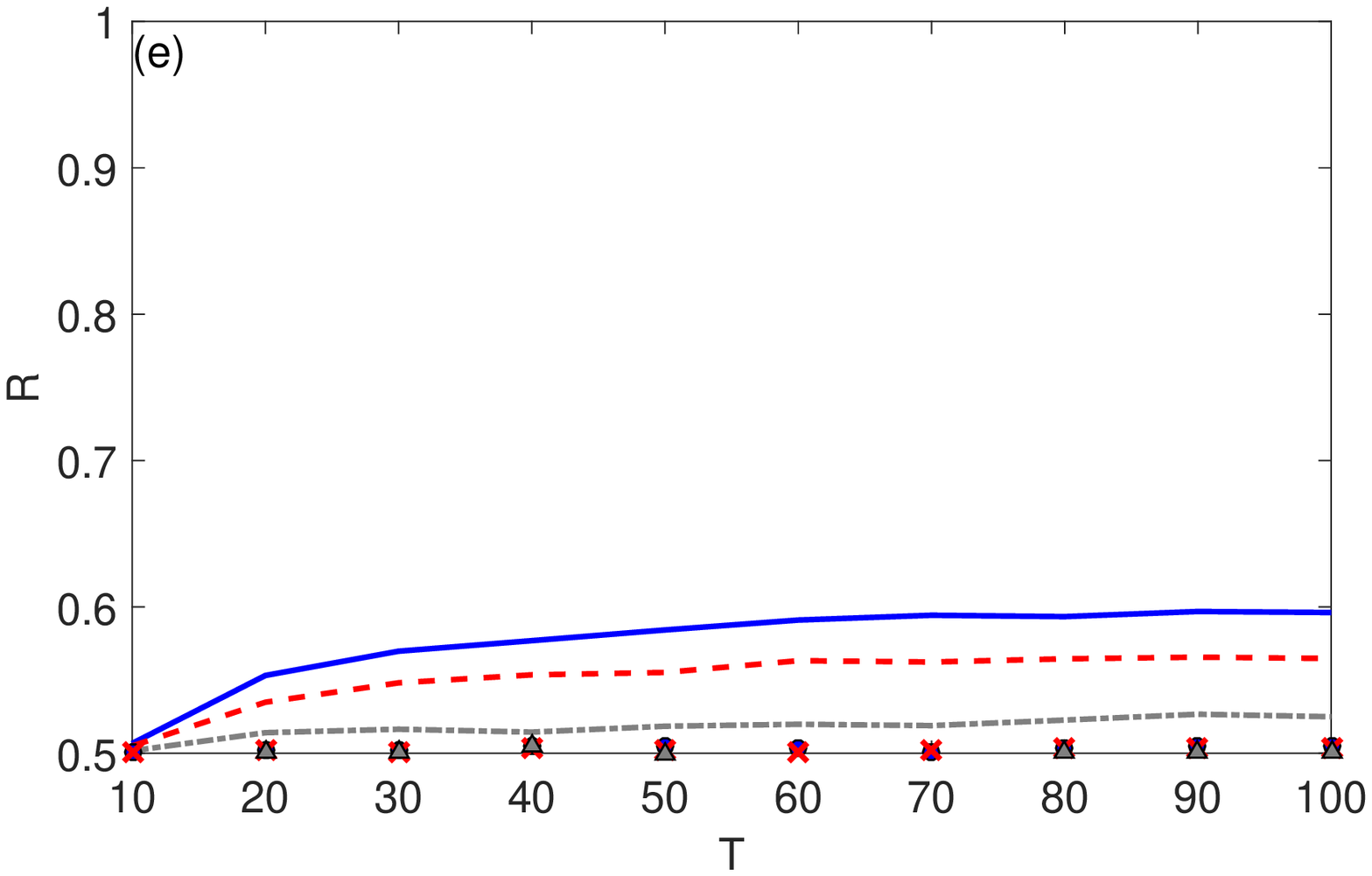}  
\end{center}
\caption{
(Color online) {\bf Model 2} {\bf (a)} A single trajectory of an
intermittent process alternating a ``slow'' phase of planar Brownian
motion without drift (dark blue) and a ``fast'' phase with drift ($\mu
= 0.5$, light green), of the mean phase duration $T=100$ (both phases
with $D = 1/2$).  Circle and square indicate the starting and ending
points.  {\bf (b)} The weighted LCH diameter $S_d(n)$ with the window
size $\tau = 10$ applied to this trajectory.  Pink shadow highlights
the false classification zones.  Dashed horizontal line shows the
empirical mean $S_d$ over that trajectory.  {\bf (c,d,e)} Recognition
score $R$ of the diameter-based discriminator $S_d(n)$ {\bf (c)}, the
volume-based discriminator $S_v(n)$ {\bf (d)}, and the TA MSD-based
discriminator {\bf (e)} as a function of the mean phase duration $T$.
Lines show the results for the case $\mu = 0.5$ with three noise
levels $\sigma_n$: $0$ (blue solid), $0.5 \sigma$ (red dashed), and
$\sigma$ (gray dash-dotted) ($\sigma$ being the empirical standard
deviation of increment calculated for each trajectory).  Symbols
correspond to the case $\mu = 0.1$ with the same levels of noise
$\sigma_n$: $0$ (circles), $0.5\sigma$ (crosses), and $\sigma$
(triangles).}
\label{Result_Bm_drift}
\end{figure*}

As the drift tends to elongate the trajectory in one direction, one
should be able to identify the drifted phase via anisotropic and
larger LCH.  To test the efficiency of the LCH method, we generate a
planar Brownian motion alternating two phases, one of which has a
small drift $\mu$ in a fixed direction.  As previously, the durations
of both phases are random and exponentially distributed variables with
the mean phase duration $T$.  Figure \ref{Result_Bm_drift} shows an
example of a single trajectory of this process, the weighted LCH
diameter $S_d(n)$, and the recognition scores for the diameter-based
and volume-based discriminators.  For a relatively strong drift, $\mu
= 0.5$ (with the one-step standard deviation $\sigma = 1$), both
estimators efficiently detect the change points.  For a much smaller
drift $\mu = 0.1$, the recognition scores are decreased but remain
good enough.  For instance, one attains the recognition score of
$80\%$ at the mean phase duration $T = 100$.  In both cases, the
diameter-based discriminator outperforms the volume-based
discriminator, and both discriminators are very robust against noise.
As expected, the local TA MSD estimator, which is {\it a priori} not
adapted to detect drift, shows a poor performance
(Fig. \ref{Result_Bm_drift}e).

\subsection{Two fractional Brownian motions}

The fractional Brownian motion $B_H(t)$ \cite{Mandelbrot1968} is a
centered Gaussian process, which is defined by its covariance
function:
\begin{equation}  \label{eq:fBm_cov}
\langle B_H(t) B_H(s)\rangle = D_H \bigl[t^{2H} + s^{2H} - |t-s|^{2H}\bigr], 
\end{equation}  
where $0 < H < 1$ is the Hurst exponent, and $D_H$ is the generalized
diffusion coefficient.  This is the long-range memory process that is
often used to model anti-persistent subdiffusive motion for $H < 1/2$
(e.g., the motion of a tracer in a visco-elastic medium with no
characteristic timescale \cite{Szymanski09,Bertseva12,Grebenkov13})
and persistent superdiffusive motion for $H > 1/2$ (e.g., active
transport of cargos on microtubules by molecular motors
\cite{Caspi00,Arcizet08,Brangwynne09,Desposito11}).

For each planar trajectory, $X$ and $Y$ coordinates were generated by
concatenating independent ``pieces'' (phases) of one-dimensional fBm
with alternating phases of subdiffusive ($H_1 = 0.35$) and
superdiffusive ($H_2 = 0.7$) motion, with $D_H = 1/2$ in both cases.
As previously, durations of phases are independent exponentially
distributed random variables, with a prescribed mean phase duration
$T$.  In other words, the numerical algorithm consists in (i)
generating a sequence of independent exponentially distributed
durations $\tau_1$, $\tau_2$, $\ldots$, from which the integer change
points are defined as $t_k = t_{k-1} + \lfloor \tau_k \rfloor$ (with
$t_0 = 0$); and (ii) generating successive phases
$\{x_{t_{k-1}+1},\ldots,x_{t_k}\}$ and
$\{y_{t_{k-1}+1},\ldots,y_{t_k}\}$ (with $k = 1,2,\ldots$) as
one-dimensional fBms, with the covariance defined in
Eq. (\ref{eq:fBm_cov}), where $H$ is equal to $H_1$ for even $k$ and
to $H_2$ for odd $k$.  The starting point of the ``piece'' $k$ is the
ending point of the ``piece'' $k-1$.  The whole trajectory is thus
composed of points
\begin{equation*}
\begin{split}
& \biggl\{ \underbrace{(x_1,y_1),\ldots, (x_{t_1},y_{t_1})}_{\textrm{``piece''~1:~ fast phase}}, 
 \underbrace{(x_{t_1+1},y_{t_1+1}), \ldots, (x_{t_2},y_{t_2})}_{\textrm{``piece''~2:~ slow phase}}, \\
& \underbrace{(x_{t_2+1},y_{t_2+1}), \ldots, (x_{t_3},y_{t_3})}_{\textrm{``piece''~3:~ fast phase}}, \ldots, \\
& \ldots, \underbrace{(x_{t_K+1},y_{t_K+1}), \ldots, (x_{N},y_{N})}_{\textrm{``piece''~K}} \biggr\}. \\
\end{split}
\end{equation*}
Note that the last ``piece'' (with index $K$) is truncated to get the
whole trajectory with $N$ points.  We emphasize that this algorithm
yields the successive phases that are independent from each other, and
there is no cross-correlation in increments along $X$ and $Y$
coordinates.  In this model, switching between two phases mimics
changes in auto-correlations of increments.

\begin{figure*}
\begin{center}
\includegraphics[width=75mm]{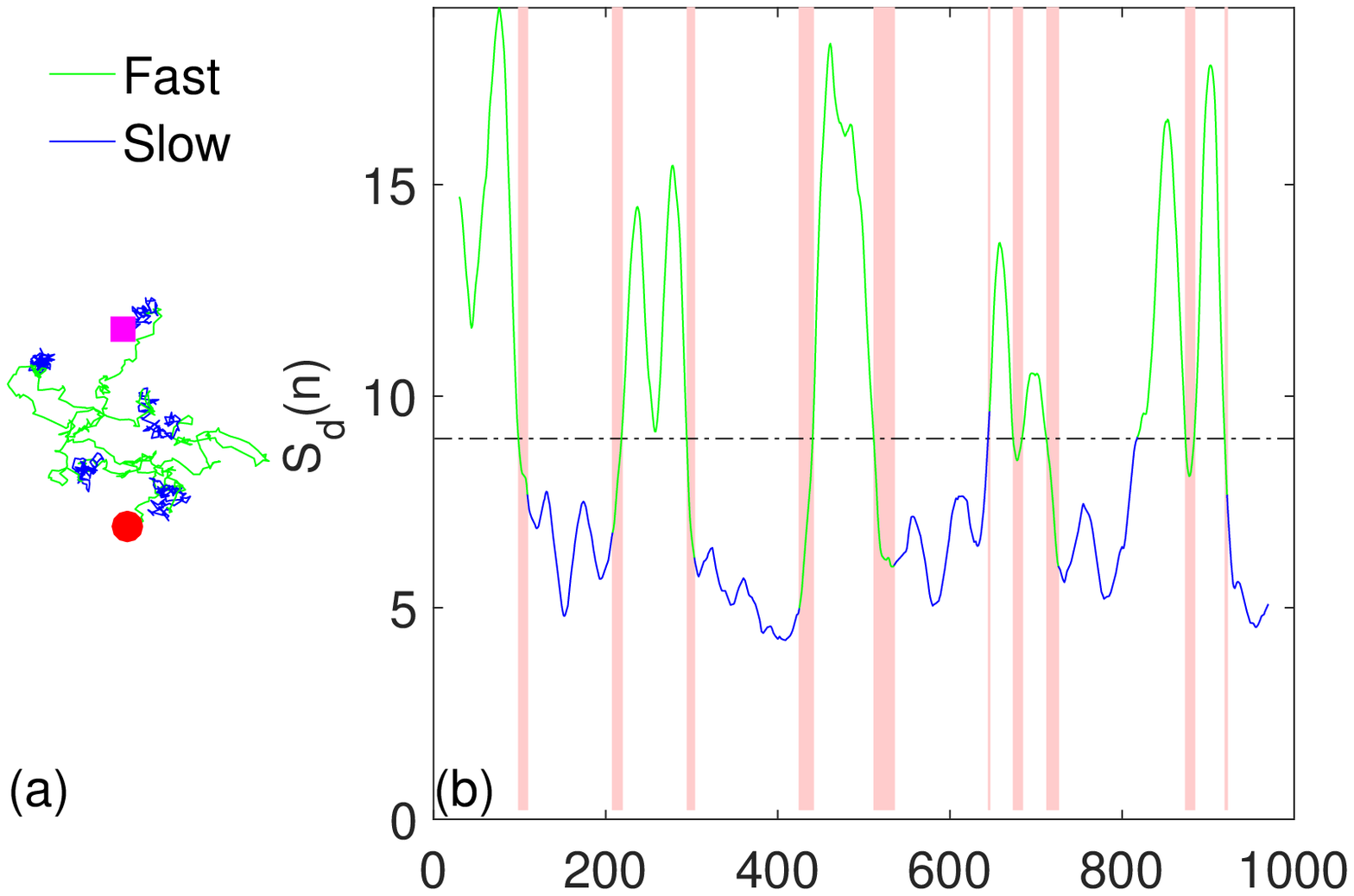} 
\includegraphics[width=75mm]{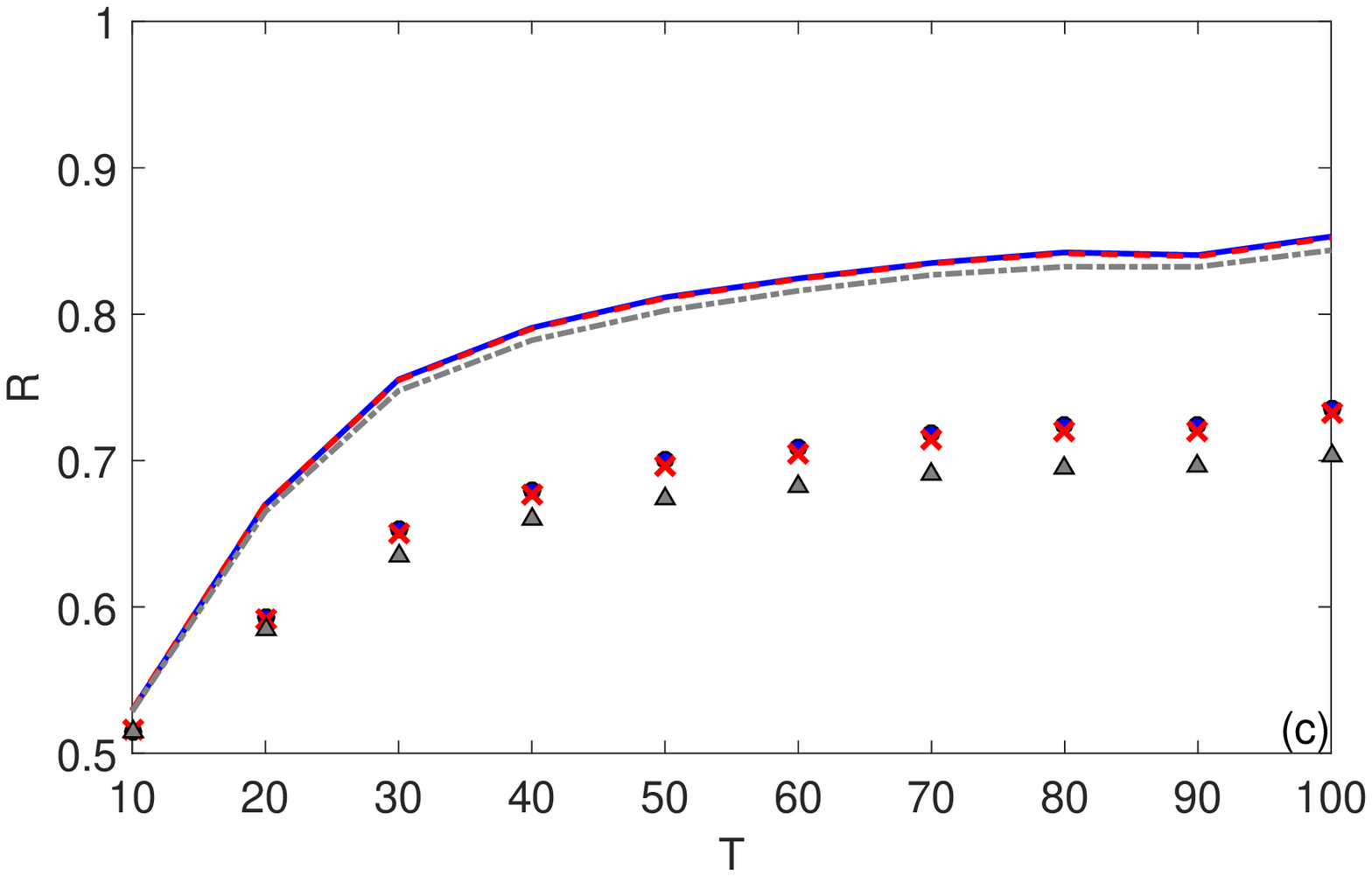}  
\includegraphics[width=75mm]{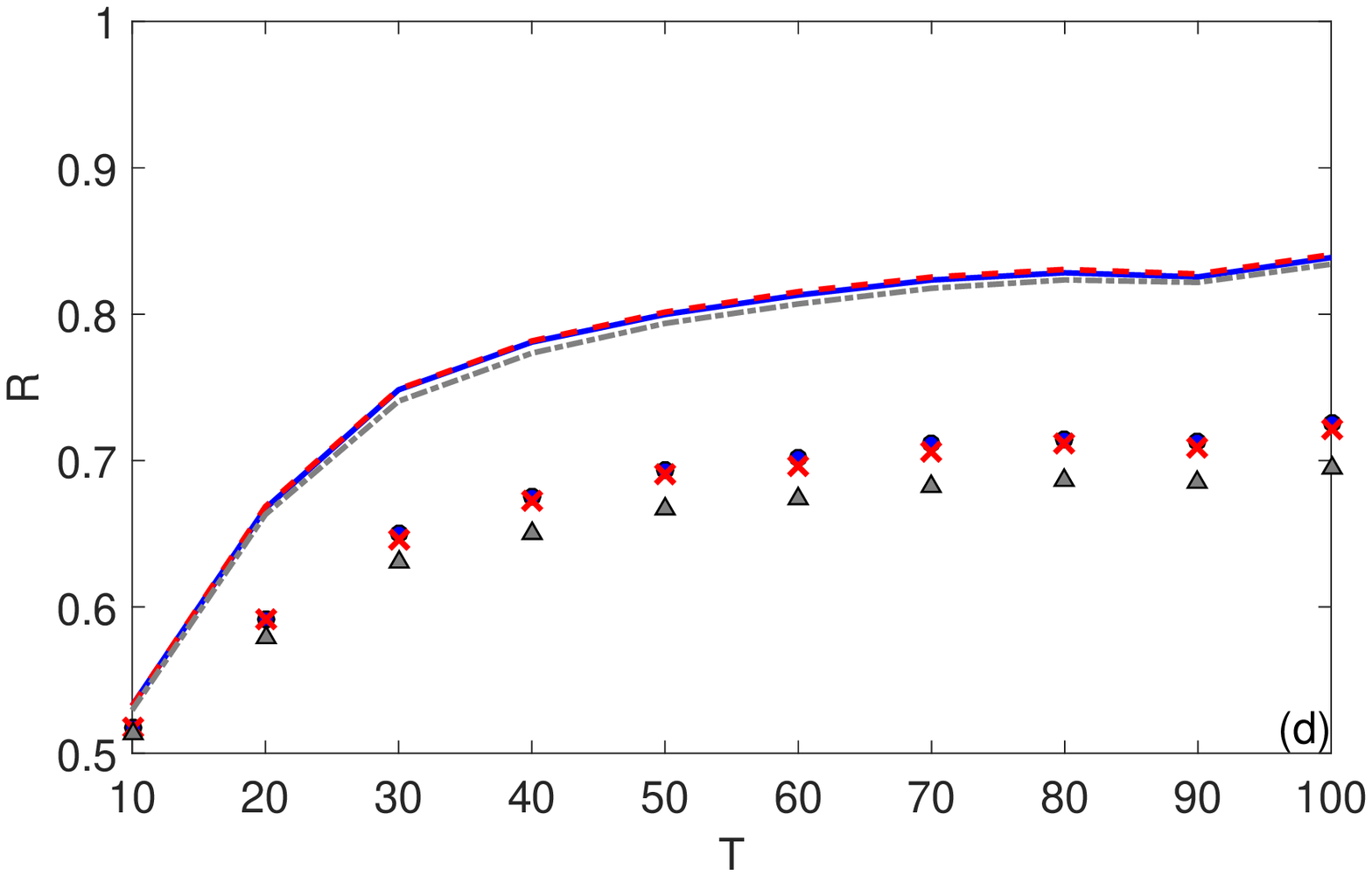}  
\includegraphics[width=75mm]{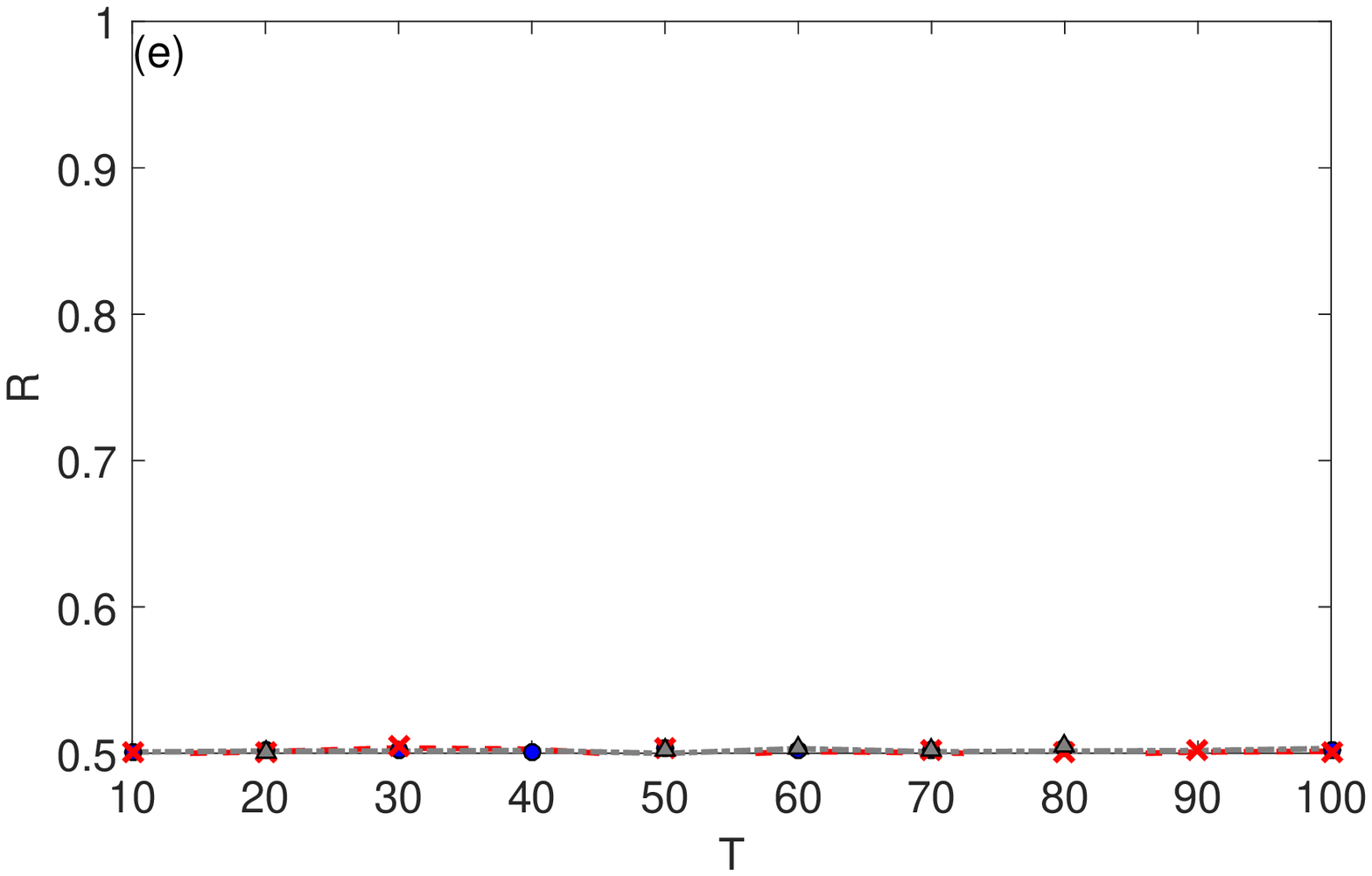}  
\end{center}
\caption{ 
(Color online) {\bf Model 3} {\bf (a)} A single trajectory of planar
fBm, alternating a ``slow'' phase ($H_1 = 0.35$, dark blue) and a
``fast'' phase ($H_2 = 0.7$, light green), of the mean phase duration
$T=100$ and $D_H = 1/2$.  Circle and square indicate the starting and
ending points.  {\bf (b)} The weighted LCH diameter $S_d(n)$ with the
window size $\tau = 10$ applied to this trajectory.  Pink shadow
highlights the false classification zones.  Dashed horizontal line
shows the empirical mean $S_d$ over that trajectory.  {\bf (c,d,e)}
Recognition score $R$ of the diameter-based discriminator $S_d(n)$
{\bf (c)}, the volume-based discriminator $S_v(n)$ {\bf (d)}, and the
TA MSD-based discriminator {\bf (e)} as a function of the mean phase
duration $T$.  Lines show the results for the case $H_1 = 0.35$, $H_2
= 0.7$ with three noise levels $\sigma_n$: $0$ (blue solid), $0.5
\sigma$ (red dashed), and $\sigma$ (gray dash-dotted) ($\sigma$ being
the empirical standard deviation of increment calculated for each
trajectory).  Symbols correspond to the case $H_1 = 0.35$, $H_2 = 0.5$
with the same levels of noise $\sigma_n$: $0$ (circles), $0.5\sigma$
(crosses), and $\sigma$ (triangles). }
\label{Result_fBm_fBm}
\end{figure*}

Figure \ref{Result_fBm_fBm} shows an example of such intermittent
trajectory, the weighted LCH diameter $S_d(n)$ with the window size
$\tau = 10$ applied to this trajectory, and the recognition score for
both the diameter-based and the volume-based classifications.  The
results are similar to that shown in Fig. \ref{Result_Bm_Bm} for
intermittent Brownian motion.  When the distinction between two phases
is lower (the example with $H_1 = 0.35$ and $H_2 = 0.5$), the
recognition scores are reduced (symbols) but remain satisfactory.
These scores are particularly favorably compared to that of the TA MSD
discriminator which cannot detect phases at all
(Fig. \ref{Result_fBm_fBm}e).  Although the conventional TA MSD
estimator may be modified to show a better performance, such a
modification would implicate additional knowledge on the model.  The
results for the case $H_1 = 0.5$ and $H_2 = 0.7$ (normal versus
superdiffusive motion) are very similar to the latter case and thus
not shown.

\subsection{Brownian motion and Ornstein-Uhlenbeck process}
\label{sec:OU}

The Ornstein-Uhlenbeck (OU) process is a Gaussian process modeling the
diffusive motion of a particle trapped by a harmonic potential.  This
is a widely used model of particle interactions (e.g., a Rouse or
bead-spring model in polymer physics \cite{Doi,deGennes}) and of the
trapping effect of optical tweezers
\cite{Kuo93,Wirtz09,Bertseva12,Grebenkov13,Grebenkov15}.  The
one-dimensional OU process can be defined as a solution of the
Langevin equation $dX_t = -kX_t dt + \sqrt{2D} dW_t$, where $k$ is the
relaxation rate (which is related to the spring constant), $D$ is the
diffusion coefficient, and $W_t$ is the standard Brownian motion.  For
each planar trajectory, $X$ and $Y$ coordinates were generated as two
independent intermittent processes, alternating a ``fast'' phase of
Brownian motion ($k = 0$) and a ``slow'' phase of Ornstein-Uhlenbeck
process ($k > 0$), with $D = 1/2$ for both phases.  Phase durations
were independent exponentially distributed random variables, with a
prescribed mean phase duration $T$.  This intermittent process can
model the motion of a particle evolving in a medium where it freely
moves until it interacts with an attracting trap, fluctuates near this
trap via the OU process, liberates itself until the next interaction,
etc.  In neurosciences, the OU process is traditionally used as a
model of the instantaneous firing rate of neurons in the neocortex
\cite{Amit97a,Amit97b}.  Recently, a change-point detection procedure 
to detect changes in the spiking activity of neurons has been proposed
\cite{Mazzucato15}.  In this section, we check the ability of the LCH
method to detect the phases of motion governed by the OU process.

\begin{figure*}
\begin{center}
\includegraphics[width=75mm]{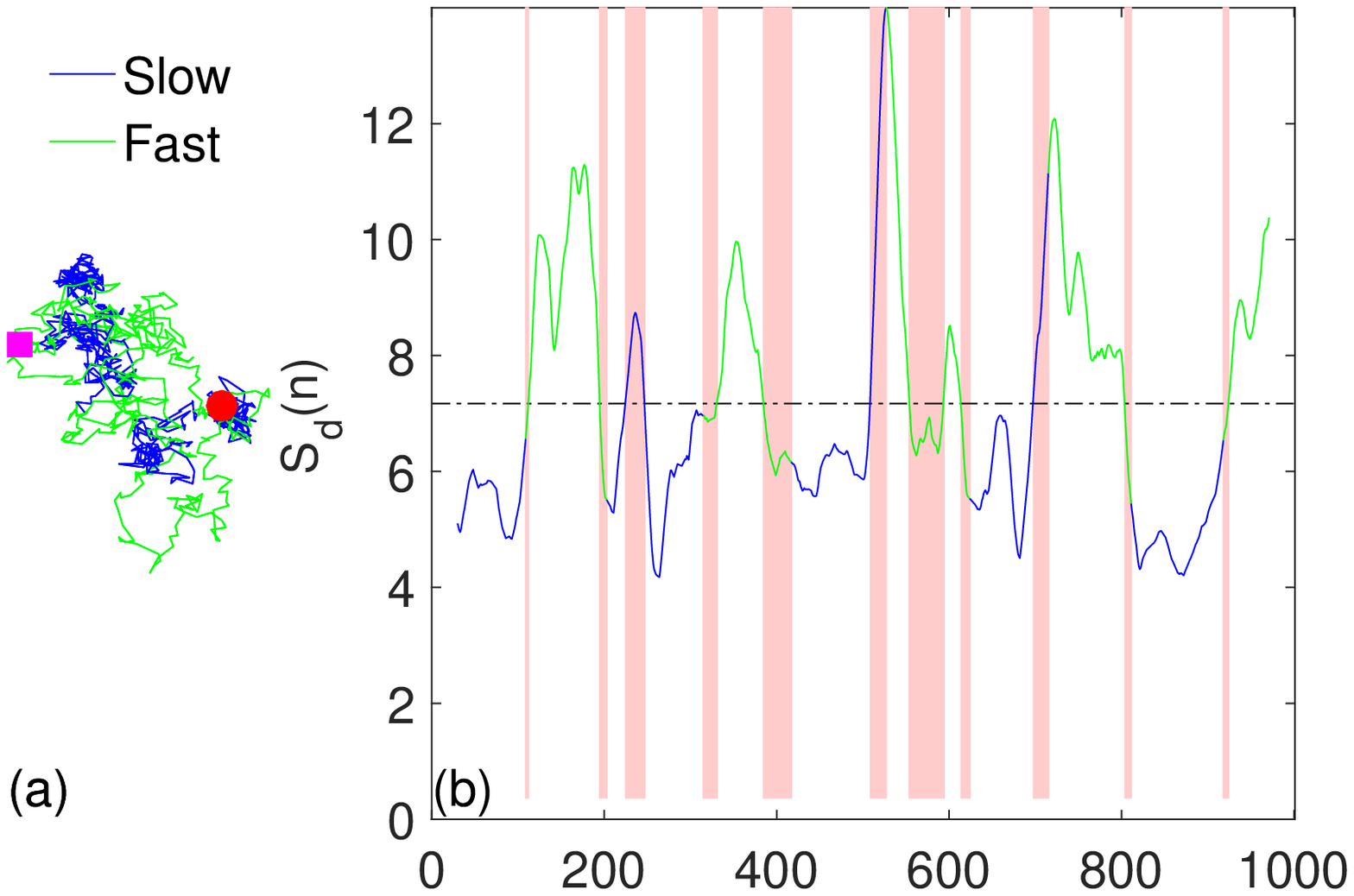} 
\includegraphics[width=75mm]{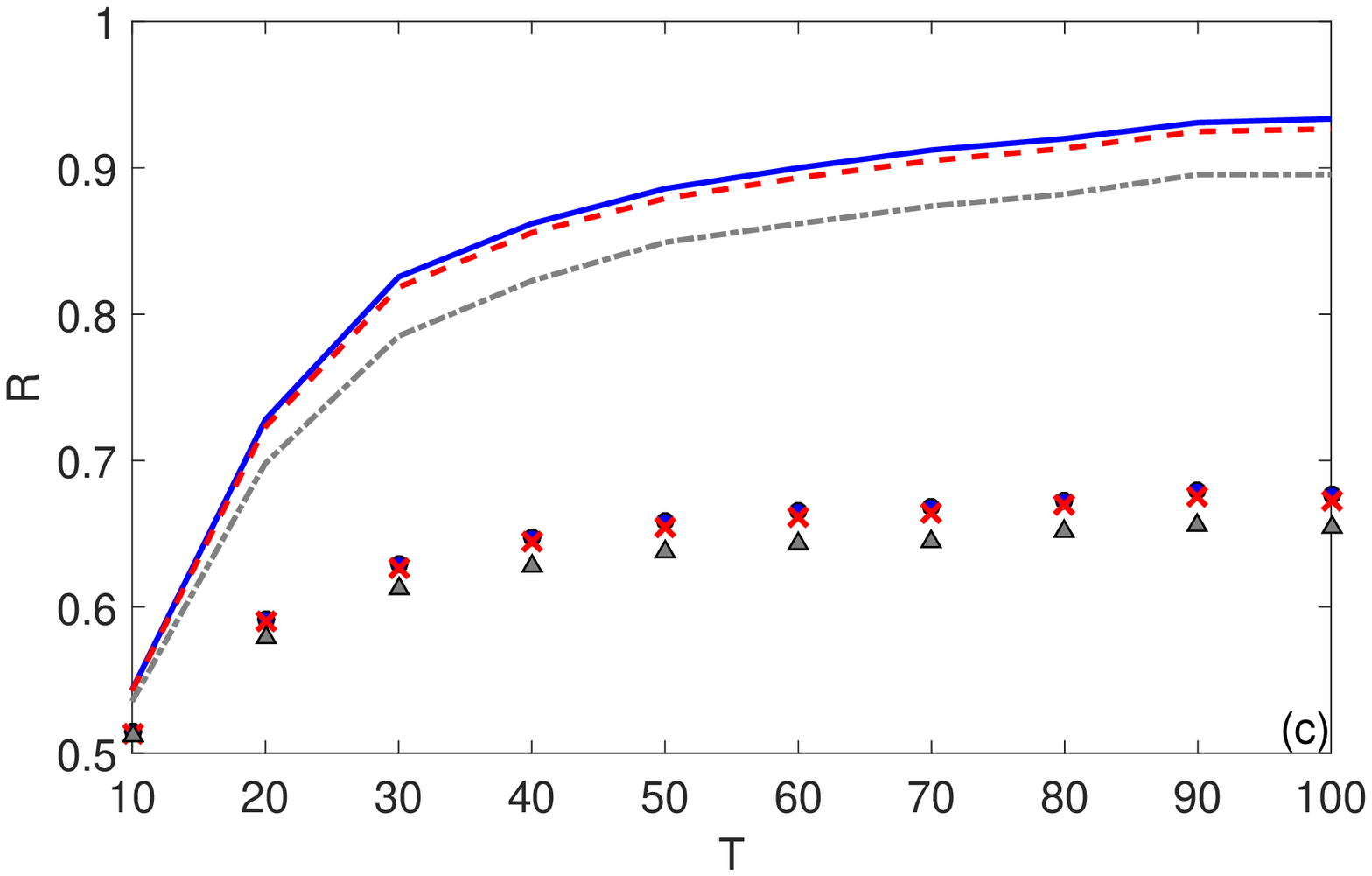}  
\includegraphics[width=75mm]{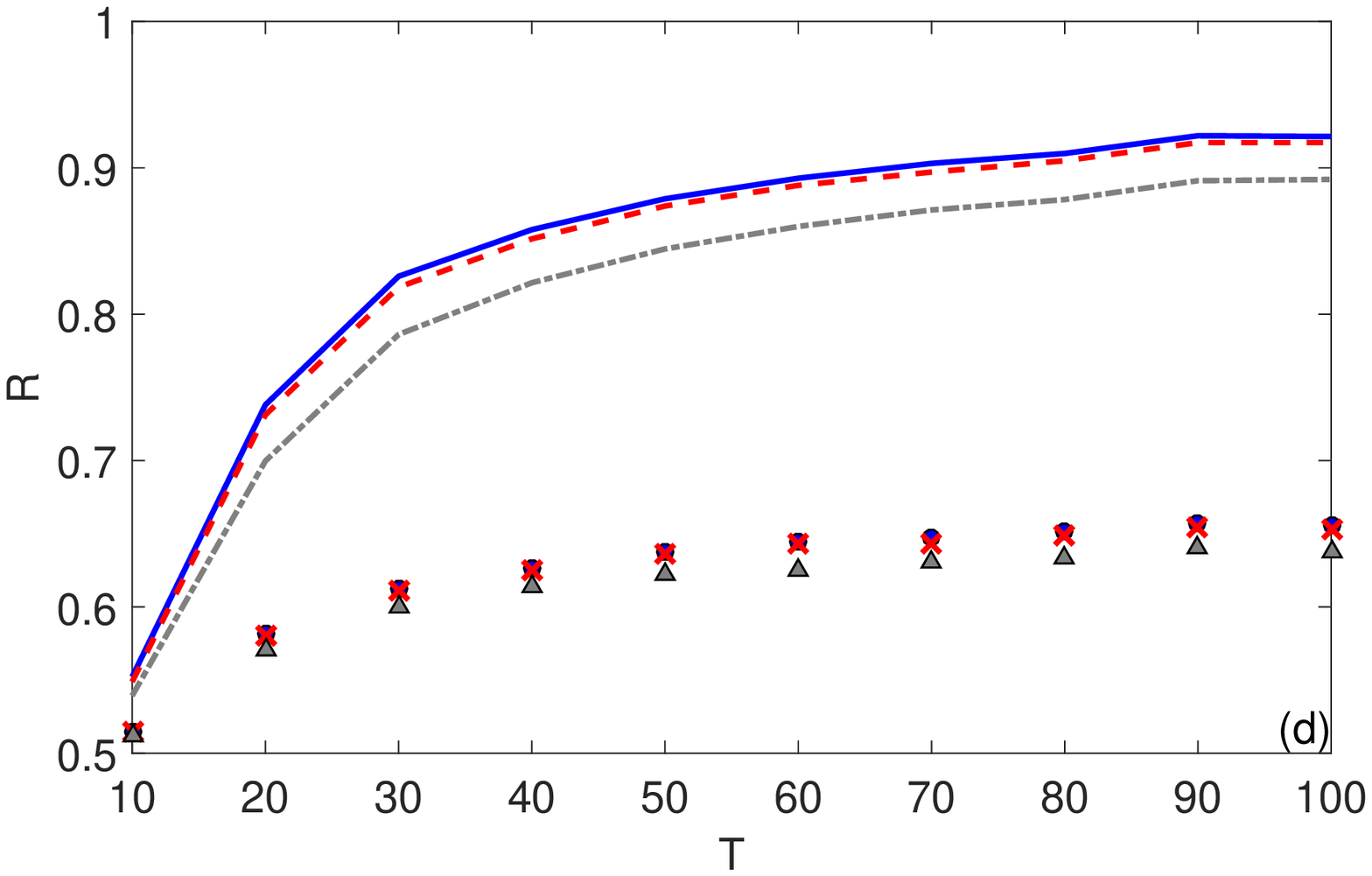}  
\includegraphics[width=75mm]{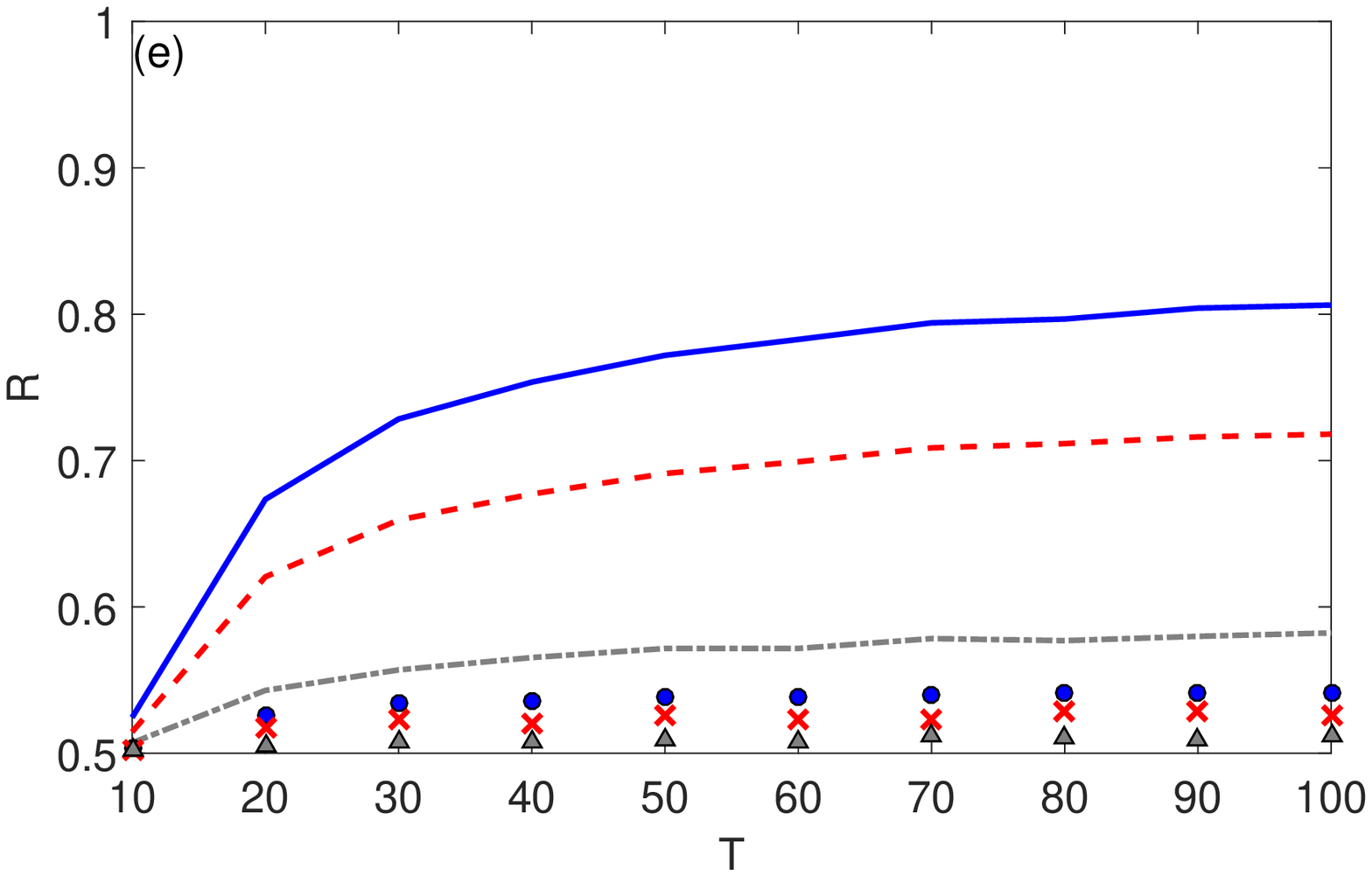}  
\end{center}
\caption{
(Color online) {\bf Model 4} {\bf (a)} A single trajectory of planar
motion, alternating a ``slow'' phase ($k = 0.1$, dark blue) and a
``fast'' phase ($k = 0$, light green), with $D = 1/2$ and the mean
phase duration $T=100$.  Circle and square indicate the starting and
ending points.  {\bf (b)} The weighted LCH diameter $S_d(n)$ with the
window size $\tau = 10$ applied to this trajectory.  Pink shadow
highlights the false classification zones.  Dashed horizontal line
shows the empirical mean $S_d$ over that trajectory.  {\bf (c,d,e)}
Recognition score $R$ of the diameter-based discriminator $S_d(n)$
{\bf (c)}, the volume-based discriminator $S_v(n)$ {\bf (d)}, and the
TA MSD-based discriminator {\bf (e)} as a function of the mean phase
duration $T$.  Lines show the results for the case $k = 1$ with three
noise levels $\sigma_n$: $0$ (blue solid), $0.5 \sigma$ (red dashed),
and $\sigma$ (gray dash-dotted) ($\sigma$ being the empirical standard
deviation of increment calculated for each trajectory).  Symbols
correspond to the case $k = 0.1$ with the same levels of noise
$\sigma_n$: $0$ (circles), $0.5\sigma$ (crosses), and $\sigma$
(triangles). }
\label{Result_Bm_OUP}
\end{figure*}

Figure \ref{Result_Bm_OUP} shows an example of such intermittent
trajectory, the weighted LCH diameter $S_d(n)$ with the window size
$\tau = 10$ applied to this trajectory, and the recognition score for
both the diameter-based and the volume-based classifications.  The
results are similar to that shown in Fig. \ref{Result_Bm_Bm} for
intermittent Brownian motion.  When $k$ is getting smaller, the OU
process becomes close to Brownian motion, and the recognition scores
are reduced (symbols).  The TA MSD discriminator shows poorer
performance (Fig. \ref{Result_Bm_OUP}e).

\subsection{Brownian motion and exponential flights}

In order to test the efficiency of the LCH method in detection of
change points between phases with distinct distributions of
increments, we consider an intermittent process, alternating a
``slow'' phase of planar Brownian motion (with $D = 1/2$) and a
``fast'' phase of two-dimensional exponential flights.  In the
``fast'' phase, an increment at each time step was generated
independently from the others as $(r\cos\theta, r\sin\theta)$, with
the exponential distribution of flight length $r$ (with the mean
length $\ell$) and the uniform distribution of angle $\theta$.  In
other words, Gaussian increments (in the ``slow'' phase) are just
replaced by such exponentially distributed increments (in the ``fast''
phase).

\begin{figure*}
\begin{center}
\includegraphics[width=75mm]{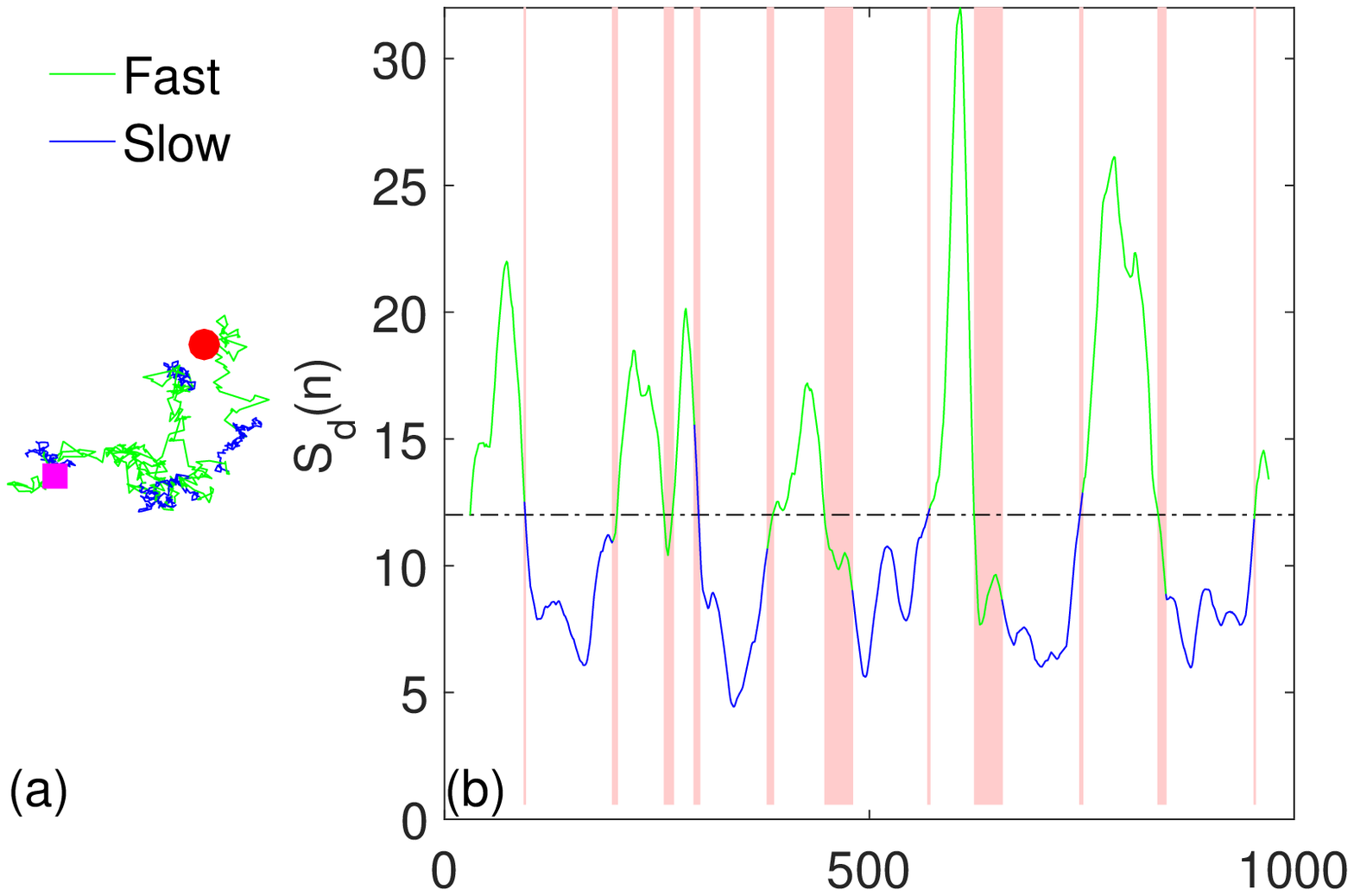} 
\includegraphics[width=75mm]{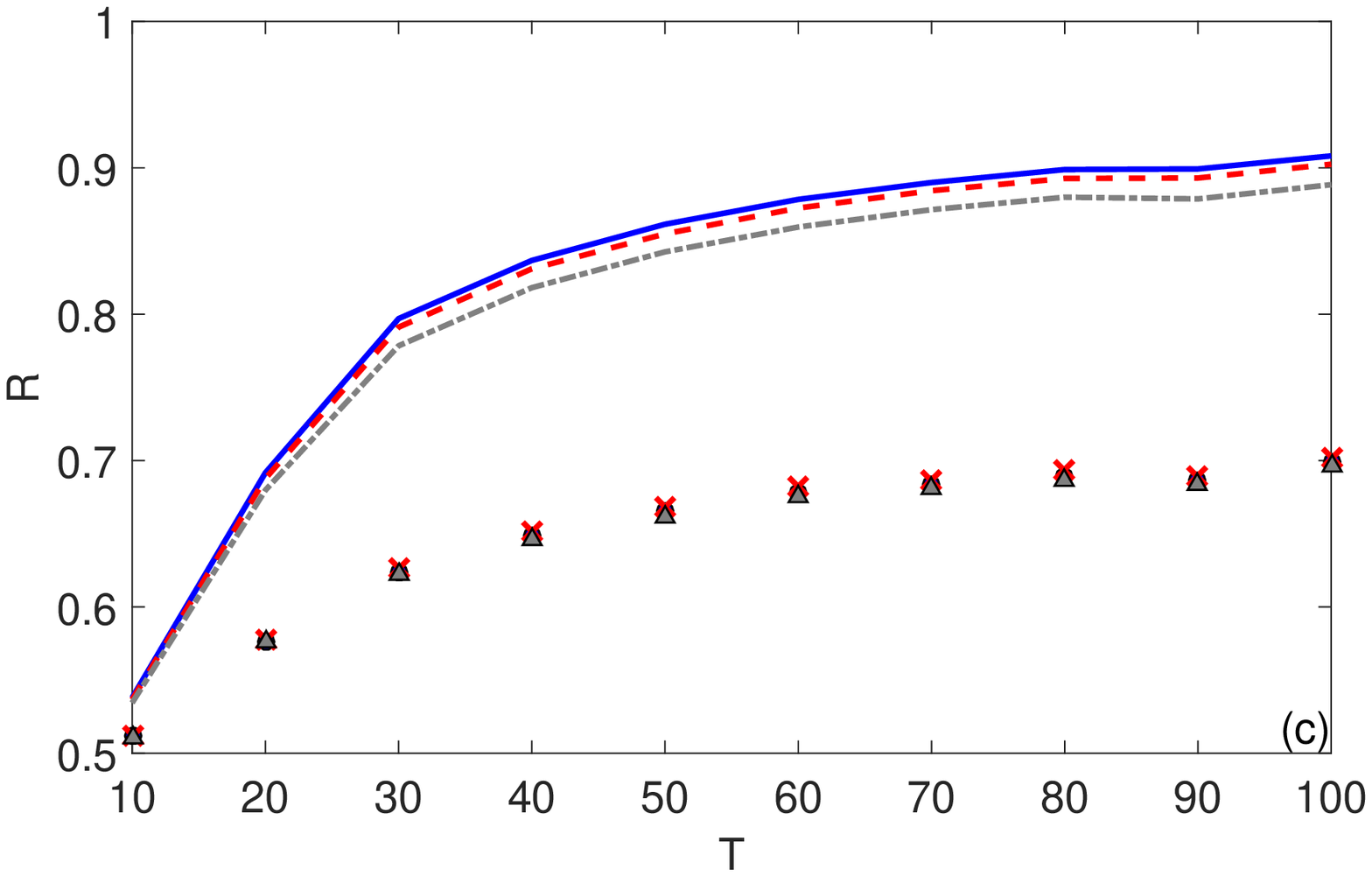}  
\includegraphics[width=75mm]{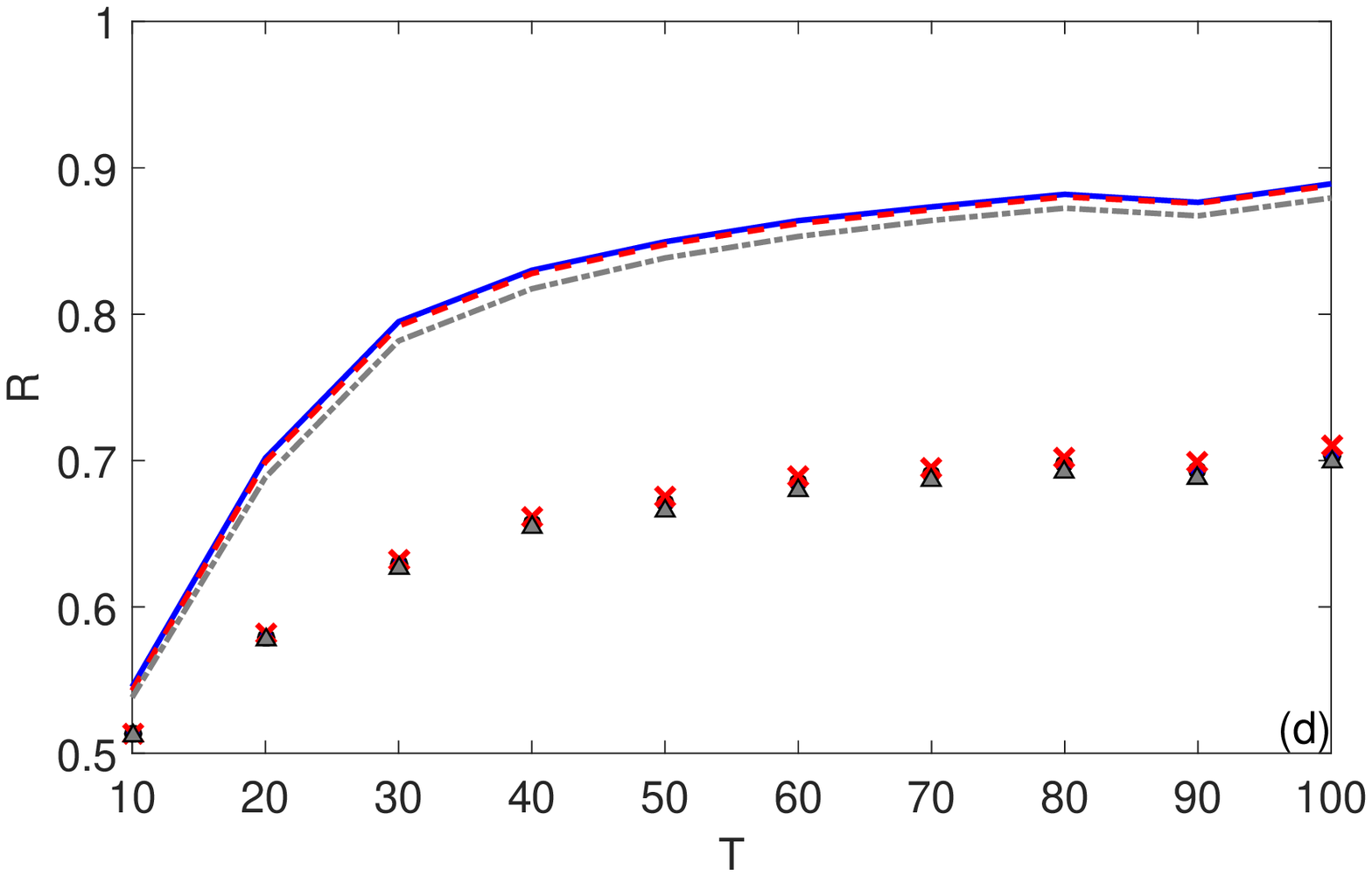}  
\includegraphics[width=75mm]{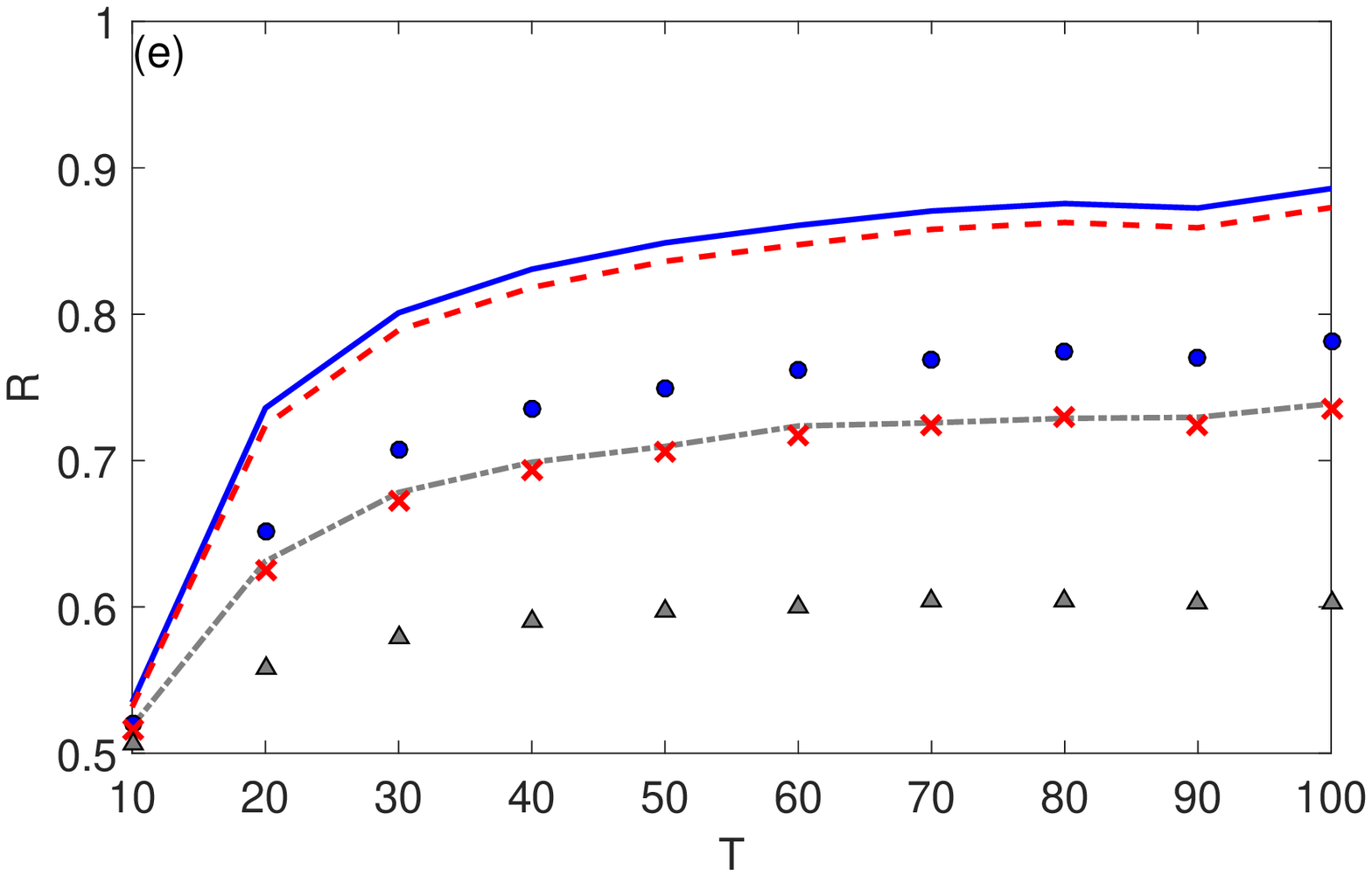}  
\end{center}
\caption{
(Color online) {\bf Model 5} {\bf (a)} A single trajectory of an
intermittent process, alternating a ``slow'' phase of planar Brownian
motion (with $D = 1/2$, dark blue) and a ``fast'' phase of
two-dimensional exponential flights (with $\ell = 3$, light green), of
the mean phase duration $T=100$.  Circle and square indicate the
starting and ending points.  {\bf (b)} The weighted LCH diameter
$S_d(n)$ with the window size $\tau = 10$ applied to this trajectory.
Pink shadow highlights the false classification zones.  Dashed
horizontal line shows the empirical mean $S_d$ over that trajectory.
{\bf (c,d,e)} Recognition score $R$ of the diameter-based
discriminator $S_d(n)$ {\bf (c)}, the volume-based discriminator
$S_v(n)$ {\bf (d)}, and the TA MSD-based discriminator {\bf (e)} as a
function of the mean phase duration $T$.  Lines show the results for
the case $\ell = 3$ with three noise levels $\sigma_n$: $0$ (blue
solid), $0.5 \sigma$ (red dashed), and $\sigma$ (gray dash-dotted)
($\sigma$ being the empirical standard deviation of increment
calculated for each trajectory).  Symbols correspond to the case $\ell
= \sqrt{2}$ with the same levels of noise $\sigma_n$: $0$ (circles),
$0.5\sigma$ (crosses), and $\sigma$ (triangles). }
\label{Result_Bm_expo}
\end{figure*}

Figure \ref{Result_Bm_expo} shows an example of such intermittent
trajectory, the weighted LCH diameter $S_d(n)$ with the window size
$\tau = 10$ applied to this trajectory, and the recognition score for
both the diameter-based and the volume-based classifications.  The
results are similar to that shown in Fig. \ref{Result_Bm_Bm} for
intermittent Brownian motion.  The second choice of the mean flight
length, $\ell = \sqrt{2}$, ensures that the variances of exponential
flights and of Gaussian jumps are equal.  In this case, two phases
differ only in the distribution of increments (Gaussian versus
exponential).  Although these two distributions are relatively close
(e.g., both distributions prohibit very large increments), the
achieved classification is reasonably good.  The performance of the TA
MSD discriminator is comparable to that of the LCH for small noises
but is reduced significantly for a larger noise.

We emphasize that our realization of exponential flights is different
from the run-and-tumble model of bacteria motion, in which case a
walker performs a ballistic motion for a random time and then
undergoes random rotations \cite{Berg1972,Berg}.  We expect that the
recognition score would be even higher for the run-and-tumble motion
(as compared to our model) because two phases are geometrically more
distinct.

\subsection{Surface-mediated diffusion}

In many biological and chemical applications, particles can adsorb to
and desorb from the surface and thus alternate between bulk and
surface diffusions.  For instance, this so-called surface-mediated
diffusion has been suggested as an efficient search mechanism for
DNA-binding proteins \cite{Richter74,Berg81,vonHippel89,Loverdo09}.
The statistics of durations of bulk and surface phases plays an
important role, in particular, in predicting the mean first passage
times \cite{Benichou10,Benichou11,Rupprecht12,Rojo13}.  For this
reason, we test the LCH method on the surface-mediated diffusion
inside a three-dimensional sphere of radius $R$.  The particle starts
from the origin of the sphere and undergoes Brownian motion in the
bulk with diffusion coefficient $D_{3d}$, until the first arrival onto
the surface.  From this moment, the particle adsorbs to the surface
and diffuses on the surface with the diffusion coefficient $D_{2d}$.
The surface diffusion occurs during a random exponentially distributed
waiting time (with the rate $\lambda$).  After desorption, the
particle is ejected into the bulk to the distance $a = 0.05 R$ from
the boundary and then resumes its bulk diffusion.  Here we consider
the equal diffusion coefficients $D_{2d} = D_{3d} = 10^{-3}$.

\begin{figure*}
\begin{center}
\includegraphics[width=75mm]{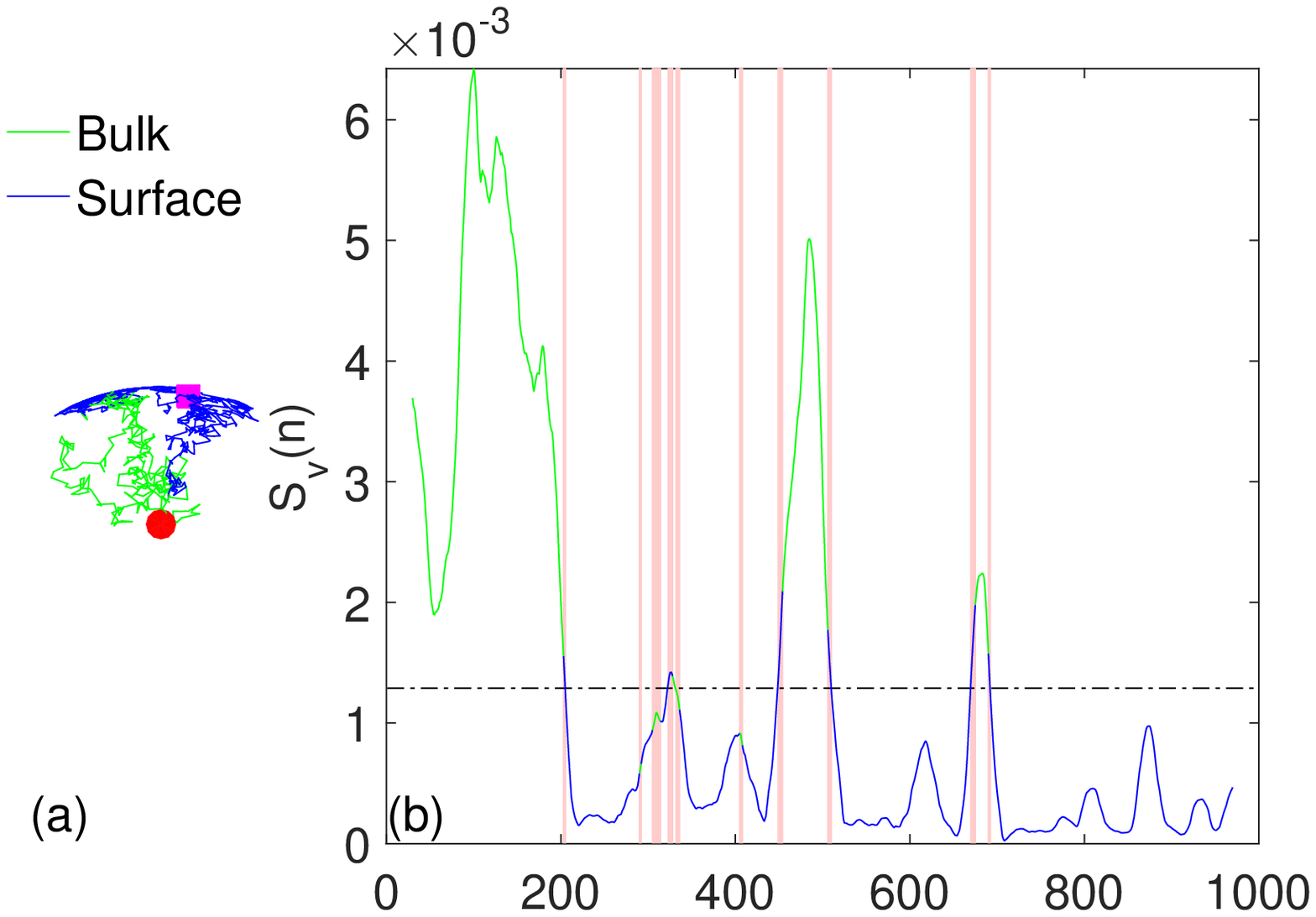}  
\includegraphics[width=75mm]{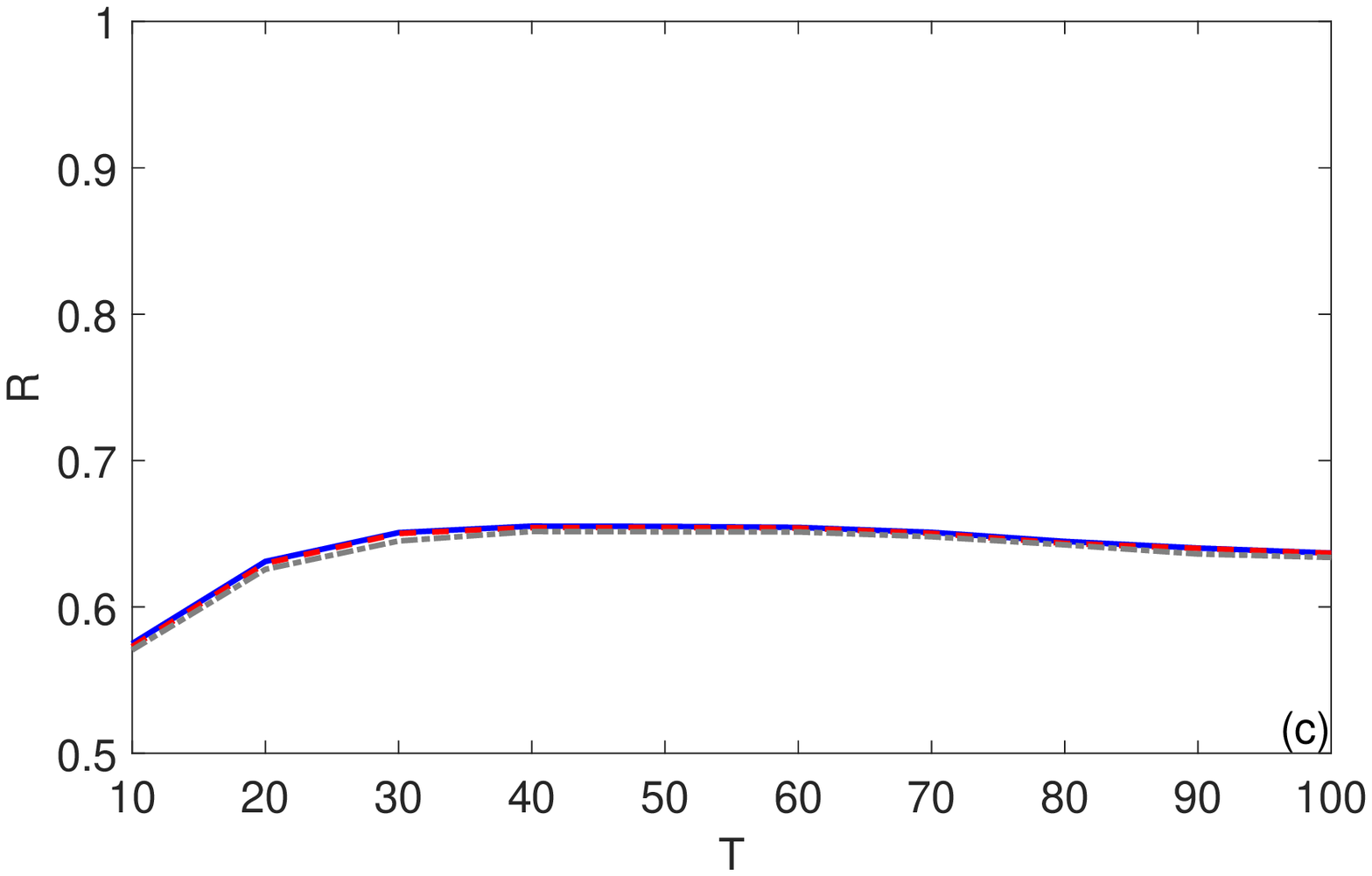}   
\includegraphics[width=75mm]{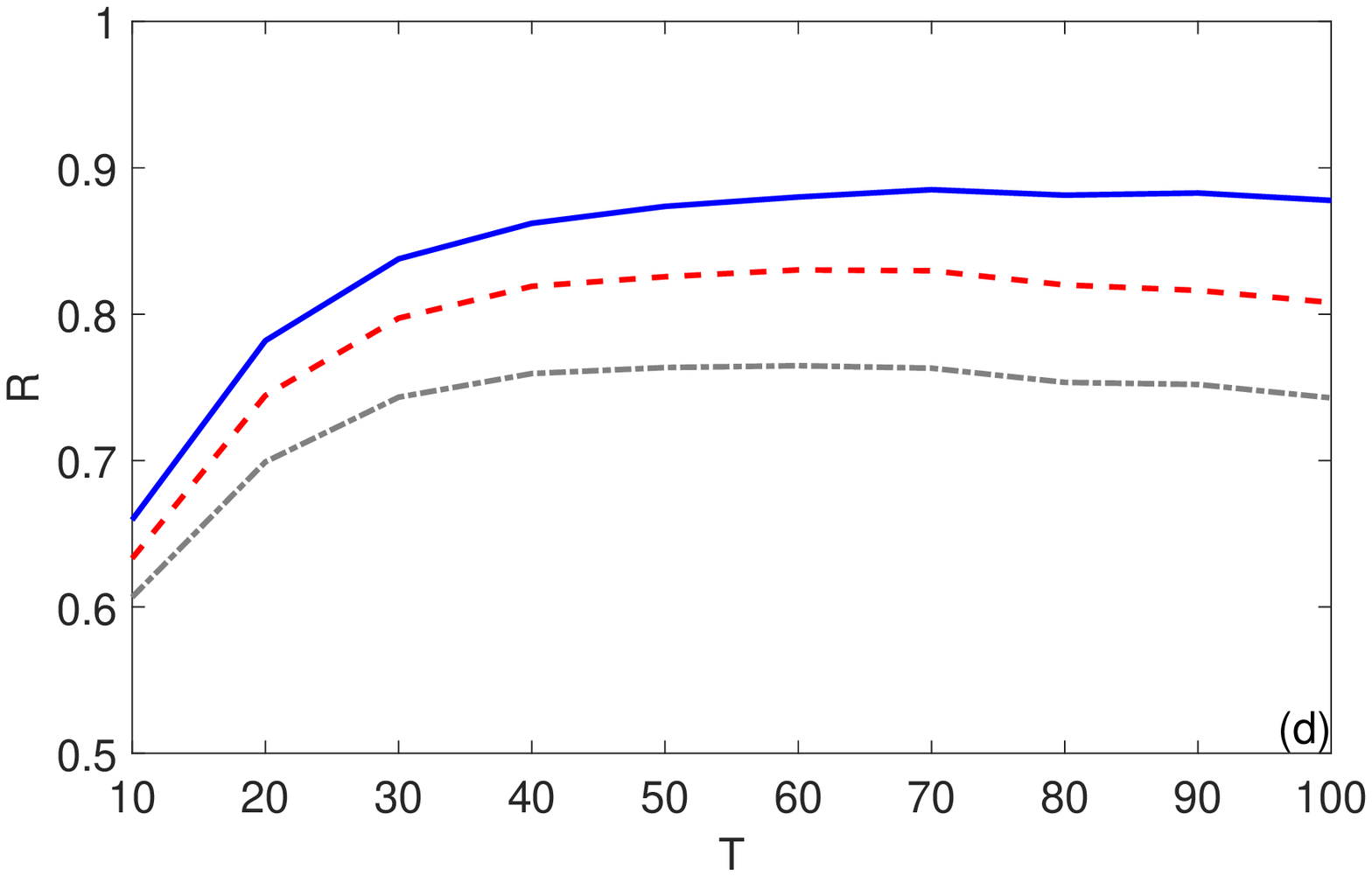}   
\includegraphics[width=75mm]{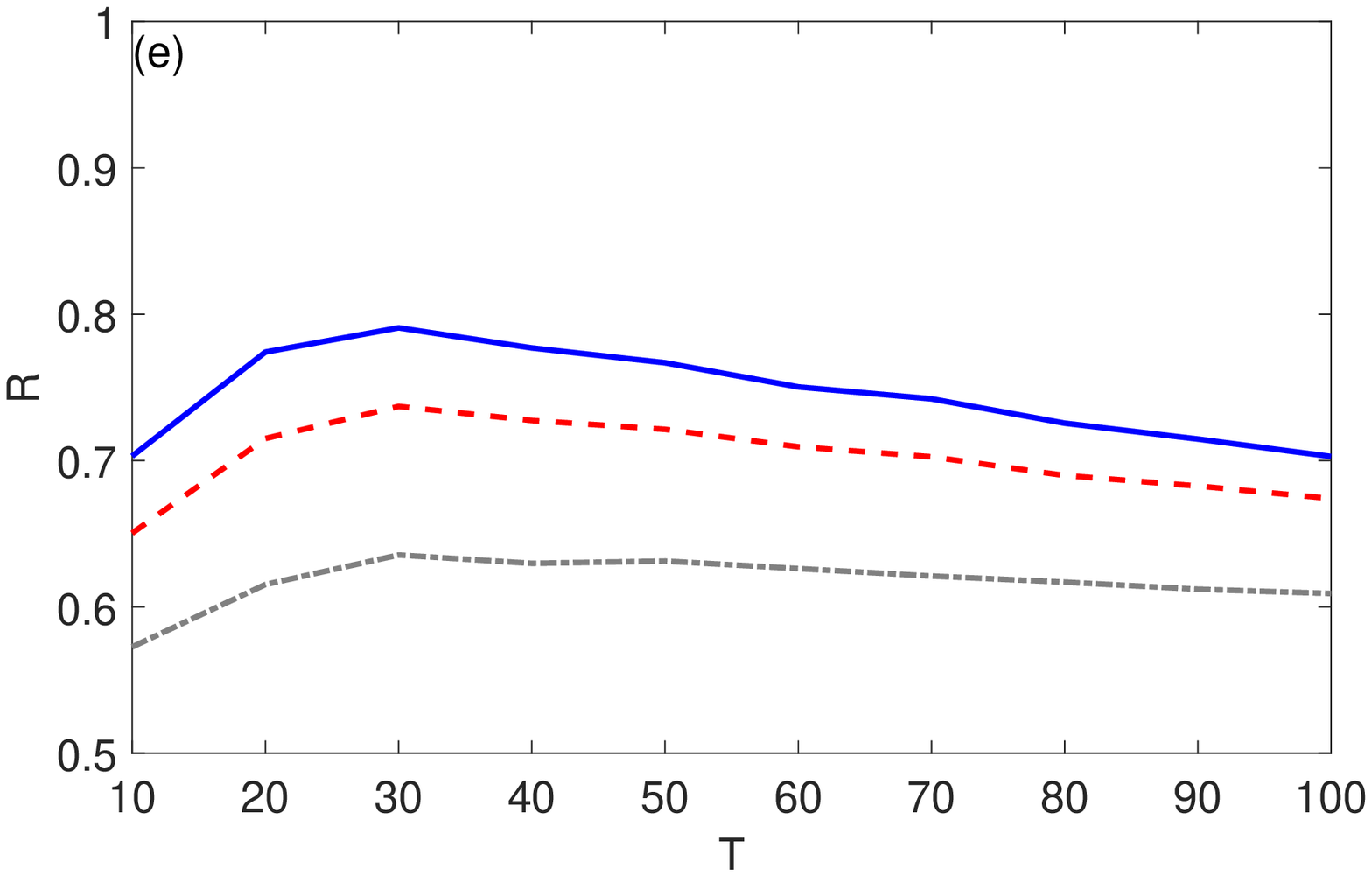}   
\end{center}
\caption{
(Color online) {\bf Model 6} {\bf (a)} A single trajectory of
surface-mediated diffusion in a three-dimensional sphere of radius $R
= 1$, alternating bulk diffusion (light green) and surface diffusion
(dark blue), with $D_{2d} = D_{3d} = 10^{-3}$ and $\lambda = 10^{-2}$.
Circle and square indicate the starting and ending points.  {\bf (b)}
The weighted LCH diameter $S_d(n)$ with the window size $\tau = 10$
applied to this trajectory.  Pink shadow highlights the false
classification zones.  Dashed horizontal line shows the empirical mean
$S_d$ over that trajectory.  {\bf (c,d,e)} Recognition score $R$ of
the diameter-based discriminator $S_d(n)$ {\bf (c)}, the volume-based
discriminator $S_v(n)$ {\bf (d)}, and the TA MSD-based discriminator
{\bf (e)} as a function of the mean duration of the surface phase $T =
1/\lambda$.  Three curves correspond to three noise levels $\sigma_n$:
$0$, $0.5 \sigma$, and $\sigma$ ($\sigma$ being the empirical standard
deviation of increment calculated for each trajectory). }
\label{Result_intermittent}
\end{figure*}

Figure \ref{Result_intermittent} shows an example of such intermittent
trajectory, the weighted LCH diameter $S_d(n)$ with the window size
$\tau = 10$ applied to this trajectory, and the recognition score for
both the diameter-based and the volume-based classifications.  The
significant difference of the surface-mediated diffusion as compared
to the earlier considered models of intermittent processes is that the
bulk phase duration is not an exponentially distributed random
variable, it is determined by the statistics of first arrivals onto
the surface.  In turn, the duration of surface diffusion can be
controlled by the desorption rate $\lambda$.  Given that the mean
surface duration is $1/\lambda$, we formally set $T = 1/\lambda$.
This distinction explains new features in the recognition scores shown
in Fig. \ref{Result_intermittent}c,d.  First, one can see that the
volume-based discriminator greatly outperforms the diameter-based one.
This is not surprising because the volume of the LCH is much more
sensitive to the dimensionality reduction than the diameter.  Second,
in both cases, the recognition score does not grow monotonously with
the mean phase duration $T$.  In fact, the mean duration of the bulk
phase, $(R^2 - (R-a)^2)/(6D)$, is fixed (and equal to $16.25$ in our
example), whereas the mean duration of the surface phase is
progressively increased.  As a consequence, as $T$ grows, it becomes
more difficult to detect the short bulk phases that results in the
decrease of the recognition score.  A similar behavior can be seen in
Fig. \ref{unequal_duration} for intermittent Brownian motion (Model 1)
when the durations of two phases differ significantly.  Note also that
the impact of noise is stronger than in other cases.  The performance
of the TA MSD discriminator is good but poorer than that of the
volume-based estimator.  This good performance is explained by the
fact that the change in dimensionality from three dimensions to two
dimensions also reduces the MSD from $6Dt$ and $4Dt$ that is captured
by TA MSD estimator.

\section{Discussion}

The numerical validation has shown that the LCH method can detect
change points between the phases that differ either by amplitudes of
increments, or by the presence of drift, or by auto-correlations
between increments, or by distribution of increments, or by
dimensionality of the explored space.  Moreover, the method is robust
against noise due to the integral-like nature of the LCH-based
estimators.  In this section, we discuss the choice of the parameters,
several limitations and future improvements of the method.

\subsection{Parameters of the method}
\label{sec:parameters}

Although the LCH method is model-free, there are two parameters to be
chosen: the window size $\tau$ and the threshold $S_d$ (or $S_v$) used
for the binary classification.  Let us discuss the choice of these
parameters in more detail.

\begin{figure}
\begin{center} 
\includegraphics[width=85mm]{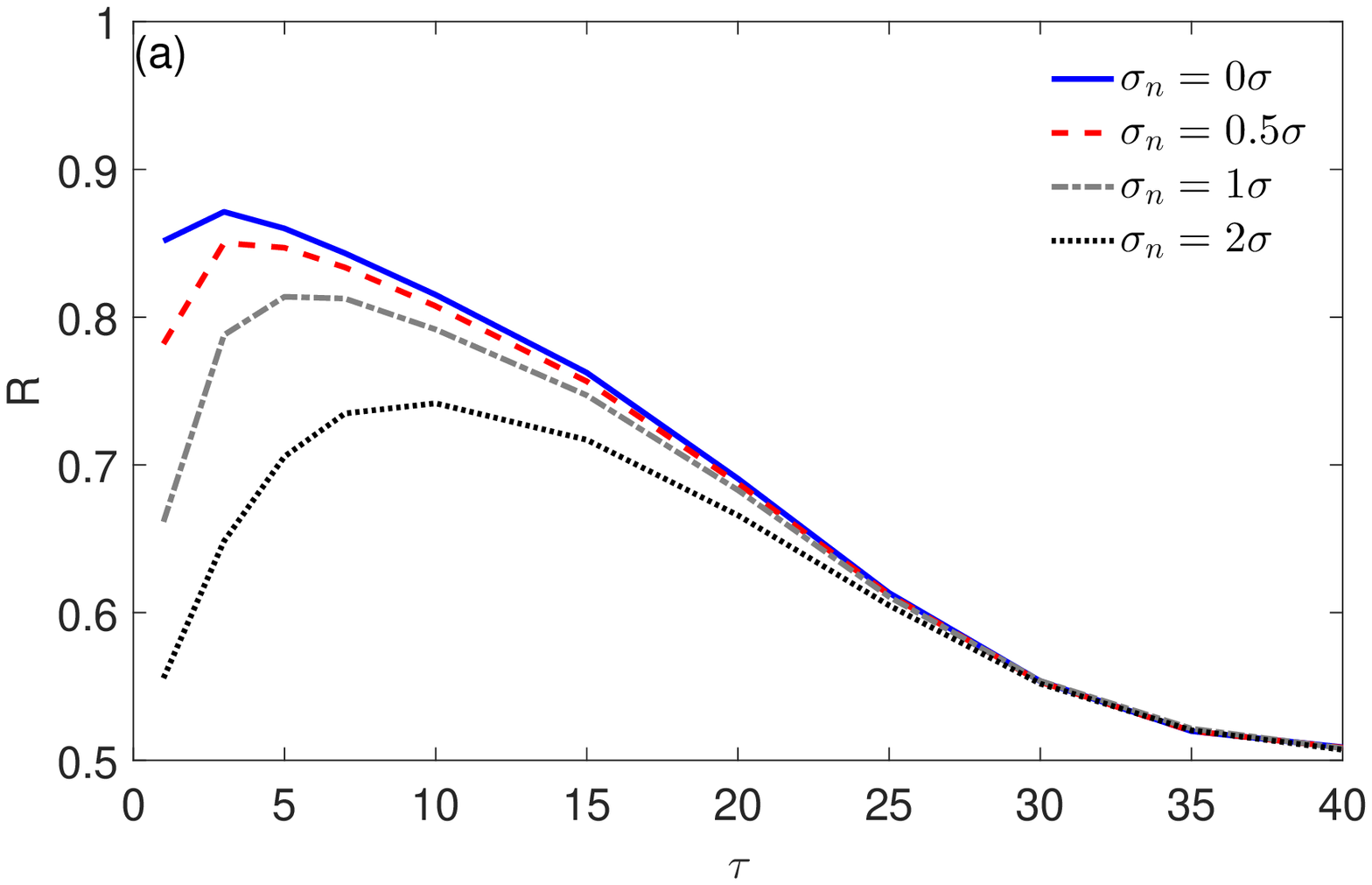} 
\includegraphics[width=85mm]{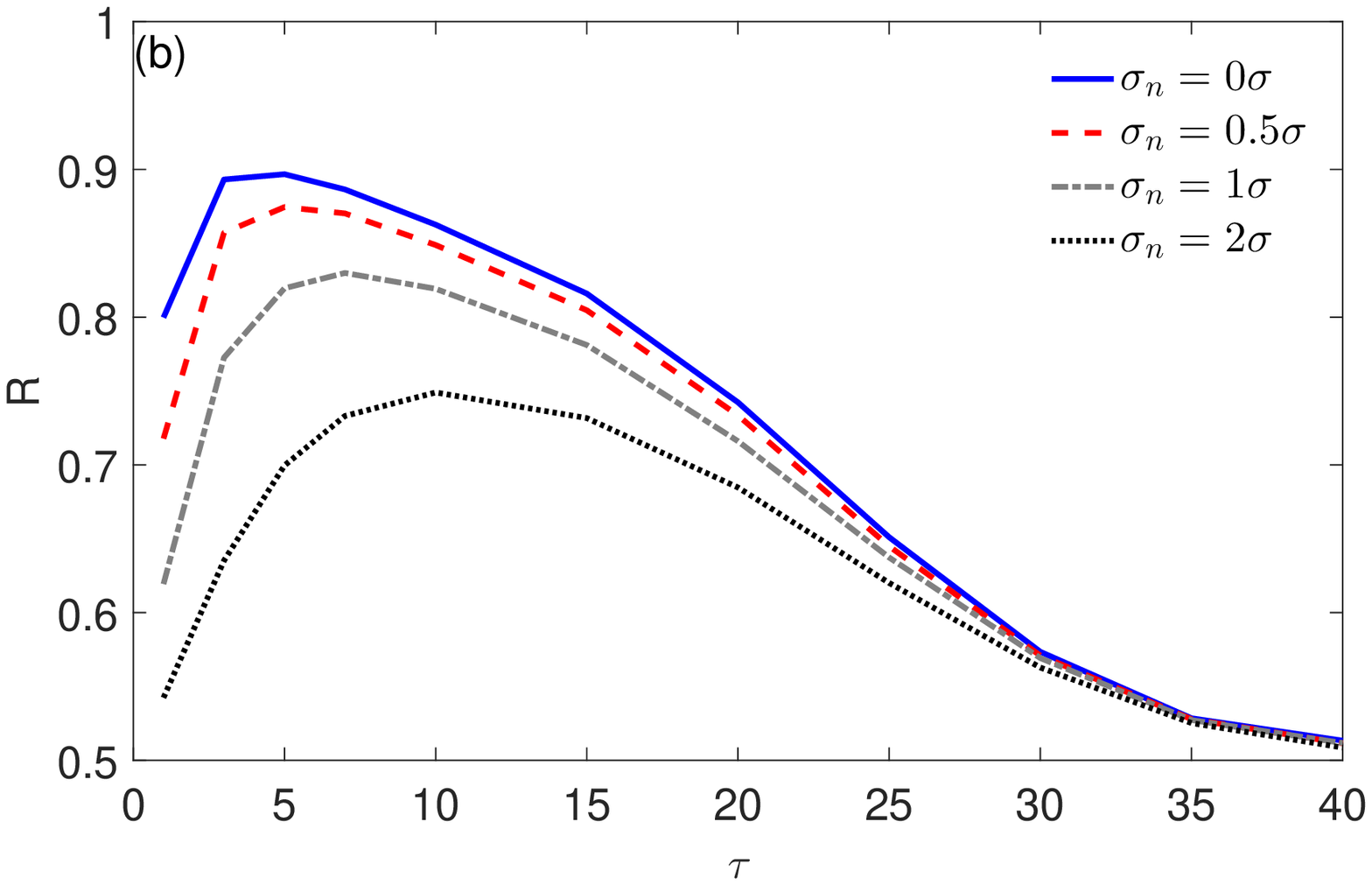} 
\end{center}
\caption{
(Color online) Effect of the window size $\tau$ on the recognition
score $R$ for intermittent Brownian motion (Model 1) with $D_1 = 1/2$,
$D_2 = 2$, and the mean phase durations $T_1 = T_2 = 40$.  {\bf (a)}
The diameter-based discriminator $S_d(n)$; {\bf (b)} the volume-based
discriminator $S_v(n)$.  Four curves correspond to four noise levels
$\sigma_n$: $0$, $0.5 \sigma$, $\sigma$, and $2\sigma$ ($\sigma$ being
the empirical standard deviation of increment calculated for each
trajectory). }
\label{Result_window_size}
\end{figure}

Throughout Sec. \ref{sec:Models_to_test}, we set $\tau = 10$.  Figure
\ref{Result_window_size} shows the effect of the window size $\tau$ on
the recognition score $R$ for intermittent Brownian motion (Model 1)
with $D_1 = 1/2$, $D_2 = 2$, and the mean phase durations $T_1 = T_2 =
40$.  First, one can see that the recognition score as a function of
$\tau$ is not monotonous, i.e., there is an optimal window size
$\tau_c$ that maximizes the recognition score.  This optimality
results from a compromise between the reactivity and the robustness of
the method.  In fact, the LCH contains too many points at large $\tau$
that leads to larger delays between actual and detected change points
and thus increases the fraction of false classifications.  In turn,
when $\tau$ is too small, the method is reactive (delays are short)
but also too sensitive to stochastic fluctuations within one phase; as
a consequence, the fraction of false classifications is also higher
due to spontaneous crossings of the discriminator $S_d(n)$ (or
$S_v(n)$) of the mean level $S_d$ (or $S_v$).  This latter effect is
drastically enhanced in the presence of noise (see how the curves with
larger noise levels $\sigma_n$ are diminished at short $\tau$).  As a
consequence, the optimal window size $\tau_c$ is increased for noisier
data.  Clearly, the $\tau_c$ should also depend on the phase duration.
One cannot therefore choose the optimal window size without {\it a
priori} knowledge about the noise and phase durations.  In practice,
the range $5\leq \tau \leq 10$ seems to be the reasonable choice of
the window size.  Note that the recognition score versus $\tau$ for
the intermittent fBm (Model 3) exhibits very similar behavior (not
shown).

\begin{figure}
\begin{center} 
\includegraphics[width=28mm]{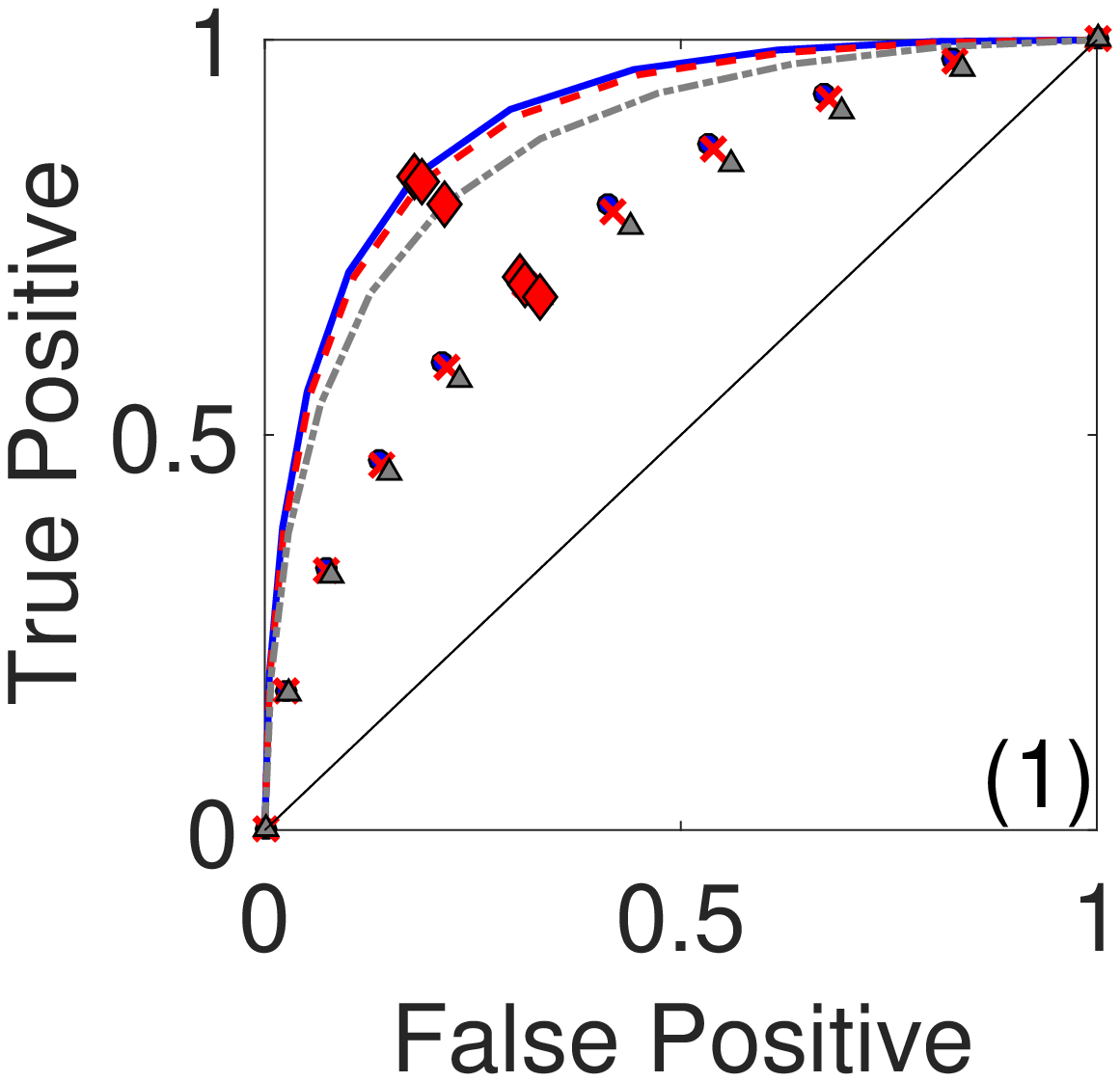} 
\includegraphics[width=28mm]{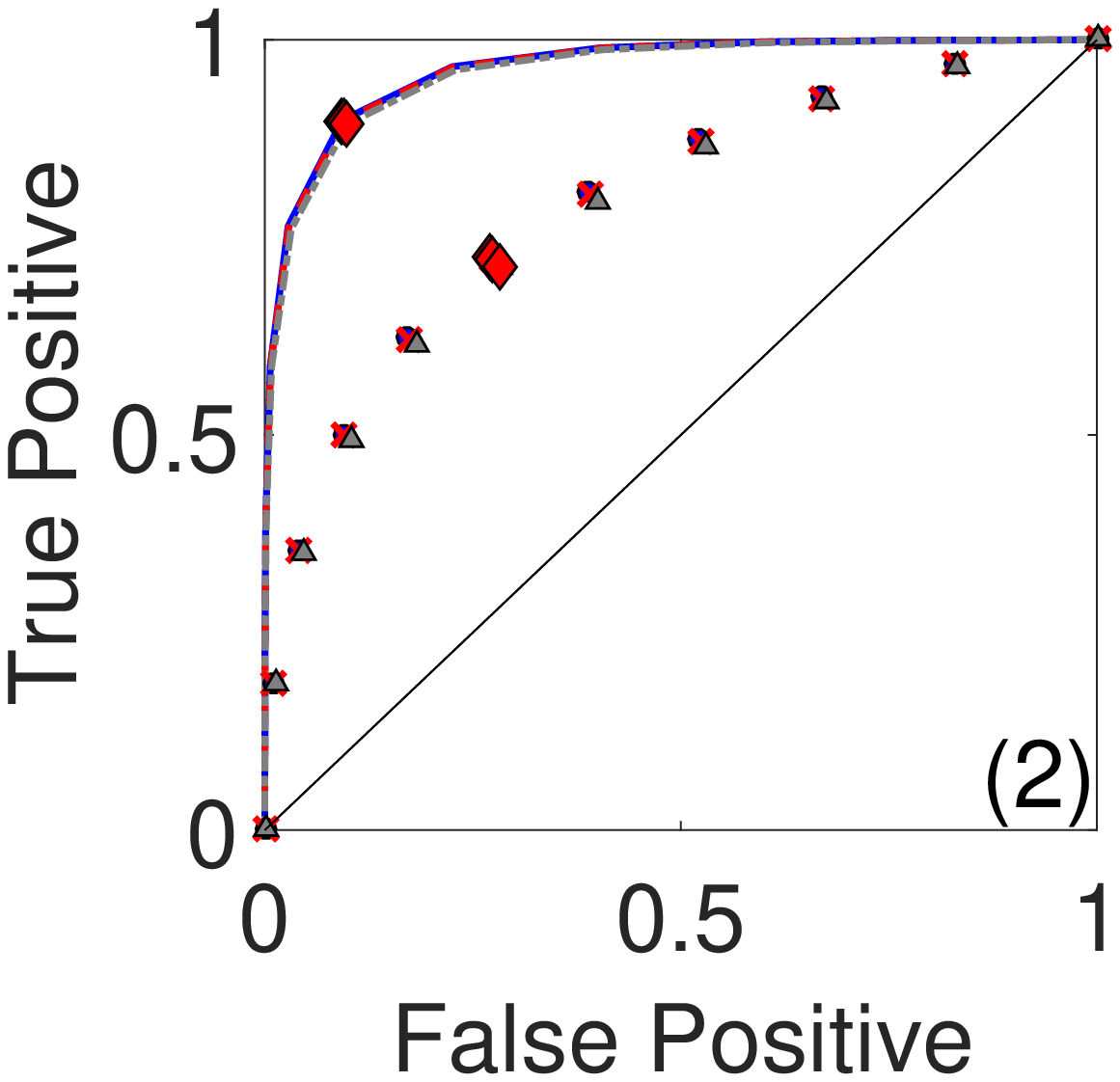} 
\includegraphics[width=28mm]{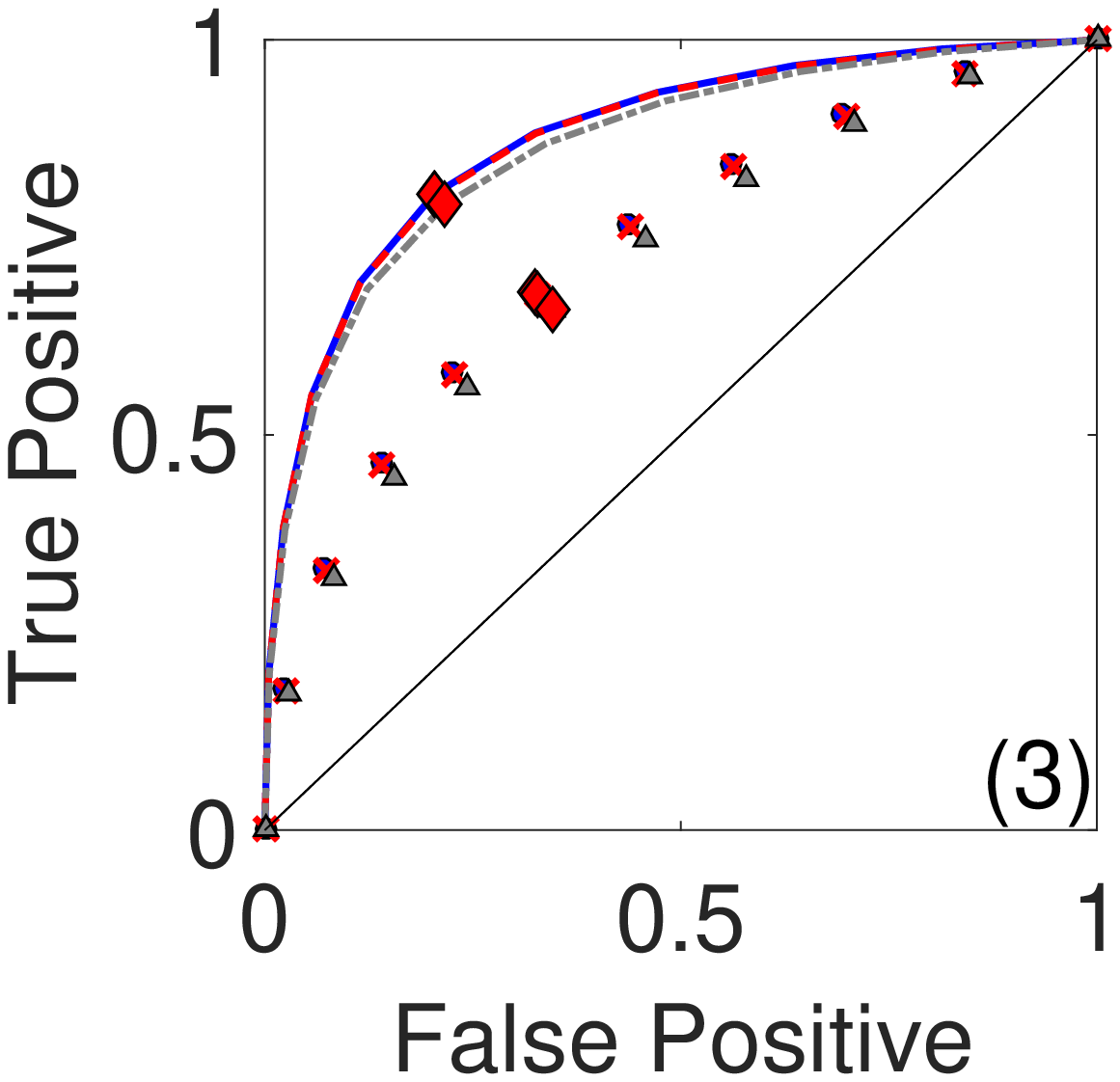} 
\includegraphics[width=28mm]{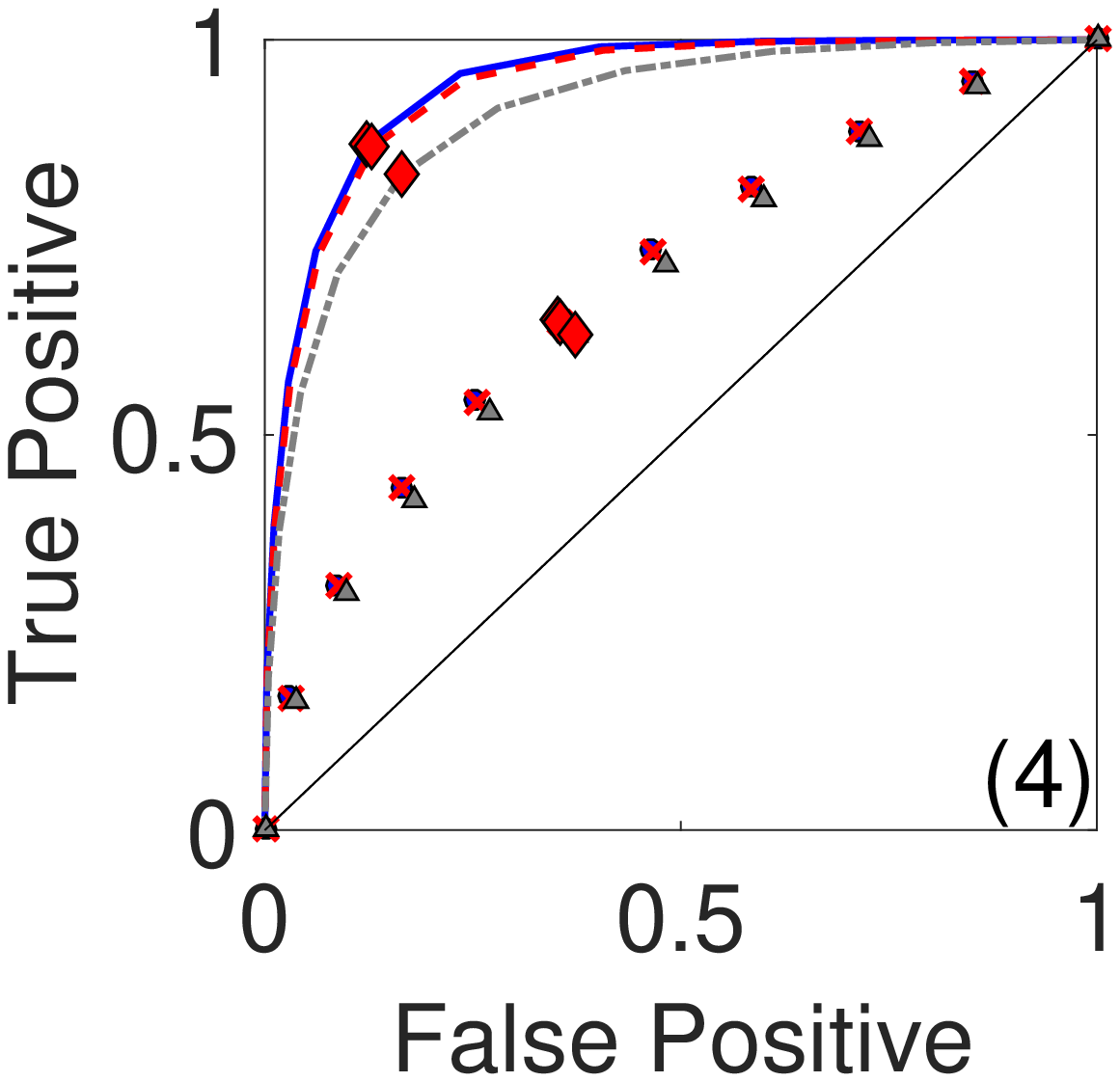} 
\includegraphics[width=28mm]{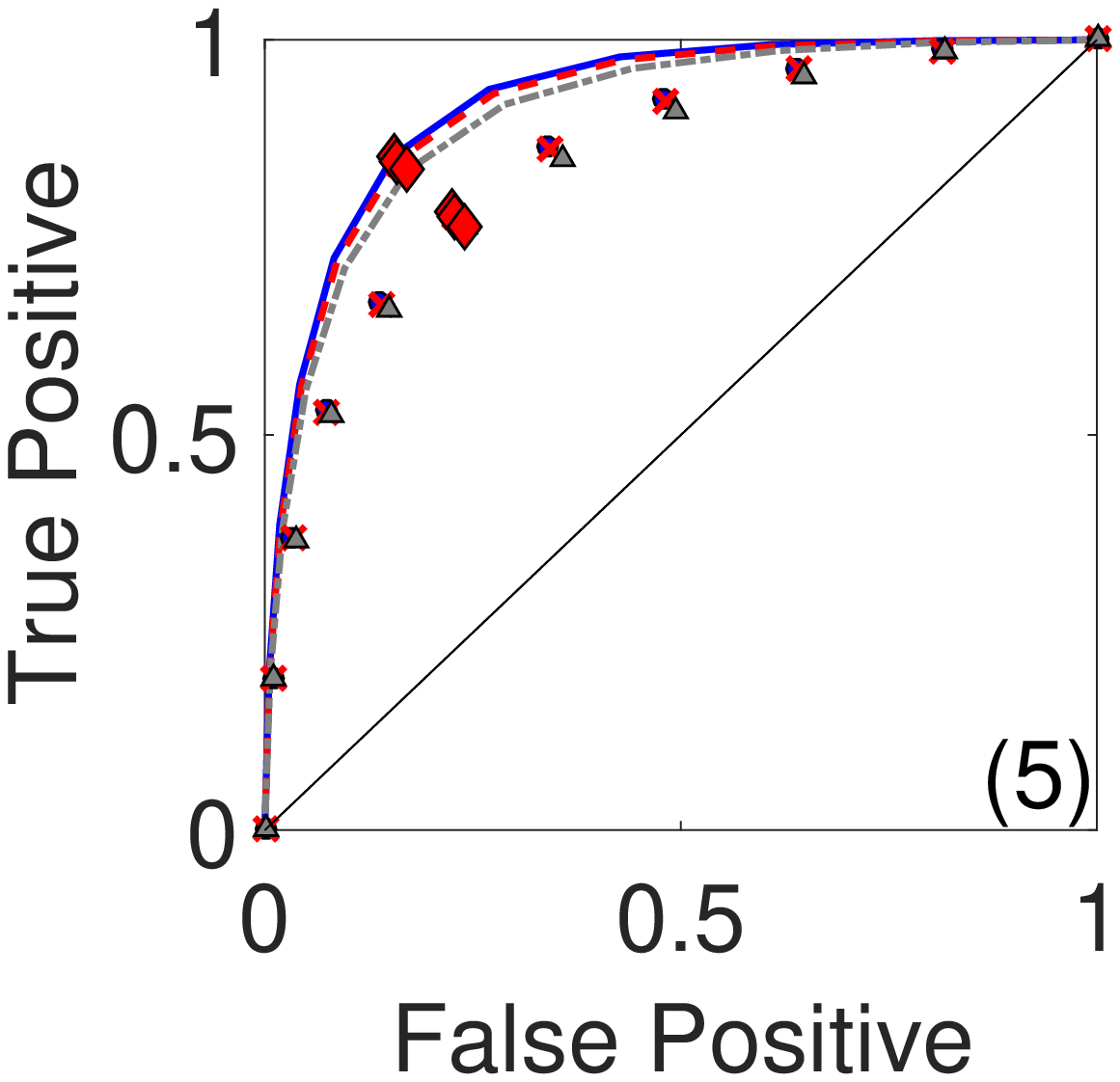} 
\includegraphics[width=28mm]{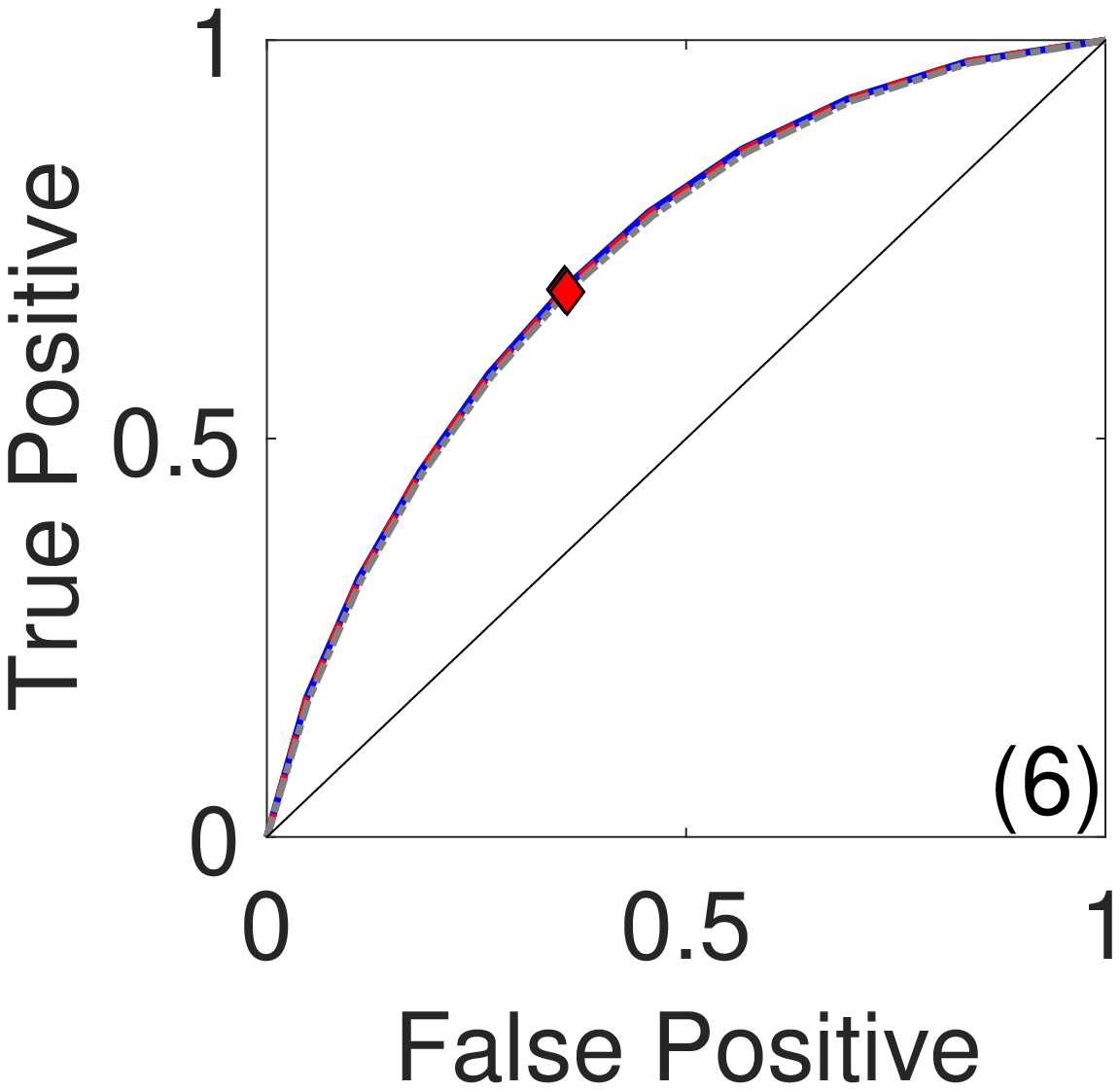} 
\includegraphics[width=28mm]{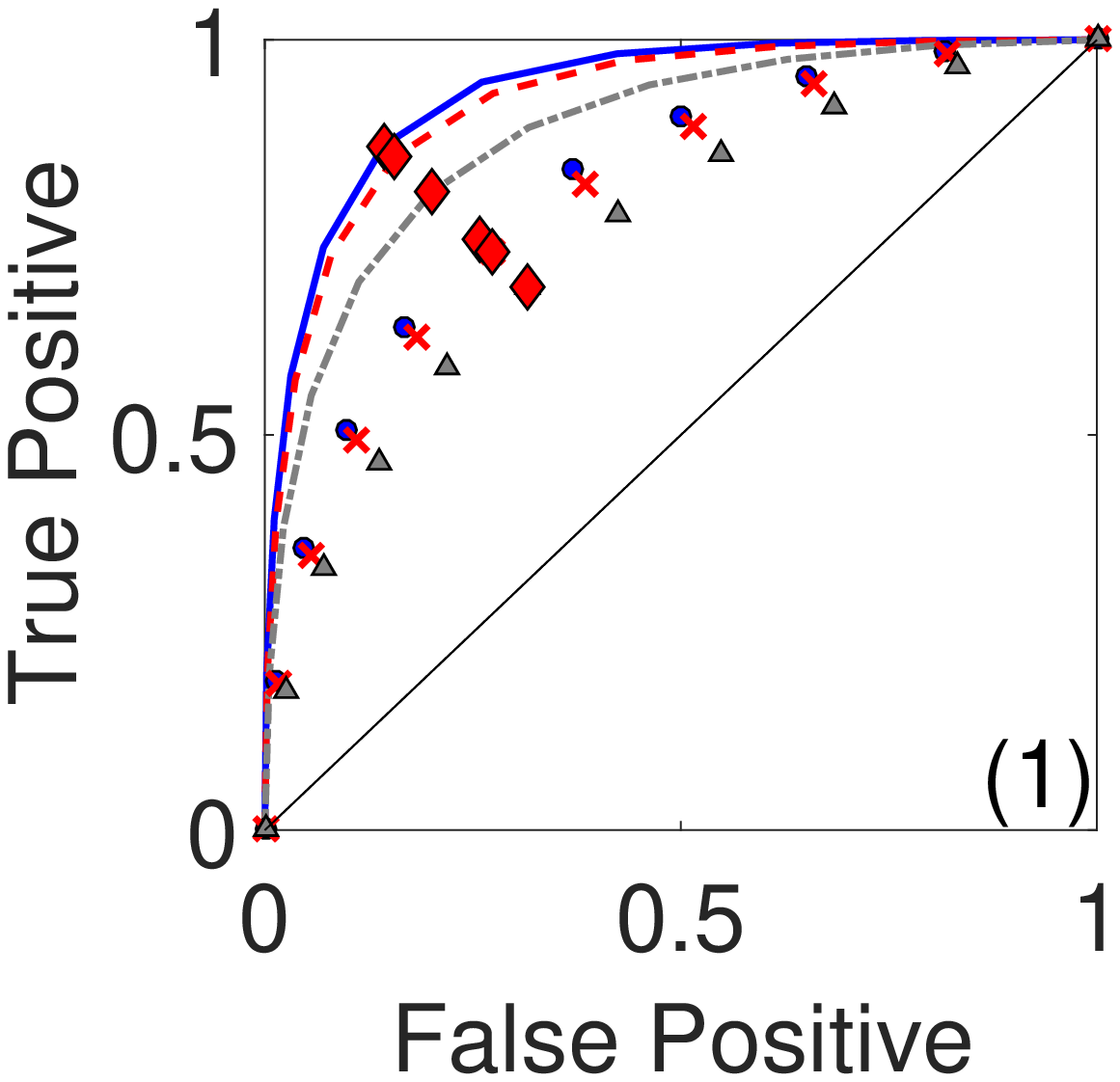} 
\includegraphics[width=28mm]{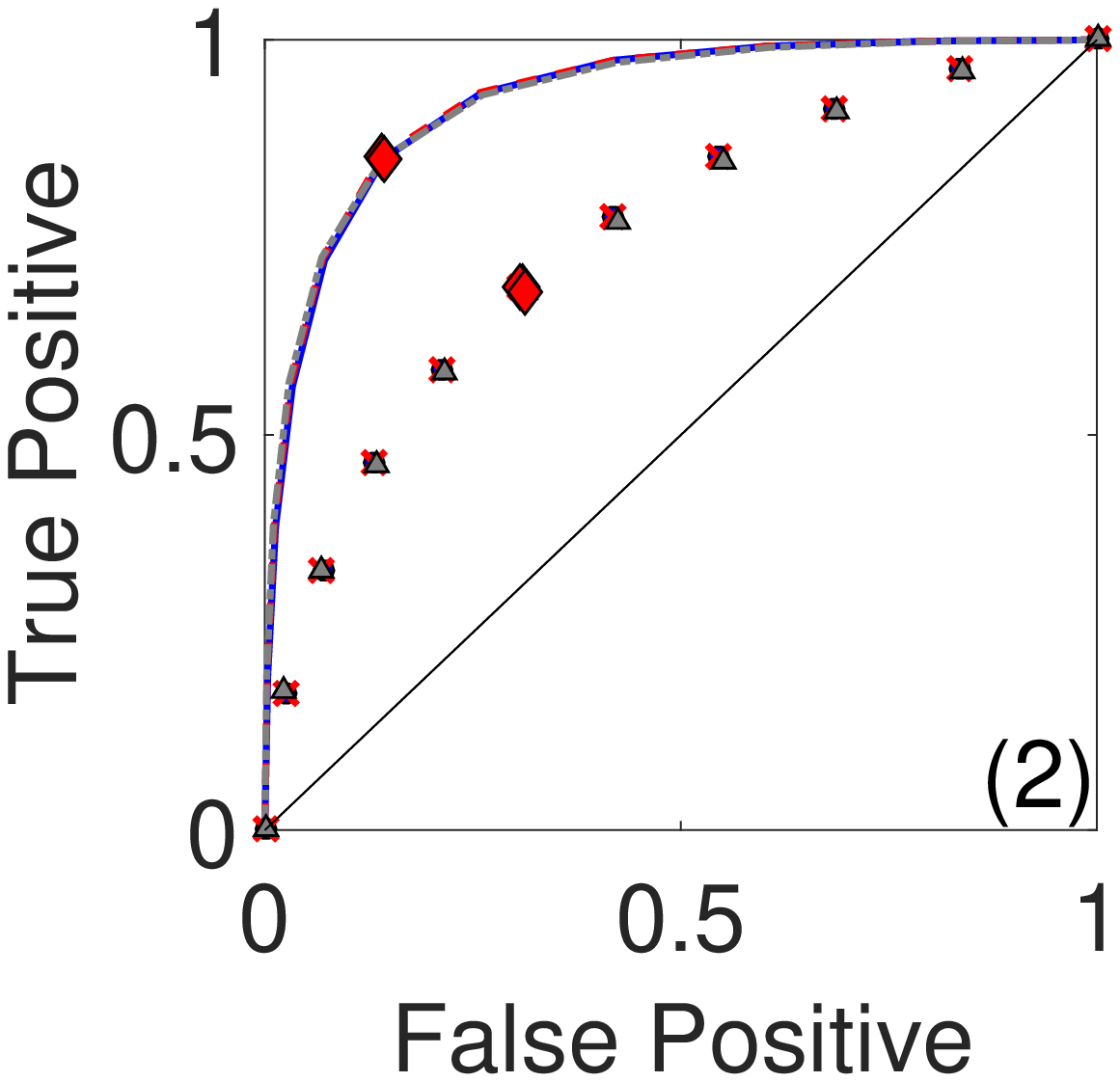} 
\includegraphics[width=28mm]{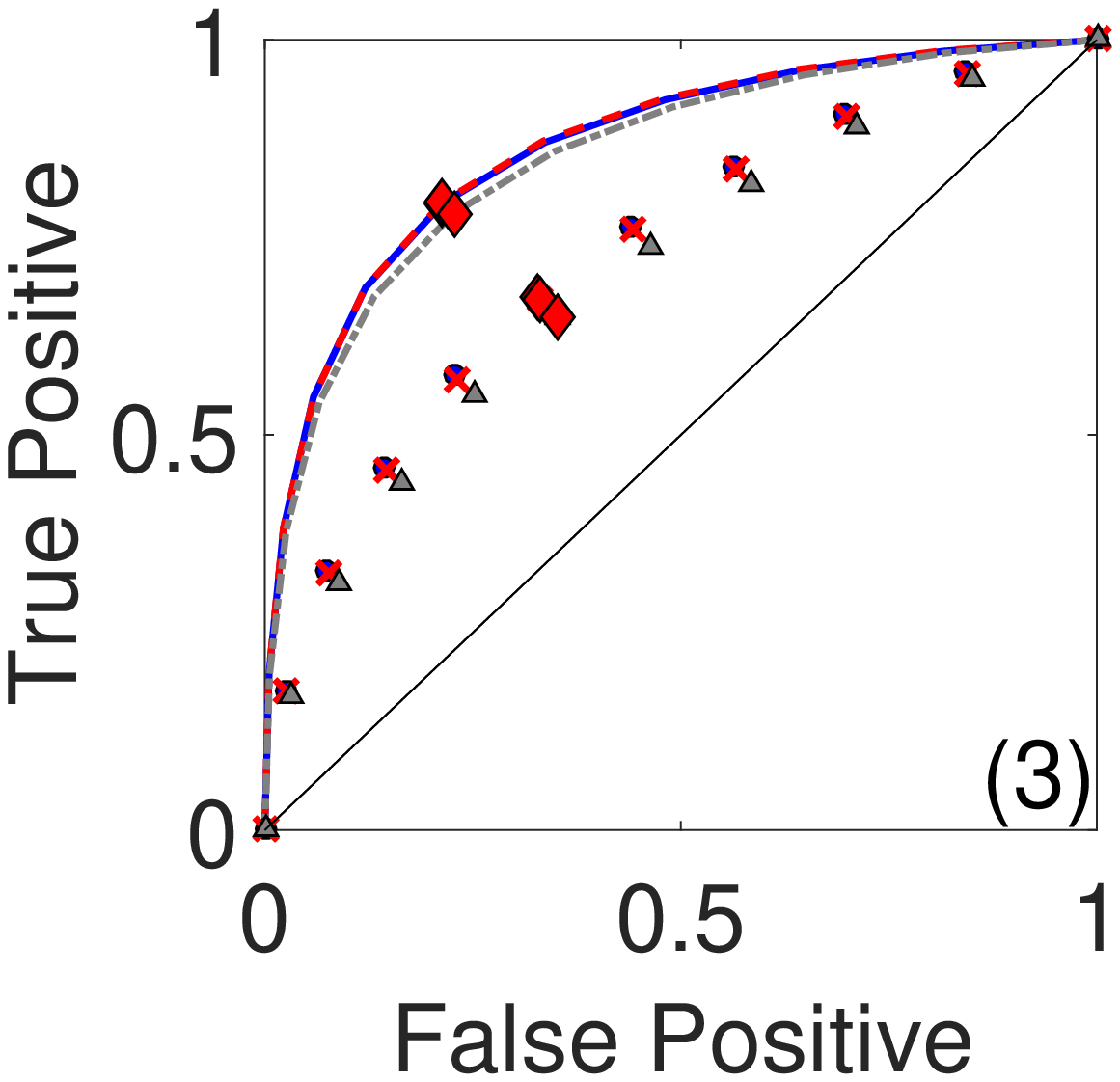} 
\includegraphics[width=28mm]{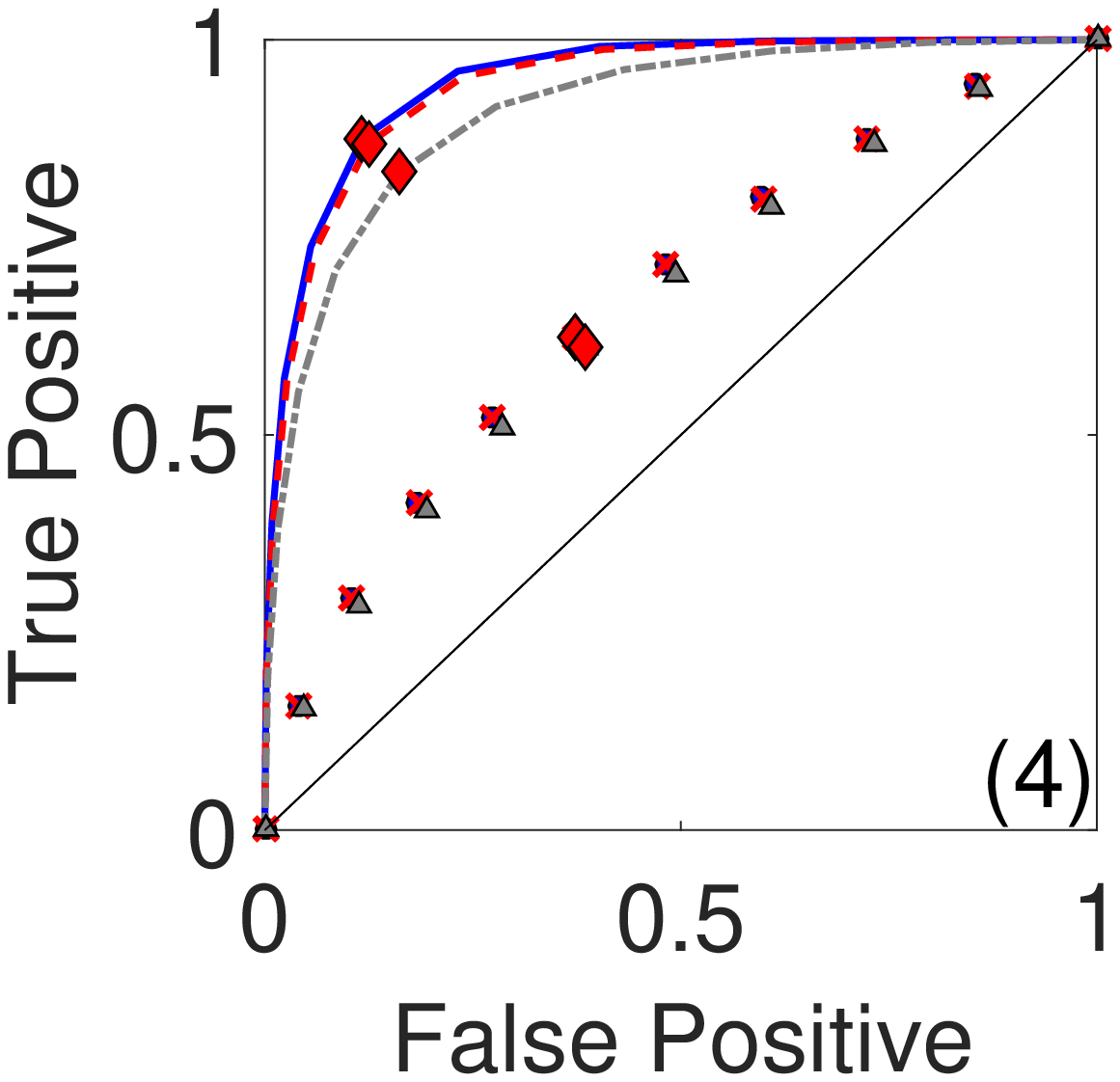} 
\includegraphics[width=28mm]{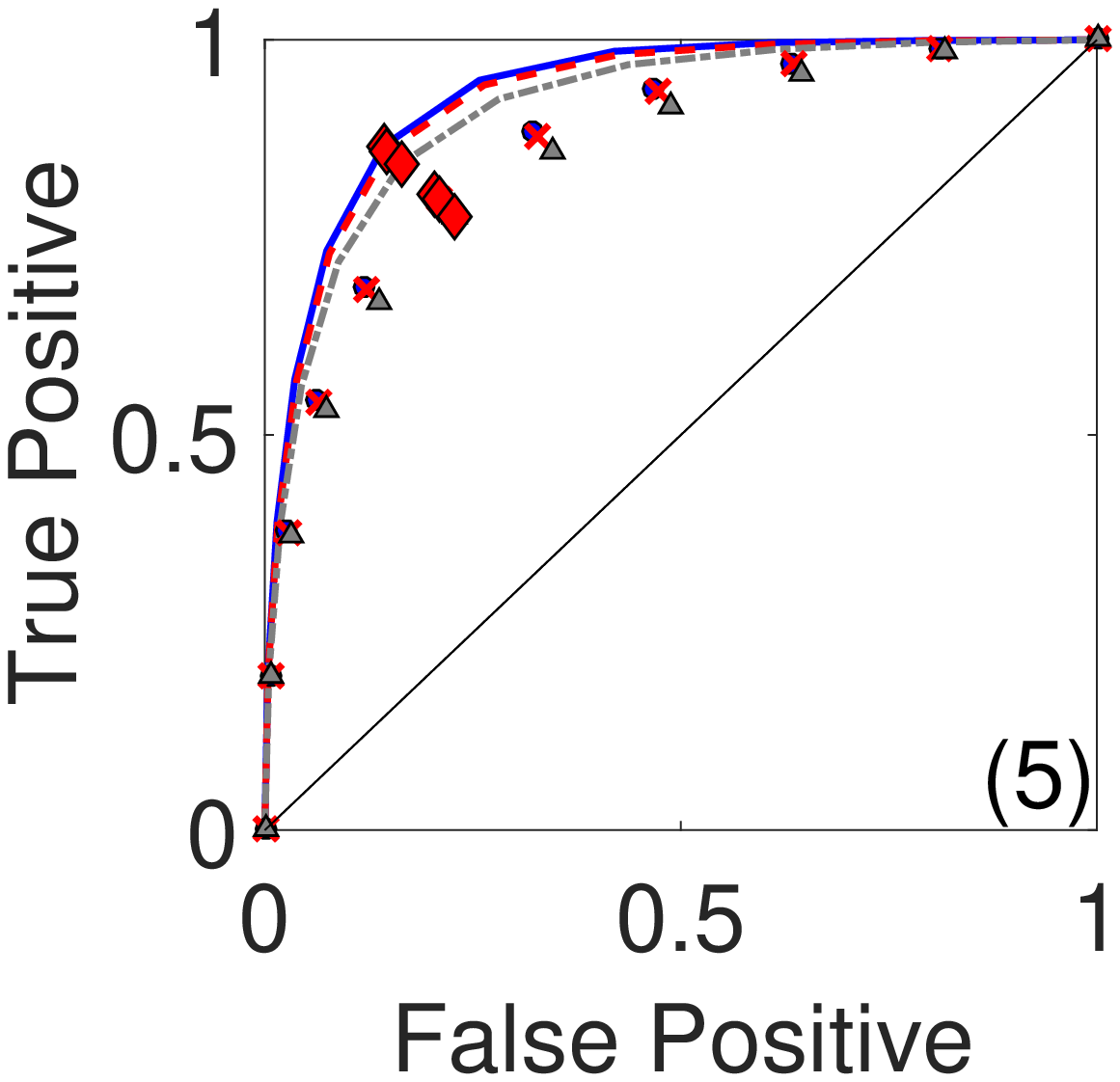} 
\includegraphics[width=28mm]{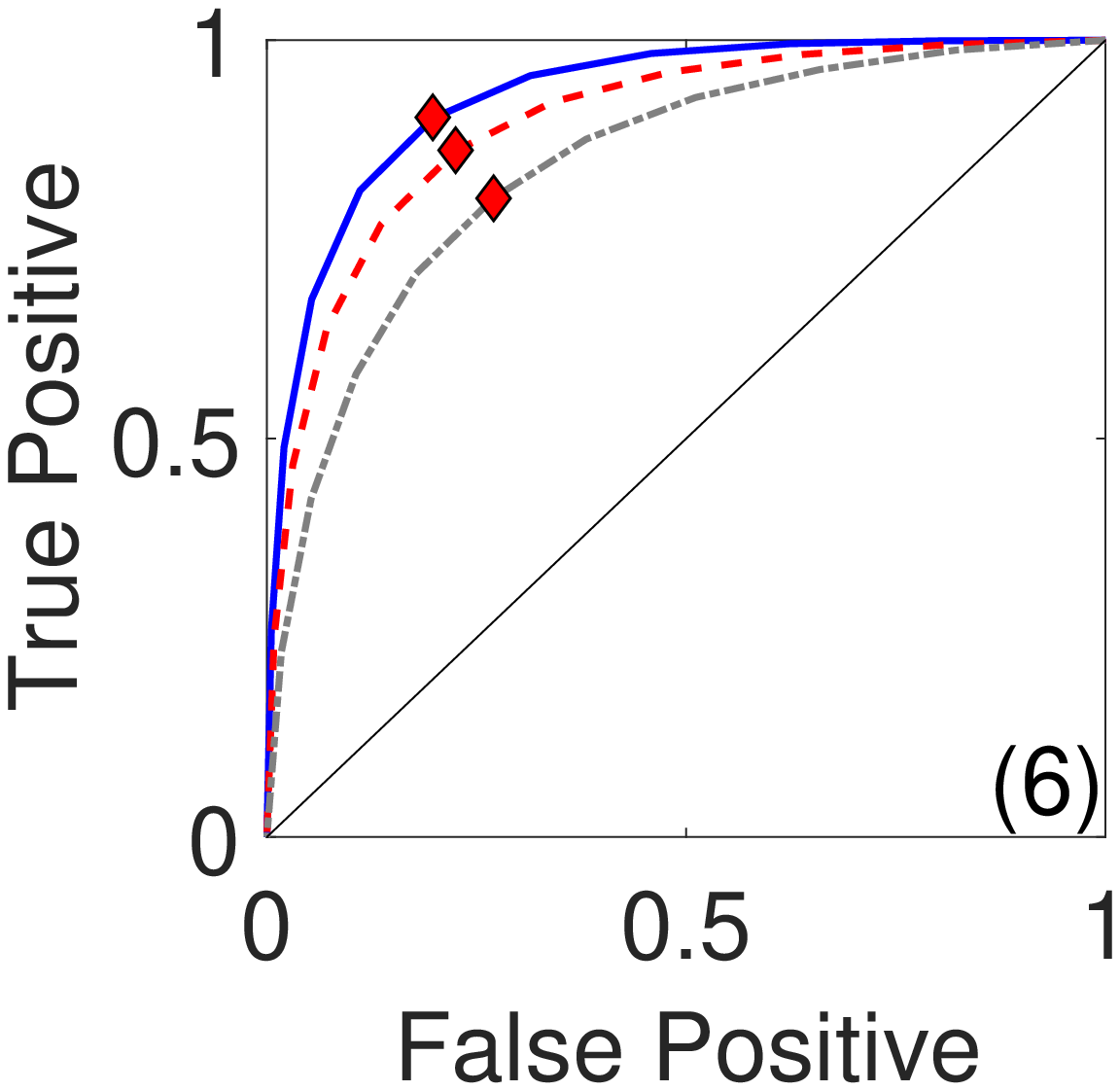} 
\end{center}
\caption{
(Color online) Effect of the threshold on the recognition quality for
six models (following the enumeration at the beginning of
Sec. \ref{sec:Models_to_test}), for the diameter-based discriminator
(top 6 plots) and the volume-based discriminator (bottom 6 plots).
The left lower corner and the right upper corner correspond to the
threshold at the maximal and the minimal values, whereas diamonds
indicate the mean value $S_d$ (or $S_v$) used as the threshold
throughout the paper.  Legend of six curves on each plot is the same
as in Figs. \ref{Result_Bm_Bm}, \ref{Result_Bm_drift},
\ref{Result_fBm_fBm}, \ref{Result_Bm_OUP}, \ref{Result_Bm_expo},
\ref{Result_intermittent} for the recognition score (three levels of
noise and two sets of model parameters, except for the Model 6). }
\label{Result_threshold}
\end{figure}

The second parameter of the method is the threshold $S_d$ (or $S_v$)
that is used to distinguish ``slow'' and ``fast'' phases.  Without
prior knowledge about the process, we chose the arithmetic mean of the
discriminator $S_d(n)$ (or $S_v(n)$) over the trajectory as the
threshold.  However, this choice is not necessarily optimal.  For
instance, one could use another mean (e.g., quadratic or harmonic), or
set the threshold to be proportional to the mean, or choose another
function or constant.  In order to justify this empirical
choice, we consider a receiver operating characteristic (ROC) curve
for both diameter-based and volume-based discriminators.  For this
purpose, we compute the true positive rate (the fraction of ``fast''
phase points that were identified as ``fast'') and the false positive
rate (the fraction of ``slow'' phase points that were identified as
``fast'') by varying the threshold from the minimal to the maximal
value of the discriminator $S_d(n)$ (or $S_v(n)$).  Figure
\ref{Result_threshold} shows the ROC curves for six considered models
and for two discriminators at the mean duration time $T = 40$.  An
ideal discriminator would yield the true positive rate at $1$ and the
false positive rate at $0$ (the left upper corner), whereas a random
discriminator would fill the diagonal.  The threshold at the minimal
value of the discriminator classifies all points as belonging to the
``fast'' phase (as $S_d(n) \geq \min_n \{S_d(n)\}$) that corresponds
to the right upper corner.  Similarly, the threshold at the maximal
value of the discriminator classifies all points as belonging to the
``slow'' phase (as $S_d(n) \leq \max_n \{S_d(n)\}$) that corresponds
to the left lower corner.  The intermediate thresholds yield a concave
ROC curve lying above the diagonal.  The diamond symbols indicate the
threshold at the mean value $S_d$ (or $S_v$) that we suggest and use
in this paper.  One can see that this value is the closest to the left
upper corner and thus the optimal choice, at least for the considered
models and sets of parameters.

The choice of the threshold is related to another important question
about the choice of the LCH-based geometric property as the
discriminator.  The local convex hull captures changes in the mutual
arrangment of points, in a somewhat similar way as our eyes do.  The
diameter and the volume are the basic geometric characteristics of the
LCH that reflect, respectively, the overall size and anisotropy of
points.  Although these characteristics appear to be natural, one can
use any function of these (or other) characteristics as well.  It is
still unclear what is the optimal function of the local convex hull to
distinguish between two phases.  In other words, among all possible
functions of the LCH, which one would yield the highest recognition
score.  We expect that the optimal choice that maximizes the
recognition score, depends on the stochastic model of both phases.

\subsection{Phase durations}
\label{sec:phase_duration}

In all examples studied in Sec. \ref{sec:Models_to_test} (except for
surface-mediated diffusion), phase durations were chosen as
independent exponentially distributed random variables with equal mean
phase durations: $T_1 = T_2 = T$.  This means that all phases have
distinct but similar durations, whose average and standard deviation
are equal to $T$.

\begin{figure}
\begin{center}
\includegraphics[width=85mm]{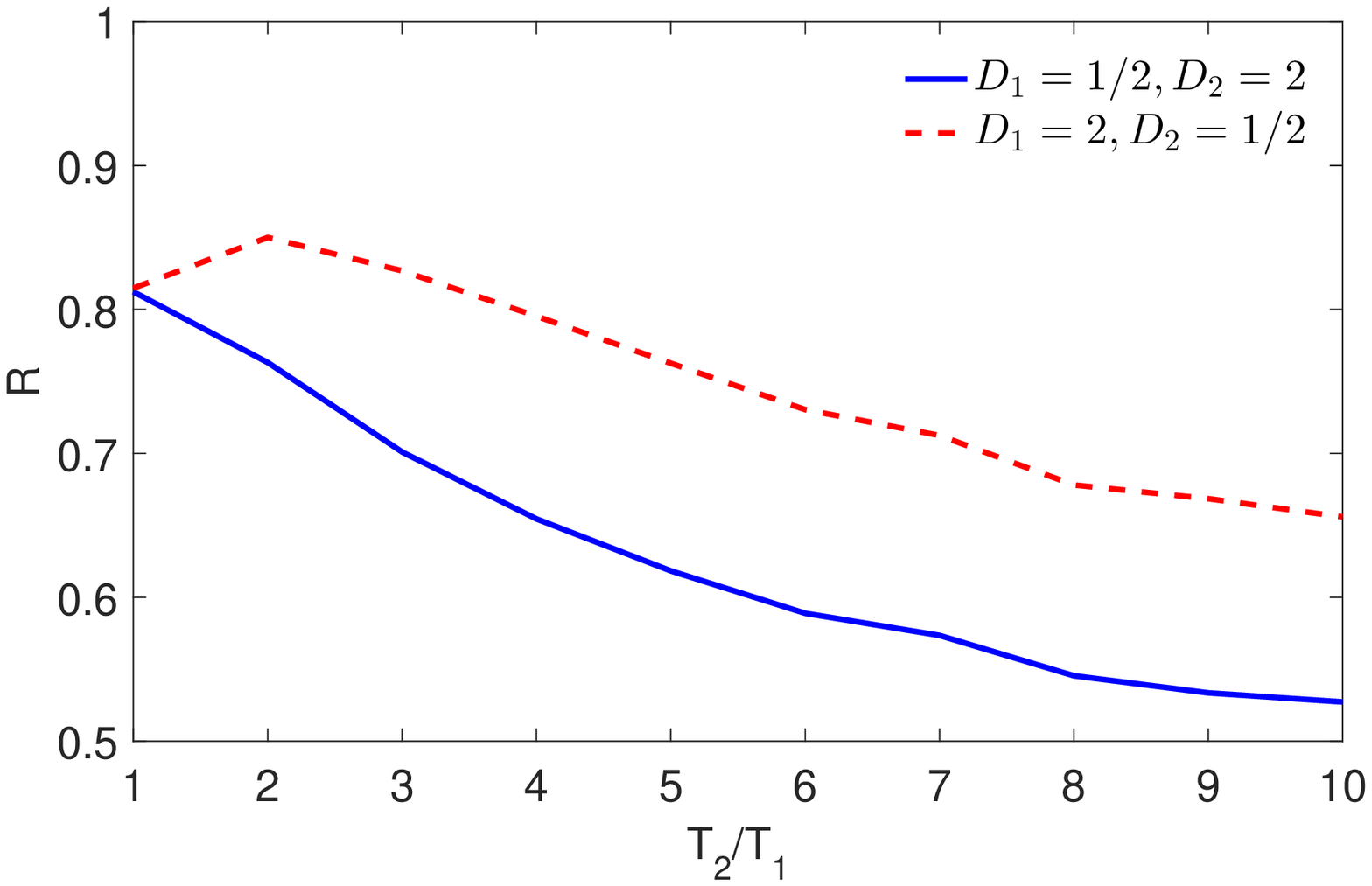} 
\includegraphics[width=85mm]{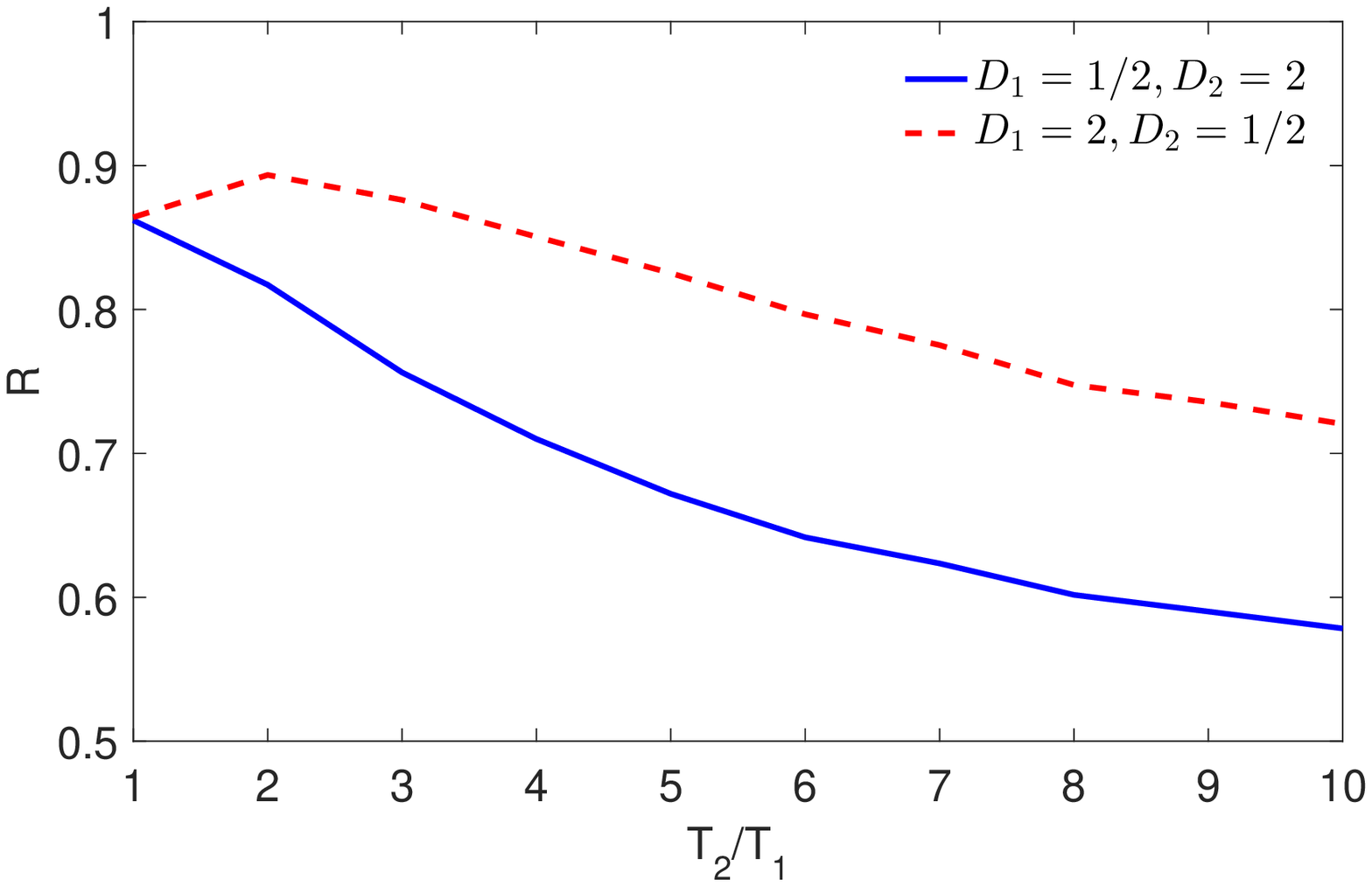} 
\end{center}
\caption{
(Color online) Effect of unequal phase durations for planar Brownian
motion alternating ``slow'' and ``fast'' phases (Model 1).
Recognition score $R$ of the diameter-based discriminator $S_d(n)$
{\bf (a)} and the volume-based discriminator $S_v(n)$ {\bf (b)} as a
function of the second phase duration $T_2$ while the first phase
duration $T_1$ is kept fixed: $T_1 = 40$.  Solid line: the first phase
is ``slow'' ($D_1 = 1/2$), the second phase is ``fast'' ($D_2 = 2$);
dashes line: the first phase is ``fast'' ($D_1 = 2$), the second phase
is ``slow'' ($D_2 = 1/2$). }
\label{unequal_duration}
\end{figure}

When two phases have significantly different durations, the detection
of the shorter phase can be problematic.  To illustrate this point, we
consider again the intermittent Brownian motion (Model 1), in which
one phase duration is kept fixed, whereas the other phase duration is
variable (in contrast to earlier figures, here the phase durations are
fixed, not exponentially distributed).  Figure \ref{unequal_duration}
shows the recognition score as a function of $T_2/T_1$, with fixed
$T_1 = 40$.  Solid line corresponds to the case, in which the first
phase is ``slow'' and the second phase is ``fast''.  As the duration
of the fast phase increases, its points start to dominate in the
geometric properties of the LCH, and thus to shift the threshold $S_d$
(or $S_v$) to higher values.  As a consequence, the detection of
shorter slow phases becomes more difficult, and the recognition score
decreases.  In the opposite case when the first phase is ``fast'' and
the second phase is ``slow'' (dashed line), the situation is slightly
different.  One can see that the recognition score first increases and
then decreases.  We conclude that short ``slow'' phases are on average
more difficult to detect than short ``fast'' phases.  In both cases,
the method is incapable of detecting the phases shorter than or
comparable to the window size $\tau$, which in turn should not be
smaller than $5-10$ steps.

If we assume that there are no phases shorter than some $T_0$
($\approx 10 - 20$), any shorter phase detected by the algorithm can
be attributed to a spontaneous crossing of $S_d(n)$ (or $S_v(n)$) of
the mean level.  The recognition quality can thus be improved by
re-classifying such inappropriate phases.  For instance, if a too
short slow phase is detected between two fast phases, the slow phase
can be re-classified into the fast one, i.e., these three consecutive
phases are merged and classified as a single fast phase.  This
post-processing correction is particularly important when one aims at
getting the statistics of slow and fast phase durations.  In fact, a
spurious short slow phase of the above example cut a long fast phase
into two shorter pieces and thus significantly affected the statistics
of phase durations.  For a better performance, such post-processing
corrections or more elaborate statistical techniques (such as an
estimation of likelihood of short phases) need to be elaborated and
tested on an application-specific basis.  Note also that even if the
corrected short phase was not spurious (and thus the correction was
erroneous), the correction may still improve the overall recognition
score, as it eliminates two false classifications related to the
delays.  We emphasize that this correction procedure has not been used
in this paper.

\section{Conclusion}

We introduced a new model-free local convex hull method for detecting
change points in a single-particle trajectory.  The LCH is constructed
on a small number of consecutive points of the trajectory and thus
reflects the dynamics {\it locally}.  In this way, the trajectory is
transformed into two time series, the diameter and the volume of the
LCH along the trajectory.  These time series are then used for a
binary classification into ``slow'' and ``fast'' phases.

The LCH method was validated on six common models of intermittent
processes: Brownian motion with two diffusivities, Brownian motion
with and without drift, fractional Brownian motion with different
Hurst exponents, Brownian motion with and without harmonic potential,
Brownian motion and exponential flights, and surface-mediated
diffusion.  For all these models, we computed the recognition score
$R$ as the fraction of successfully classified points.  We showed that
$R$ grows with the mean phase duration and reaches the values as high
as $90\%$ at $T = 100$ when two phases were quite distinct.  We
analyzed how distinct two phases should be for a robust
classification.  We also showed that, due to the integral-like
character of the LCH, the method is much more robust against noise
than conventional methods such as, e.g., time averaged mean square
displacement.  In particular, the TA MSD discriminator has
outperformed the LCH discriminators only in the case of two
alternating Brownian motions and only without noise.  In all other
cases, the performance of LCH estimators was much higher.  The LCH
method, being based on purely geometric features of the trajectory,
can be applied to a wide range of relevant intermittent processes such
as animal foraging, active/passive motion in the cell or motion
exhibiting a change in dimensionality.  The LCH method also has the
potential to be successfully applied to detect changes in
instantaneous firing rates in neurons.

\begin{acknowledgments}
DG acknowledges the support under Grant No. ANR-13-JSV5-0006-01 of the
French National Research Agency.
\end{acknowledgments}

\appendix
\section{Some properties of the convex hull}
\label{sec:Aproperties}

The convex hull of points of a random trajectory is a highly
nontrivial geometric functional of this trajectory.  Here we summarize
several rigorous results for both Brownian motion and simple random
walk.

Takacs showed that the mean perimeter of the convex hull of standard
Brownian motion $W_t$ in the plane is
\begin{equation}
\E\{\vol_1(\Conv(W_t))\} = \sqrt{16\pi D t} ,
\end{equation}
whereas El Bahir obtained the mean area
\cite{Takacs1980,Majumdar2010a}
\begin{equation}
\E\{\vol_2(\Conv(W_t))\} = \pi D t .
\end{equation}
More general results for a standard Brownian motion $W_t$ in $\R^d$
read \cite{Eldan2014}
\begin{eqnarray}
\label{eq:volConv}
\E\{\vol_d(\Conv(W_t))\} &=& \frac{(\pi D t)^{d/2}}{\Gamma(d/2+1)^2}, \\
\label{eq:areaConv}
\E\{\vol_{d-1}(\Conv(W_t))\} &=& \frac{2(4\pi D t)^{(d-1)/2}}{\Gamma(d)} . 
\end{eqnarray}
Extensions to a set of $n$ Brownian paths have been provided in
\cite{Majumdar2010a}.  The perimeter of the planar convex hull in the
case of confinement to a semi-infinite domain presents nontrivial
behavior as a function of distance from the starting point to the
boundary \cite{Chupeau2015a}.

These formulas are exact for continuous Brownian motion.  However,
experimental trajectories are discretized so that
Eqs. (\ref{eq:volConv}, \ref{eq:areaConv}) can only be valid
asymptotically when the number of points in the convex hull is large.
In turn, our LCH estimators are based on a relatively small number of
points, and Eqs. (\ref{eq:volConv}, \ref{eq:areaConv}) are thus not
applicable.  Spitzer and Widom considered a two-dimensional discrete
random walk, modeled as a sum of independent random variables in the
complex plane, and derived the average perimeter of the related convex
hull \cite{Spitzer1961},
\begin{equation}
\langle L_N\rangle=2\sum_{k=1}^N\frac{\E(\vert x_k+iy_k \vert)}{k}
\end{equation}
(here $x_k+iy_k$ is the position of the walker in the complex plane
after $k$ steps), with the concentration inequality \cite{Snyder93}
\begin{equation}
\P\{\vert L_N-\langle L_N\rangle\vert \geq \varepsilon\}\leq 2\exp\left(-\frac{\varepsilon^2}{8\pi^2N}\right).
\end{equation}
Some other properties were reported in \cite{Snyder93}.   The
asymptotic behavior of the mean perimeter and the mean area of the
convex hull over planar random walks was investigated in
\cite{Grebenkov17}.  The obtained formulas are even applicable to a
moderate number of jumps that makes them suitable for the analysis of
the local convex hulls.  The distributions of the area and of the
perimeter of the convex hull were computed numerically by a
sophisticated large-deviation approach in \cite{Claussen15}.  The
properties of the convex hull of Gaussian samples and of
$d$-dimensional fractional Brownian motion were analyzed in
\cite{Davydov2011a,Davydov2012}.


\begin{thebibliography}{50}


\bibitem{Tolic04}		I. M. Toli\'c-N{\o}rrelykke, E.-L. Munteanu, G. Thon, L. Oddershede, and K. Berg-S{\o}rensen,
				``Anomalous Diffusion in Living Yeast Cells'',
				Phys. Rev. Lett. {\bf 93}, 078102 (2004).

\bibitem{Golding06}		I. Golding and E. C. Cox,
				``Physical Nature of Bacterial Cytoplasm'',
				Phys. Rev. Lett. {\bf 96}, 098102 (2006).

\bibitem{Wilhelm08}		C. Wilhelm,
				``Out-of-Equilibrium Microrheology inside Living Cells'',
				Phys. Rev. Lett. {\bf 101}, 028101 (2008).

\bibitem{Szymanski09}		J. Szymanski and M. Weiss,
				``Elucidating the Origin of Anomalous Diffusion in Crowded Fluids'',
				Phys. Rev. Lett. {\bf 103}, 038102 (2009).

\bibitem{Metzler09}		R. Metzler, V. Tejedor, J.-H. Jeon, Y. He, W. H. Deng, S. Burov, and E. Barkai,
				``Analysis of Single Particle Trajectories: From Normal to Anomalous Diffusion'',
				Acta Phys. Pol. B {\bf 40}, 1315-1331 (2009).

\bibitem{Sackmann10}		E. Sackmann, F. Keber, and D. Heinrich,
				``Physics of Cellular Movements'',
				Ann. Rev. Conden. Matt. Phys. {\bf 1}, 257-276 (2010).

\bibitem{Jeon11}		J.-H. Jeon, V. Tejedor, S. Burov, E. Barkai, C. Selhuber-Unkel, K. Berg-S{\o}rensen, L. Oddershede, and R. Metzler,
				``In Vivo Anomalous Diffusion and Weak Ergodicity Breaking of Lipid Granules'',
				Phys. Rev. Lett. {\bf 106}, 048103 (2011).

\bibitem{Bertseva12}		E. Bertseva, D. S. Grebenkov, P. Schmidhauser, S. Gribkova, S. Jeney, and L. Forr\'o,
				``Optical Trapping Microrheology in Cultured Human Cells'',
				Eur. Phys. J. E {\bf 35}, 63 (2012).

\bibitem{Bressloff13}		P. C. Bressloff and J. M. Newby,
				``Stochastic models of intracellular transport'',
				Rev. Mod. Phys. {\bf 85}, 135-196 (2013).








\bibitem{Gal2013a} 		N. Gal, D. Lechtman-Goldstein, and D. Weihs,
				``Particle tracking in living cells: A review of the mean square displacement method and beyond'',
				Rheol. Acta {\bf 52}, 425-443 (2013).

\bibitem{Metzler14}		R. Metzler, J.-H. Jeon, A. Cherstvy, and E. Barkai,
				``Anomalous diffusion models and their properties: non-stationarity, non-ergodicity, 
				and ageing at the centenary of single particle tracking'',
				Phys. Chem. Chem. Phys. {\bf 16}, 24128-24164 (2014).

\bibitem{Kepten15}		E. Kepten, A. Weron, G. Sikora, K. Burnecki, and Y. Garini,
				``Guidelines for the Fitting of Anomalous Diffusion Mean Square Displacement Graphs from Single Particle Tracking Experiments'',
				PLoS ONE {\bf 10}, e0117722 (2015).



\bibitem{Condamin2008} 		S. Condamin, V. Tejedor, R. Voituriez, O. B\'enichou, and J. Klafter,
				``Probing microscopic origins of confined subdiffusion by first-passage observables'',
				Proc. Nat. Ac. Sci. USA {\bf 105}, 5675 (2008).

\bibitem{Kenwright2012} 	D. Kenwright, A. Harrison, T. Waigh, P. Woodman, and V. J. Allan,
				``First-passage-probability analysis of active transport in live cells'',
				Phys. Rev. E {\bf 86}, 031910 (2012).

\bibitem{Tejedor2010} 		V. Tejedor, O. B\'enichou, R. Voituriez, R. Jungmann, F. Simmel, C. Selhuber-unkel, L. B. Oddershede, and R. Metzler,
				``Quantitative analysis of single particle trajectories: mean maximal excursion method'',
				Biophys. J. {\bf 98}, 1364 (2010).

\bibitem{Thiel2013} 		F. Thiel, F. Flegel, and I. M. Sokolov,
				``Disentangling sources of anomalous diffusion'',
				Phys. Rev. Let. {\bf 111}, 010601 (2013).

\bibitem{Meroz2013} 		Y. Meroz, I. M. Sokolov, and J. Klafter,
				``Test for determining a subdiffusive model in ergodic systems from single trajectories'',
				Phys. Rev. Let. {\bf 110}, 090601 (2013).



\bibitem{Meroz2015} 		Y. Meroz and I. M. Sokolov,
				``A toolbox for determining subdiffusive mechanisms'',
				Phys. Rep. {\bf 573}, 1-29 (2015).


\bibitem{Burnecki2012d} 	K. Burnecki, E. Kepten, J. Janczura, I. Bronshtein, Y. Garini, and A. Weron,
				``Universal algorithm for identification of fractional Brownian motion. A case of telomere subdiffusion'',
				Biophys. J. {\bf 103}, 1839-1847 (2012).


\bibitem{Magdziarz2009} 	M. Magdziarz, A. Weron, K. Burnecki, and J. Klafter,
				``Fractional brownian motion versus the continuous-time random walk: A simple test for subdiffusive dynamics'',
				Phys. Rev. Let. {\bf 103}, 180602 (2009).

\bibitem{Weiss2013} 		M. Weiss,
				``Single-particle tracking data reveal anticorrelated fractional Brownian motion in crowded fluids'',
				Phys. Rev. E {\bf 88}, 010101 (2013).

\bibitem{Ernst2014} 		D. Ernst, J. K\"ohler, and M. Weiss,
				``Probing the type of anomalous diffusion with single-particle tracking'',
				Phys. Chem. Chem. Phys. {\bf 16}, 7686 (2014).



\bibitem{Magdziarz2011} 	M. Magdziarz and A. Weron,
				``Anomalous diffusion: Testing ergodicity breaking in experimental data'',
				Phys. Rev E {\bf 84}, 051138 (2011).

\bibitem{Lanoiselee2016} 	Y. Lanoisel\'ee and D. S. Grebenkov,
				``Revealing nonergodic dynamics in living cells from a single particle trajectory'',
				Phys. Rev. E {\bf 93}, 052146 (2016).


\bibitem{Chen2012} 		B. Chen and Y. Hong,
				``Testing for the Markov Property in Time Series'',
				Eco. Theo. {\bf 28}, 130-178 (2012).

\bibitem{Benichou14}		O. B\'enichou and R. Voituriez,
				``From first-passage times of random walks in confinement to geometry-controlled kinetics'',
				Phys. Rep. {\bf 539}, 225-284 (2014).





\bibitem{Berg1972} 		H. C. Berg and D. A. Brown,
				``Chemotaxis in Escherichia coli analysed by three-dimensional tracking'',
				Nature {\bf 239}, 500-504 (1972).

\bibitem{Berg}			H. C. Berg,
				{\it E. coli in Motion} 
				(Springer-Verlag, New York, 2004).


\bibitem{Tailleur2008} 		J. Tailleur and M. E. Cates,
				``Statistical mechanics of interacting run-and-tumble bacteria'',
				Phys. Rev. Let. {\bf 100}, 218103  (2008).

\bibitem{Benichou2011} 		O. B\'enichou, C. Loverdo, M. Moreau, and R. Voituriez,
				``Intermittent search strategies'',
				Rev. Mod. Phys. {\bf 83}, 81-109 (2011).







\bibitem{Richter74}		P. H. Richter and M. Eigen, 
				``Diffusion controlled reaction rates in spheroidal geometry: Application to repressor-operator 
				association and membrane bound enzymes'',
				Biophys. Chem. {\bf 2}, 255-263 (1974).

\bibitem{Berg81}		O. G. Berg, R. B. Winter, and P. H. von Hippel,
				``Diffusion-driven mechanisms of protein translocation on nucleic acids. 1. Models and theory'',
				Biochemistry {\bf 20}, 6929 (1981).

\bibitem{vonHippel89}		P. H. von Hippel and O. G. Berg,
				``Facilitated target location in biological systems'',
				J. Biol. Chem. {\bf 264}, 675-678 (1989).


\bibitem{Loverdo09}		C. Loverdo, O. B\'enichou, R. Voituriez, A. Biebricher, I. Bonnet, and P. Desbiolles,
				``Quantifying Hopping and Jumping in Facilitated Diffusion of DNA-Binding Proteins'',
				Phys. Rev. Lett. {\bf 102}, 188101 (2009).





\bibitem{Wong04}		I. Y. Wong, M. L. Gardel, D. R. Reichman, E. R. Weeks, M. T. Valentine, A. R. Bausch, and D. A. Weitz,
				``Anomalous Diffusion Probes Microstructure Dynamics of Entangled F-Actin Networks'',
				Phys. Rev. Lett. {\bf 92}, 178101 (2004).

\bibitem{Fodor16}		E. Fodor, H. Hayakawa, P. Visco, and F. van Wijland,
				``Active cage model of glassy dynamics'',
				Phys. Rev. E {\bf 94}, 012610 (2016).


\bibitem{Shoup82}		D. Shoup and A. Szabo,
				``Role of diffusion in ligand binding to macromolecules and cell-bound receptors'',
				Biophys. J. {\bf 40}, 33 (1982).

\bibitem{Zwanzig91}		R. Zwanzig and A. Szabo,
				``Time dependent rate of diffusion-influenced ligand binding to receptors on cell surfaces'',
				Biophys. J. {\bf 60}, 671 (1991).

\bibitem{Saxton96}		M. J. Saxton,
				``Anomalous diffusion due to binding: a Monte Carlo study'',
				Biophys. J. {\bf 70}, 1250-1262 (1996).

\bibitem{Bongrand99}		P. Bongrand,
				``Ligand-receptor interactions'',
				Rep. Prog. Phys. {\bf 62}, 921-968 (1999).

\bibitem{Michelman09}		A. Michelman-Ribeiro, D. Mazza, T. Rosales, T. J. Stasevich, H. Boukari, V. Rishi, C. Vinson, J. R. Knutson, and J. G. McNally,
				``Direct Measurement of Association and Dissociation Rates of DNA Binding in Live Cells'',
				Biophys. J. {\bf 97}, 337 (2009).

\bibitem{Torreno-Pina14}	J. A. Torreno-Pina, B. M. Castro, C. Manzo, S. I. Buschow, A. Cambi, and M. F. Garcia-Parajo,
				``Enhanced receptor-clathrin interactions induced by N-glycan-mediated membrane micropatterning'',
				Proc. Nat. Acad. Sci. USA {\bf 111}, 11037-11042 (2014).



\bibitem{Calderon10}		C. Calderon,
				``Detection of Subtle Dynamical Changes Induced by Unresolved 'Conformational Coordinates' 
				in Single-Molecule Trajectories via Goodness-of-Fit Tests'',
				J. Phys. Chem. B {\bf 114}, 3242-3253 (2010).

\bibitem{Galanti16}		M. Galanti, D. Fanelli, and F. Piazza,
				``Conformation-controlled binding kinetics of antibodies'',
				Scient. Rep. {\bf 6}, 18976 (2016).


\bibitem{Saxton94}		M. Saxton,
				``Single-particle tracking: models of directed transport'',
				Biophys. J. {\bf 67}, 2110-2119 (1994).

\bibitem{Caspi00}		A. Caspi, R. Granek, and M. Elbaum, 
				``Enhanced diffusion in active intracellular transport'', 
				Phys. Rev. Lett. {\bf 85}, 5655-5658 (2000).

\bibitem{Arcizet08}		D. Arcizet, B. Meier, E. Sackmann, J. O. R\"adler, and D. Heinrich,
				``Temporal Analysis of Active and Passive Transport in Living Cells'',
				Phys. Rev. Lett. {\bf 101}, 248103 (2008).

\bibitem{Brangwynne09}		C. P. Brangwynne, G. H. Koenderink, F. C. MacKintosh, and D. A. Weitz, 
				``Intracellular transport by active diffusion'',
				Trends Cell Biol. {\bf 19}, 423-427 (2009).

\bibitem{Katrukha2017}		E. A. Katrukha, M. Mikhaylova, H. X. van Brakel, P. M. van Bergen en Henegouwen, A. Akhmanova, 
				C. C. Hoogenraad, and L. C. Kapitein,
				``Probing cytoskeletal modulation of passive and active intracellular dynamics using nanobody-functionalized quantum dots'',
				Nat. Comm. {\bf 8}, 14772 (2017).


\bibitem{Levitz05}		P. Levitz,
				``Random flights in confining interfacial systems'',
				J. Phys. Condens. Matter {\bf 17}, S4059 (2005).

\bibitem{Benichou10} 		O. B\'enichou, D. S. Grebenkov, P. Levitz, C. Loverdo, and R. Voituriez, 
				``Optimal Reaction Time for Surface-Mediated Diffusion'',
				Phys. Rev. Lett. {\bf 105}, 150606 (2010).

\bibitem{Benichou11} 		O. B\'enichou, D. S. Grebenkov, P. Levitz, C. Loverdo, and R. Voituriez, 
				``Mean first-passage time of surface-mediated diffusion in spherical domains'',
				J. Stat. Phys. {\bf 142}, 657-685 (2011).


\bibitem{Rupprecht12} 		J.-F. Rupprecht, O. B\'enichou, D. S. Grebenkov, and R. Voituriez, 
				``Kinetics of active surface-mediated diffusion in spherically symmetric domains'',
				J. Stat. Phys. {\bf 147}, 891-918 (2012).

\bibitem{Rojo13}		F. Rojo, C. E. Budde Jr., H. S. Wio, and C. E. Budde,
				``Enhanced transport through desorption-mediated diffusion'',
				Phys. Rev. E {\bf 87}, 012115 (2013).







\bibitem{Adams2007} 		R. P. Adams and D. J. MacKay,  
				``Bayesian online changepoint detection'',
				ArXiv preprint arXiv:0710.3742 (2007).

\bibitem{Barber2011} 		D. Barber, A. T. Cemgil, and S. Chiappa,
				``Bayesian Time Series Models'',
				Cambridge University Press,  1-26 (2011).

\bibitem{Turkcan2013} 		S. T\"urkcan and J.-B. Masson,
				``Bayesian decision tree for the classification of the mode of motion in single-molecule trajectories'',
				PloS ONE {\bf 8}, 0082799 (2013).

\bibitem{Masson2014} 		J.-B. Masson, P. Dionne, C. Salvatico, M. Renner, C. G. Specht, A. Triller, and M. Dahan,
				``Mapping the energy and diffusion landscapes of membrane proteins at the cell surface using 
				high-density single-molecule imaging and Bayesian inference: Application to the multiscale 
				dynamics of glycine receptors in the neuronal membrane'',
				Biophys. J. {\bf 106}, 74-83  (2014).

\bibitem{Bosch2014} 		P. J. Bosch, J. S. Kanger, and V. Subramaniam,
				``Classification of dynamical diffusion states in single molecule tracking microscopy'',
				Biophys. J. {\bf 107}, 588-598 (2014).

\bibitem{Ruggieri2016} 		E. Ruggieri and M. Antonellis,
				``An exact approach to Bayesian sequential change point detection'',
				Comput. Stat. Data Anal. {\bf 97}, 71-86 (2016).

\bibitem{Hinsen2016} 		K. Hinsen and G. R. Kneller,
				``Communication: A multiscale Bayesian inference approach to analyzing subdiffusion in particle trajectories'',
				J. Chem. Phys. {\bf 145}, 151101 (2016).






\bibitem{Grebenkov2011}		D. S. Grebenkov,
				``Time-averaged quadratic functionals of a Gaussian process'',
				Phys. Rev. E {\bf 83}, 061117 (2011).




\bibitem{ORourke}		J. O'Rourke,
				{\it Computational Geometry in C},
				2nd Ed. (Cambridge Tracts in Theoretical Computer Science,
				Cambridge University Press, Cambridge, 1998).

\bibitem{deBerg}		M. de Berg, O. Cheong, M. van Kreveld, and M. Overmars,
				{\it Computational Geometry: Algorithms and Applications},
				Third ed. (Springer-Verlag, Berlin-Heidelberg, 2008).




\bibitem{Barber96}		C. B. Barber, D. P. Dobkin, and H. T. Huhdanpaa, 
				``The Quickhull Algorithm for Convex Hulls'',
				ACM Trans. Math. Software {\bf 22}, 469-483 (1996).




\bibitem{Worton1995} 		B. J. Worton,
				``A convex hull based estimator of homerange size'',
				Biometrics {\bf 51}, 1206-1215 (1995).

\bibitem{Getz2004a} 		W. M. Getz and C. C. Wilmers,
				``A local nearest neighbor convex-hull construction of home ranges and utilization distribution'',
				Ecography {\bf 27}, 03835 (2004).

\bibitem{Randon-Furling2009a} 	J. Randon-Furling, S. N. Majumdar, and A. Comtet,
				``Convex hull of N planar Brownian motions: Exact results and an application to ecology'',
				Phys. Rev. Let. {\bf 103}, 140602 (2009).

\bibitem{Normant1991a} 		F. Normant and C. Tricot,
				``Method for evaluating the fractal dimension of curves using convex hulls'',
				Phys. Rev. A {\bf 43}, 6518-6525 (1991).




\bibitem{Bidaux99}		R. Bidaux, J. Chave, and R. Vocka,
				``Finite time and asymptotic behaviour of the maximal excursion of a random walk'',
				J. Phys. A: Math. Gen. {\bf 32}, 5009-5016 (1999).




\bibitem{Voisinne10}		G. Voisinne, A. Alexandrou, and J.-B. Masson,
				Biophys. J. {\bf 98}, 596 (2010).

\bibitem{Berglund10}		A. J. Berglund,
				Phys. Rev. E {\bf 82}, 011917 (2010).









\bibitem{Mandelbrot1968} 	B. B. Mandelbrot and J. W. Van Ness,
				``Fractional Brownian motions, fractional noises and applications'',
				SIAM Rev. {\bf 10} 422-437 (1968).


\bibitem{Grebenkov13}		D. S. Grebenkov, M. Vahabi, E. Bertseva, L. Forro, and S. Jeney,
				``Hydrodynamic and subdiffusive motion of tracers in a viscoelastic medium'',
				Phys. Rev. E {\bf 88}, 040701R (2013).

\bibitem{Desposito11}		M. A. Desposito, C. Pallavicini, V. Levi, and L. Bruno,
				``Active transport in complex media: Relationship between persistence and superdiffusion'',
				Physica A {\bf 390}, 1026-1032 (2011).


\bibitem{Doi}			M. Doi and S. F. Edwards, 
				{\it The Theory of Polymer Dynamics}
				(Clarendon Press, 1986).

\bibitem{deGennes}		P.-G. de Gennes,
				{\it Introduction to Polymer Dynamics}
				(Cambridge University Press, Cambridge, 1990).


\bibitem{Kuo93}			S. C. Kuo and M. P. Sheetz,
				``Force of single kinesin molecules measured with optical tweezers'',
				Science {\bf 260}, 232-234 (1993).

\bibitem{Wirtz09}		D. Wirtz, 
				``Particle-Tracking Microrheology of Living Cells: Principles and Applications'', 
				Ann. Rev. Biophys. {\bf 38}, 301-326 (2009).

\bibitem{Grebenkov15}		D. S. Grebenkov,
				``First exit times of harmonically trapped particles: a didactic review'',
				J. Phys. A {\bf 48}, 013001 (2015).



\bibitem{Amit97a}		D. J. Amit and N. Brunel,
				``Model of global spontaneous activity and local structured activity during delay periods in the cerebral cortex'',
				Cereb Cortex. {\bf 7}, 237-252 (1997).

\bibitem{Amit97b}		D. J. Amit and N. Brunel,
				``Dynamics of a recurrent network of spiking neurons before and following learning'',
				Network: Comput. Neural Syst. {\bf 8}, 373-404 (1997).


\bibitem{Mazzucato15}		L. Mazzucato, A. Fontanini, and G. La Camera,
				``Dynamics of multistable states during ongoing and evoked cortical activity'',
				J. Neurosci. {\bf 35}, 8214-8231 (2015). 



\bibitem{Takacs1980} 		L. Tak\'acs,
				``Expected perimeter length'',
				Amer. Math. Month. {\bf 87}, 5-6 (1980).

\bibitem{Majumdar2010a} 	S. N. Majumdar, A. Comtet, and J. Randon-Furling,
				``Random Convex Hulls and Extreme Value Statistics'',
				J. Stat. Phys. {\bf 138}, 1-55 (2010).

\bibitem{Eldan2014} 		R. Eldan,
				``Volumetric properties of the convex hull of an n-dimensional Brownian motion'',
				Elec. Jour. Prob. {\bf 19}, 1-34  (2014).

\bibitem{Chupeau2015a} 		M. Chupeau, O. B\'enichou, and S. N. Majumdar,
				``Convex hull of a Brownian motion in confinement'',
				Phys. Rev. E {\bf 91}, 050104 (2015).


\bibitem{Spitzer1961} 		F. Spitzer and H. Widom,
				``The circumference of a convex polygon'',
				Proc. Amer. Math. Soc. {\bf 12}, 506-509 (1961).


\bibitem{Grebenkov17}		D. S. Grebenkov, Y. Lanoisel\'ee, and S. N. Majumdar,
                                ``Mean perimeter and mean area of the convex hull over planar random walks''
				(accepted to J. Stat. Mech.; see online at ArXiv1706.08052).



\bibitem{Claussen15}		G. Claussen, A. K. Hartmann, and S. N. Majumdar,
				``Convex hulls of random walks: Large-deviation properties'',
				Phys. Rev. E {\bf 91}, 052104 (2015).

\bibitem{Snyder93} 		T. L. Snyder and J. M. Steele,
				``Convex hulls of random walks'',
				Proc. Am. Math. Soc. {\bf 117}, 1165-1173 (1993).


\bibitem{Davydov2011a} 		Y. Davydov,
				``On convex hull of Gaussian samples'',
				Lit. Math. Jour. {\bf 51}, 171-179 (2011).

\bibitem{Davydov2012} 		Y. Davydov,
				``On convex hull of d-dimensional fractional Brownian motion'',
				Stat. Prob. Let. {\bf 82}, 37-39 (2012).


\end{thebibliography}
\end{document}